\documentclass[a4paper,notitlepage,showpacs,superscriptaddress,nofootinbib,10pt]{revtex4-1}
\usepackage{amsmath,amssymb,bm,natbib}
\usepackage{epsfig}
\usepackage{epstopdf}
\usepackage{graphicx}
\usepackage{multirow}
\usepackage{slashed}
\usepackage{color}
\usepackage{soul}
\makeatletter

\newcommand{\Rmnum}[1]{\expandafter\@slowromancap\romannumeral #1@}
\makeatother
\begin{document}
\def\bib{\bibitem}
\def\be{\begin{equation}}
\def\ee{\end{equation}}
\def\beq{\begin{equation}}
\def\eeq{\end{equation}}
\def\beqar{\begin{eqnarray}}
\def\eeqar{\end{eqnarray}}
\def\barr{\begin{array}}
\def\earr{\end{array}}
\def\lsim{\:\raisebox{-0.5ex}{$\stackrel{\textstyle<}{\sim}$}\:}
\def\gsim{\:\raisebox{-0.5ex}{$\stackrel{\textstyle>}{\sim}$}\:}
\def\tilh{\tilde{h}}
\def\and{\qquad {\rm and } \qquad}
\def\vev{{\it v.e.v. }}
\def\p{\partial}
\def\ga{\gamma^\mu}
\def\slp{p \hspace{-1ex}/}
\def\sleps{ \epsilon \hspace{-1ex}/}
\def\slk{k \hspace{-1ex}/}
\def\slq{q \hspace{-1ex}/\:}
\def\prl#1{Phys. Rev. Lett. {\bf #1}}
\def\prd#1{Phys. Rev. {\bf D#1}}
\def\plb#1{Phys. Lett. {\bf B#1}}
\def\npb#1{Nucl. Phys. {\bf B#1}}
\def\mpl#1{Mod. Phys. Lett. {\bf A#1}}
\def\ijmp#1{Int. J. Mod. Phys. {\bf A#1}}
\def\zp#1{Z. Phys. {\bf C#1}}
\def\etal{ {\it et al.} }
\def\ie{ {\it i.e.} }
\def\eg{ {\it e.g.} }
\def\sbar{ \overline{s} }
\def\thmin{\theta_0}
\def\cmin{\cos \theta_0}
\def\pepebar{P_{e} P_{\overline{e}}}
\def\eebar{$e^+e^-~$}
\def\eegz{$e^+e^- \to \gamma Z$}
\def\ggz{$\gamma\gamma Z~$}
\def\gzz{$\gamma ZZ~$}
%\begin{document}
%\thispagestyle{empty}
%\setcounter{page}{0}
%\renewcommand{\thefootnote}{\fnsymbol{footnote}}

\title{Top Yukawa coupling measurement with indefinite CP Higgs in
$e^+e^-\to t\bar{t}\Phi$
}
\author{{\bf B. Ananthanarayan}}
\email{anant@cts.iisc.ernet.in}
\affiliation{
Centre for High Energy Physics, 
Indian Institute of Science, 
Bangalore 560 012, India} 
\author{{\bf Sumit K. Garg}}
\email{aslv15@gmail.com}
\affiliation{
Department of Physics and IPAP, Yonsei University,
 Seoul 120-749, Korea}
\author{{\bf C. S. Kim}}
\email{cskim@yonsei.ac.kr} 
\affiliation{
Department of Physics and IPAP, Yonsei University,
 Seoul 120-749, Korea}
\author{{\bf Jayita Lahiri}}
\email{jayita@cts.iisc.ernet.in} 
\affiliation{
Centre for High Energy Physics, 
Indian Institute of Science, 
Bangalore 560 012, India}
\author{{\bf P. Poulose}}
\affiliation{ 
Department of Physics,
Indian Institute of Technology Guwahati,
Guwahati 781 039, India} 
\email{poulose@iitg.ernet.in}

%%ALIAS=ILC2=arXiv:0709.1893%%
%%ALIAS=review=hep-ph/0507011%%

\begin{abstract}

We consider the issue of the top quark Yukawa coupling measurement in a model independent and general case with the inclusion of CP-violation 
in the coupling. Arguably the best process to study this coupling is the associated production of Higgs boson along with a $t\bar t$ pair in 
a machine like the International Linear Collider (ILC). While detailed analyses of the sensitivity of the measurement assuming a Standard 
Model (SM) - like coupling are available in the context of ILC, conclude that the coupling could be pinned down at about 10\% level with modest 
luminosity, our investigations show that the scenario could be different in case of a more general coupling. The modified Lorentz structure 
resulting in a changed functional dependence of the cross section on the coupling, along with the difference in the cross section itself leads to 
considerable deviation in the sensitivity. Our studies with an ILC of center of mass energies of 500 GeV, 800 GeV and 1000 GeV show that 
moderate CP-mixing in the Higgs sector could change the sensitivity to about 20\%, while it could be worsened to 75\% in cases which could 
accommodate more dramatic changes in the coupling. While detailed considerations of the decay distributions point to a need for a relook at the analysis 
strategy followed for the case of SM such as for a model independent analysis of the top quark Yukawa coupling measurement. This study strongly suggests that, a joint analysis of the CP properties and the Yukawa coupling measurement would be the way forward at the ILC and that caution must be excercised in the measurement of the Yukawa couplings and the conclusions drawn from it.

\end{abstract}

\maketitle

%\tableofcontents
%%%%%%%%%%%%%%%%%%%%%%%%%%
%%%%%%%%%%%%%%%%%%%%%%%%%
\section{Introduction}

The recent discovery of a Higgs-like particle by the Large Hadron Collider (LHC), weighing about  125\,GeV$/c^2$ \cite{cms,atlas,Moroind1,Moroind2,Moroind3,Moroind4,Moroind5}, 
quite positively indicates that the Higgs mechanism is at work to bring in electroweak symmetry breaking, thus providing mass to the elementary particles. While one of 
the parameters, namely the mass of the new particle, is somewhat precisely determined in different  detection channels, and by two independent experiments, one need to go a 
long way before establishing the full identity of this particle, in terms of its couplings to other particles, as well as in terms of its constitution. At the same time, the 
reasons to look beyond the SM will not be diminished, even if the new resonance has all the properties as expected within the SM. For example, concerning the Higgs sector, 
one will still need to cure the quadratically divergent quantum corrections to the mass of the standard Higgs boson. The other reasons include, existence of the dark 
matter, the existence of neutrino masses, baryon asymmetry, etc, which clearly indicates the need to look beyond the SM. Indeed, the newly discovered particle, with 
the understanding of its properties will provide the much needed handle in the search beyond the SM. 
%in our opinion, the era of serious search for signals beyond the SM has only begun.
With the limitations of a hadronic machine, the LHC may not be able to exhaustively
study the properties of the new resonance\footnote{Refer to \cite{Maltoni:2002jr,Ellis:2013yxa,Chang:2014rfa,Nishiwaki:2013cma} for some of the studies of top-Higgs Yukawa couplings
in the context of LHC.}. On the other hand, the clean environment of the proposed International Linear Collider (ILC) ~\cite{ILC1,ILC2}, which
is an $e^{+}e^{-}$ collider, will help carry out precision experiments
on elementary particles and establish their properties, including that of the purported Higgs Boson, which here we shall denote by $\Phi$.  The possibility of beam 
polarization could significantly enhance the sensitivity of ILC in general, and also to probe beyond the SM signals~\cite{polarizationreview}.

\begin{figure}[h]
\begin{center}
\includegraphics[width=0.45\textwidth]{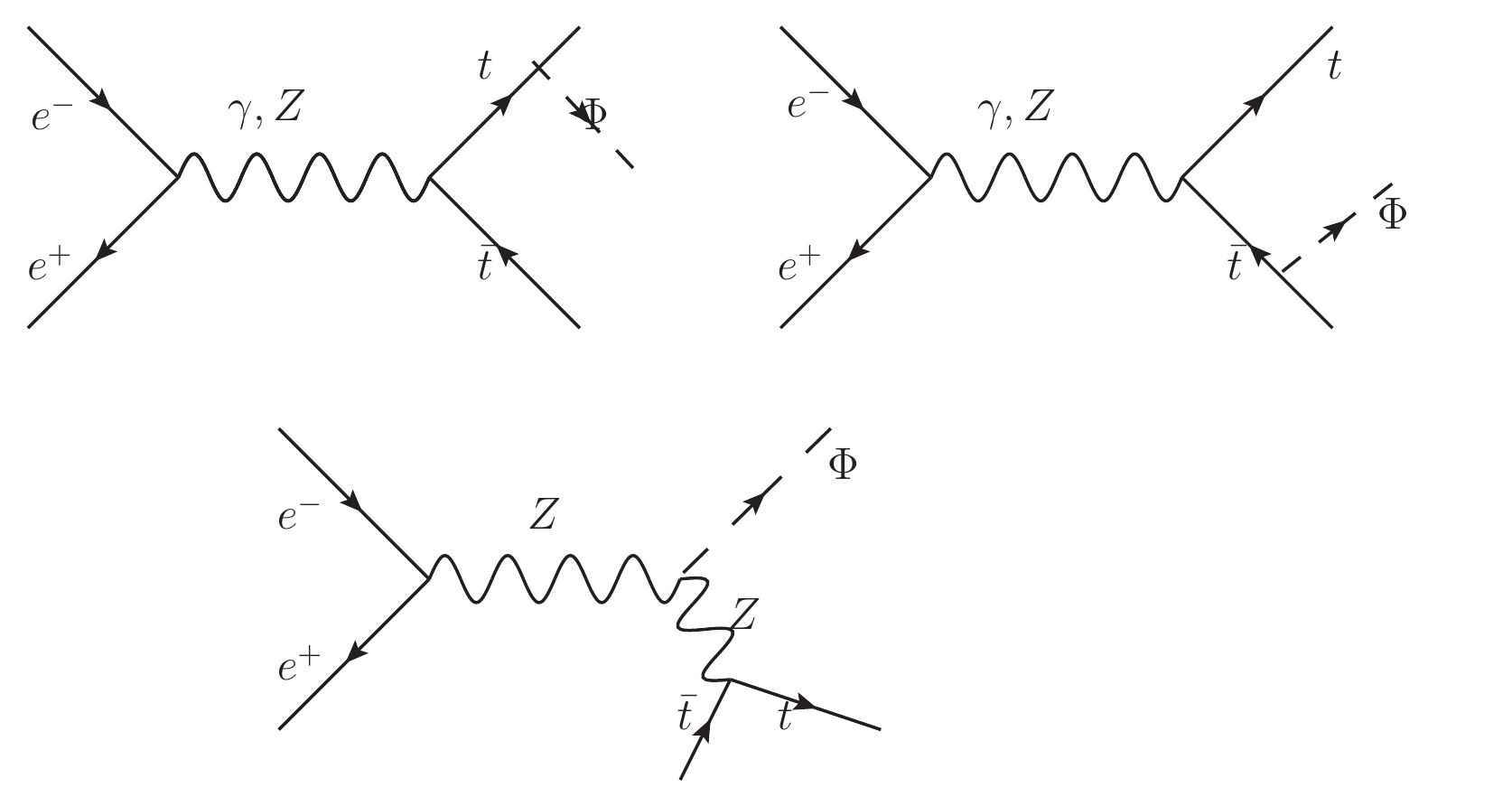}
\end{center}
\caption{Feynman diagrams contributing to  the process $e^{-}e^+ \rightarrow t\bar{t}\Phi$ in Standard Model. }
\label{fdiag}
\end{figure}

The process that is of interest to us in this work is:
\begin{equation}\label{ourprocess}
e^+e^-\to t \bar{t}\Phi.
\end{equation}
 This process is, by definition, key to the 
measurement of the top quark Yukawa coupling to the Higgs. The Feynman diagrams, at tree level, corresponding  to this process are given in Fig.~\ref{fdiag}. This include 
the Higgs-strahlung process (with $ZZ\Phi$ coupling),  the contribution of which to the total cross section is about 5\%.
We invite the attention of the reader to Ref.~\cite{Baer:1999ge,AG1,AG2}
and references therein for some early work on this topic.  More recently, the process
has attracted renewed attention in Refs.~\cite{KS1,KS2}, which present a detailed analysis of the background involved and the feasibility of ILC to measure the top 
Yukawa coupling, concluding that the ILC at 800 GeV center of mass energy is best suited for such measurement.  The issue of threshold corrections due to $t\bar t$ bound state 
effects, more relevant at center of mass(c.m.) energy close to 500\,GeV, are considered in some detail in Ref.~\cite{YIUF,YITFKSY}, which also carried out detailed and sophisticated 
analysis, including considerations of the signal and backgrounds in various channels, with the help of prototype ILC event generators, and making use of the beam polarization. 
Restricted within the SM, these studies have established that the top quark Yukawa coupling could be measured with an accuracy of about 10\%. The direct measurement of the 
coupling, using the semi-leptonic decays of the $W$'s coming from the top quarks is considered in Ref.~\cite{TM}, which concludes that the accuracy of the 
determination of the coupling is about 30\%. 
The guiding relation between the limit with which Yukawa coupling ($g_t^M$) can be measured,  and that of the cross section ($\sigma$)is \cite{LCtopyuk1000},
\begin{equation}
\frac{\Delta g_t^M}{g_t^M}=\left(\frac{\sigma/g_t}{\left| d\sigma/dg_t\right|}\right)_{g_t=g_t^M}~\frac{\Delta \sigma}{\sigma}.
\label{eq:delta_g}
\end{equation} 
where, $g_t$ is considered as a variable, and $g_t^{\rm M}$ is the top yukawa coupling for the model under consideration, which
is to be evaluated at top mass, $m_t=173.5$ GeV. It may be noted that, while $\Delta \sigma$ depends on the details of the experimental efficiency of isolating the signal 
over the background, apart from detection efficiencies relevant to the final states involved, the pre-factor, $\left(\frac{\sigma/g_t}{\left| d\sigma/dg_t\right|}\right)_{g_t=g_t^M}$ 
depends crucially on how the cross section depends on $g_t$. It is easily imagined in a beyond-the-SM (BSM) scenario with multiple Higgs fields participating in electroweak symmetry 
breaking (EWSB), to have the functional dependence of the cross section on the Yukawa coupling different from that in the SM \cite{Han:1999xd}. For example, cases with CP-mixed Higgs bosons will have a $t\bar{t}\Phi$ 
coupling with different Lorentz structure than that of the SM, which can complicate the dependence \footnote{For an early review, please see \cite{Atwood:2000tu}.}. 
 {\it A priori}, 
 it is not clear what kind of differences are expected in the prefactor in the presence of anomalous couplings, compared to the SM case. Neither is it clear, if the signal 
 significance remains the same in both cases. It is the purpose of this work to attempt to answer this question, without getting into the detailed analysis. 

While, it has been reported by the LHC collaborations, that the new resonance is very  likely a scalar, the spin and parity studies so far are very limited. Investigating the case of a scalar-pseudoscalar admixture should involve a different strategy that what has been adopted so far. Most of the multi-Higgs models, including the two 
Higgs doublet model (2HDM), and the Minimal, and Next-to Minimal Superpsymmetric Standard Models (MSSM and NMSSM) are some of the popular extensions of the SM with CP-odd as 
well as CP-even scalar particles in their CP-conserving versions. Inclusion of CP violation in the Higgs sector of such models allow CP-mixed physical scalars.  
%To our knowledge, in the Yukawa coupling measurement studies, beam polarization has not received sufficient attention. (Please check paper of Juste, plen0043, ALCPG workshop.)
Phenomenology of such possibility has been considered in the literature \cite{2HDMcpv, MSSMcpv,MSSMcpvPP,Carena:2001bg}. In a recent study ~\cite{DDGMR} it has been pointed out that the ILC 
is  an ideal setting to probe the CP nature of the Higgs Boson in the process considered here. It is clear that the scalar and pseudo-scalar parts of a CP-mixed Higgs Boson 
will couple differently to different polarization combinations of the top quark and top antiquark. Suitable observables involving top quark polarization, such as polarization 
asymmetry, could therefore probe the CP-nature of the Higgs boson through this process. With the combined use of total cross section, its energy dependence, 
the polarization asymmetry of the top quark, and the up-down asymmetry of the antitop with respect to the top quark - electron plane,  Ref.~\cite{GHMRS} has shown that the 
CP properties can be efficiently probed through the same process of $t\bar{t}\Phi$ production at ILC. The other process being scrutinized to investigate the CP nature of the Higgs 
is  $e^+e^-\to Z \Phi$, with the Higgs boson decaying to $\tau$ lepton pairs~\cite{BBS,HJ,RR}. A general case of model independent effective anomalous couplings is studied by 
Ref.~\cite{AguilarSaavedra}.

In our recent work~\cite{Ananthanarayan:2013cia}, we have studied
the possibility of fingerprinting the departure from the CP-even case in decay distributions of 
the process in Eq.(\ref{ourprocess}). In this study,  two definite scenarios, 
which we denote as Model I and Model II, have been considered.  
Model I corresponds to the minimal extension of the 
SM with one additional pseudo-scalar 
degree of freedom, which mixes with the SM scalar to form the physical Higgs 
Boson~\cite{DDGMR}. This model is  characterized by one free 
parameter, and the $t\bar t$ coupling to scalar and pseudo scalar are related by a sum rule that the sum of the squares of the couplings is the same as that 
of the square of the SM coupling. Similarly, the $ZZ\Phi$ coupling is scaled by the same parameter that scales the scalar coupling to $t\bar t$, as the gauge bosons 
do not couple to pseudo scalars  at tree level. Model II is a more realistic case, similar to the CP-violating 2HDM model, which has some 
essential features that make it quite different from Model I. In particular, there is no theoretical constraint (sum rule) on the parameters, and the scaling of the 
$ZZ\Phi$ coupling is not necessarily scaled by the same parameter that scales the scalar coupling to $t\bar t$, which is true in general in multi-Higgs models with more 
than one Higgs field transforming non-trivially under $SU(2)_L$. Confining ourselves to some reasonable ranges of these parameters guided by the experimental indications 
that the new resonance is close to a CP-even case, we considered the effect of CP-violating Higgs Boson 
in the decay spectrum of both the top quark as well as the Higgs boson itself, noting that the decay distributions are the
spin analyzers of the parent particle. In view of the remarks above, it is a natural extension of this work to understand the impact of such CP-indefinite Higgs 
boson on the top quark Yukawa coupling measurement. In the present work we address this issue. In order to perform numerical analysis, 
we have used the integrated Monte-Carlo and event generation package {\sc WHIZARD}~\cite{WHIZARD}, which incorporates the SM, as well 
as some of its popular extensions. Further,{\sc WHIZARD} also allows to incorporate any new model described by a Lagrangian, through an interface \cite{Interface} 
generated using {\tt FeynRules}~\cite{FeynRules}. In particular, for our analysis, we have considered a suitably modified version of the model including the anomalous 
$tt\Phi$ coupling as prescribed by Model I and Model II mentioned above.

The scheme of this paper is as follows.  In Section~\ref{formalism}  we first
introduce and describe the basic structure of an indefinite CP Higgs
sector in the two scenarios mentioned above. In Section~\ref{process} we describe the processes we consider.  
In Section~\ref{numresults} we present the results of our analysis.  In Section~\ref{discussion}
we present a discussion and our conclusions.

\section{Top Yukawa Coupling}\label{formalism}
%\section{CP-mixed Higgs}\label{cpmixed_higgs}

After the discovery of Higgs boson studying its properties in much detail
is the foremost task which will further establish Higgs symmetry breaking mechanism. The Standard Model contains one
Higgs doublet which by acquiring vacuum expectation value (vev) results in a CP-even physical scalar field.   The heaviness of the top quark makes it possible to study the fermion-Higgs interactions at the production level, augmenting to the possibilities through the fermionic 
decay channels.
In SM the strength of Yukawa couplings to fermions at tree level is given by 
\begin{equation}
 g_{f\bar{f}\Phi} = M_f/v
\end{equation}
where $v = 246$GeV is the vacuum expectation value and $M_f$ is the mass of fermion. For a top quark of mass $M_t = 174$GeV, the Yukawa
coupling is given by $y_{t\bar{t} \Phi}=0.71$. QCD and weak corrections to this coupling are estimated 
to be about 10\%. While this is so, there are compelling reasons to look beyond the SM, as indicated in the Introduction. Many of the scenarios proposed to go beyond the SM have more complex Higgs sector, offering possibilities of more than one physical Higgs bosons. In the 
CP-conserving scenarios, some of these are scalar particles, and some are pseudoscalar particles, while in the CP-violating cases, the scalar-pseudoscalar mixing could result in 
CP-indefinite Higgs bosons. For example, in one of the minimal extensions of the Higgs sector beyond the SM, an additional $SU(2)_L$ doublet field is introduced in the Two-Higgs Doublet Model (2HDM), resulting in five physical Higgs bosons, 
one of which is CP-odd neutral particle. In such multi-Higgs scenarios, the Yukawa coupling is in general expected to be different from that of the SM case. While in the CP-conserving cases, there could maximally be a scaling of the Yukawa 
coupling, compared to its SM value, in the CP-violating cases, there is a different 
Lorentz structure involved in the coupling. Such a CP-indefinite state will also couple differently to the gauge bosons. Concerning the present study,  we shall focus on the $t\bar t\Phi$ production at ILC, relevant to which,  both the 
$t\bar t \Phi$ as well as $ZZ\Phi$ couplings take a form,
which may be parametrized as follows~\cite{AguilarSaavedra, DDGMR}.
\begin{eqnarray}
g_{t\bar{t} \Phi}&=&-i \frac{e}{s_W} \frac{m_t}{2 M_W} (a+i b \gamma_5) \nonumber \\
g^{\mu\nu}_{ZZ\Phi}&=&-ic \frac{e M_Z}{s_W c_W} g^{\mu\nu}.
\label{eq:generic_coupl_t}
\end{eqnarray}
Here,
$s_W\equiv \sin\theta_W(=\sqrt{1-c_{W}^{2}})$, where $\theta_W$ is the Weinberg angle.  
In the SM with only one scalar Higgs Boson ($h$), the parameters take values $a=1$, 
$b=0$ and $c=1$.  In specific models, all these parameters may not be independent. In the following we will describe two simple scenarios that are phenomenologically viable, and can be linked to some of the popular extensions of the SM.

\subsection{Model I}
One of the most simple and straightforward extensions of the Higgs sector of the SM to incorporate CP violation is to imagine the presence of an additional pseudoscalar degree of freedom ($A$), which mixes with the scalar degree of freedom to produce a physical states:

\begin{equation}
\left(\begin{array}{c}
\Phi \\ \chi \end{array}\right) = 
\left(\begin{array}{cc}\cos\theta &\sin\theta\\ \sin\theta&\cos\theta\end{array}\right)
\left(\begin{array}{c}h\\A\end{array}\right).
\end{equation}
Considering $\Phi$ as the 125 GeV resonance\footnote{ The orthogonal combination,$\chi$ could be imagined to be heavy, and therore does not interfere with the phenomenology at the energy scales we consider.}, being under investigation here, the parameters $a$ and $b$ above are identified as $a=\cos\theta$ and $b=\sin\theta$, which are now constrained  by
\begin{equation}
a^2+b^2=1.\label{model1constr}
\end{equation}
Since the SM gauge Boson, $Z$ does not couple to the pseudo-scalar degree
of freedom, we have $c=a$ in this scenario. The down-type quarks as well as the charged
leptons will also have the same coupling structure as that of the up-type quarks, so
that, for example, the $b$-quark coupling becomes
\begin{eqnarray}
g_{\Phi bb}&=&-i \frac{e}{s_W} \frac{m_b}{2 M_W} (a+i b \gamma_5).
\label{eq:generic_coupl_b}
\end{eqnarray}
In the following, we call this scenario as Model I. The advantage here is that there is only one additional parameter to deal with, making it very friendly to perfom phenomenological investigations. The disadvantage is that, most of the realistic extensions of the SM with multi-Higgs scenarios have more complex Higgs sector, which do not support the above constraint. 

\subsection{Model II}

In more realistic extentions of the SM, the Higgs sector allows  a more relaxed assignment of the parameter values compared to the scenario in Model I above. 

In a  completely model independent approach we can treat parameters $a, b$ to be independent of each other. 
Some specific cases of this scenario, are the 2HDM and  MSSM 
with CP violation where there are two Higgs doublet fields, leading to two neutral scalar bosons and one neutral pseudoscalar boson 
which could mix with each other.  Thus the physical mass eigenstates are given as:
\begin{equation}
\left(\begin{array}{c} \phi_1\\[1mm]  \phi_2\\[1mm] A\end{array}\right)=
{\cal O}_{3\times 3}~
\left(\begin{array}{c}H_1\\[1mm] H_2\\[1mm] H_3\end{array}\right),
\end{equation}
where $\phi_1$ and $\phi_2$ are the scalar gauge eigenstates, and  $A$ is the pseudoscalar gauge eigenstate \cite{CPsuperH}.
This, in effect, removes the restricting relations between the parameters $a,~~b$ and $c$.
For ready reference we take the example of MSSM case (or 2HDM) with CP violation in the Higgs
sector. The couplings of the Higgs Boson with the 
fermions and the gauge Bosons, where $\tan\beta$ is the ratio of the vev's of the two Higgs 
fields are given by:
\begin{eqnarray}
{\rm top~quark:}&&~~a_u=~{\cal O}_{2i}~/~\sin\beta,~~~~~~~b_u=-{\cal O}_{3i}~\cot\beta\nonumber\\
{\rm bottom~ quark / \tau -lepton:}&&~~a_d=~{\cal O}_{1i}~/~\cos\beta,~~~~~~~b_d=-{\cal O}_{3i}~\tan\beta\nonumber\\
{\rm gauge~ Bosons:}&&~~c=~{\cal O}_{1i}~\cos\beta+{\cal O}_{2i}~\sin\beta,
\label{eq:cpvmssm_coupl}
\end{eqnarray}
where we have introduced the subscripts $u$ and $d$ on the parameters $a$ and $b$ to
denote the up-type and down-type quarks, respectively.  The mixing matrix elements 
satisfy the normalization conditions:
\begin{equation}
{\cal O}_{1i}^2+{\cal O}_{2i}^2+{\cal O}_{3i}^2=1.
\end{equation}

We call this scenario as Model II in the rest of this article.

The lightest of the Higgs Bosons, $H_1$ could be assumed to be the discovered 125 GeV
resonance (denoted as $\Phi$) , while $H_2$ and $H_3$ are considered to be heavy
enough to be out of LHC's range investigated so far.

As mentioned in the Introduction, it will be a little premature to make conclusions regarding the CP properties of the new resonance from the LHC results so far. It can accomodate some amount of sclar-pseudoscalar mixing within specific models like 2HDM, MSSM, and NMSSM with CP-violating Higgs sector. However, the amount of mixing in these models are highly constrained mainly by the results from flavour sector, as well as the atomic edm measurements \cite{Brodetal, Higgcision}. At the same time, it is possible to have large values for the parameters, $a$ and $b$ even when the CP-odd component of the physical Higgs boson is small \cite{Ananthanarayan:2013cia}.
In the present work, we do not intend to consider any specific model, as the viability of such models, and restrictions on their parameters depend on many constraints outside the considerations of the Higgs sector itself. Instead, we will take a model independent approach, letting the relevant parameters ($a$, $b$ and $c$) rather free, but at the same time keeping them within a small range, without any further justification. We call this scenario as Model II in the rest of this article.

In the following sections, we will  analyze the effect of the anomalous couplings, employed through the scenarios presented above, in the top quark Yukawa coupling measurements.  

\section{Analysis Methodology}\label{process}

The process under scrutiny is the associated production of Higgs Boson with $t\bar{t}$ pair in  
$e^{+}e^-$ collision. As explicitly shown in the Feynman diagrams in Fig.~\ref{fdiag}, this process proceeds through Higgs radiation off the top quarks, or through the Higgs-strahlung off the Z boson. Assuming the $eeZ/\gamma$ couplings to be standard, the process receive contributions from new physics through the anomalous couplings, $tt\Phi$ and $ZZ\Phi$. 
Keeping in focus the main goal of this study, namely, understanding the role of anomalous $tt\Phi$ and $ZZ\Phi$ couplings in the determination of top quark Yukawa couplings, 
we will follow the analysis strategy adopted in the proposed top quark Yukawa measurements, as in Refs.~\cite{YIUF,YITFKSY}. 
The strategy there was to use Eq.~\ref{eq:delta_g} to determine the sensitivity of the coupling. It involves determination of two quantities: 

\medskip

(i) The prefactor, \(\left(\frac{\sigma/g_t}{|d\sigma/dg_t|}\right)_{g_t=g_t^M}\), is determined from the slope of $\sigma$ vs. $g_t$ curve. The cross section, $\sigma_{t\bar{t}\Phi} 
\propto g_t^2$ when the contribution of Higgs-strahlung off the Z boson is neglected, leading to prefactor value of 1/2. The inclusion of the Higgs-strahlung modifies 
this by about 4\% to 0.52 ~\cite{YIUF,YITFKSY}. In general, the prefactor is determined by the functional dependence of the cross section on the Yukawa coupling.

\medskip
 
(ii) The other factor in Eq.~\ref{eq:delta_g}, $\Delta \sigma/\sigma = \sqrt{S + B}/S$, where $S$ is the number of signal events, and $B$ is the number of background events. Getting the best (smallest) value 
 for $\frac{\Delta \sigma}{\sigma}$, is a matter of experimental efficiency,  and suitable choice of observables at hand to reduce the background over the signal.

\medskip  
 
We will first assume 
 that the strategy adopted by Ref.~\cite{YIUF,YITFKSY} to reduce the background, and enhance the signal significance through the kinematic cuts on suitable variables considered therein are  acceptable as it is 
 even in the presence of anomalous couplings. Further, we will consider various kinematic distributions comparing the case of SM and the case of anomalous couplings  with the parameters
 assuming different values. Such a comparison is expected to indicate if the above assumptions regarding signal significance is realistic or not. This is important, as a detailed study of the signal and background, event reconstruction, machine and detector efficiency specific to ILC, etc, ar beyond the scope of this study, and therefore not attempted.
Rather, we will be satisfied with a  qualitative analysis of the various distributions of the signal with anomalous couplings. 

Considering the signal process, we first discuss the different final states possible through different decay channels of $tt\Phi$. We note that, while the top quak decays almost 100\% into $Wb$, the Higgs boson of mass 125 GeV could go through $b\bar{b}$, $W W^{*}$ and 
$\tau^{+}\tau^{-}$ with branching fractions (BR) of 57.7\%, 21.5\% and 6.32\%, respectively in the SM. In our analysis we consider the $b\bar b$ channel alone. This leaves the following distinct final states, depending on the decay chanel of $W$:

\begin{itemize}

\item Pure Hadronic mode: In this mode both the W's decay hadronically (BR=45.6\%), resulting in 4 jets + 4b's in the final state.  \\

 $e^- e^+ \rightarrow t \bar{t} \Phi \rightarrow W^+ W^- b \bar{b}\Phi \rightarrow q_1\bar q_1' q_2\bar q_2' bb \bar{b}\bar{b}$;

 \item  Semileptonic mode: In this mode one of the $W$'s decays hadronically,  while the other decays leptonically (BR=43.9\%), resulting in 2jets + 4b's + 1lepton + $E_{\rm missing}$  in the final state;\\
 
  $e^- e^+ \rightarrow t \bar{t} \Phi \rightarrow W^+ W^- b \bar{b}\Phi \rightarrow q_1\bar q_1' l \nu bb \bar{b}\bar{b}$

 \item Leptonic mode: In this mode both the $W$'s decay leptonincally (BR=10.5\%), resulting in 4b's + 2 lepton + $E_{\rm missing}$ in the final state;\\
 
  $e^- e^+ \rightarrow t \bar{t} \Phi \rightarrow W^+ W^- b \bar{b}\Phi \rightarrow l\bar{l} \nu \bar{\nu} bb \bar{b}\bar{b}$\\

\end{itemize}

In the leptonic decays of the $W$, we have included only the channels with electrons and muons, keeping aside the tau decay channel. We have also assumed that $b$-tagging  can be performed with high efficiency, and thus $b$-jets are distinguished from the lighter quark jets.  In our further discussion we include the hadronic and semileptonic modes, leaving out the purely leptonic channel, owing to its very small BR, and two missing particles in the final state. This is the strategy followed in other studies of top quark Yukawa coupling measurement through the same process. 

Coming to the background, $t\bar t Z$ and $t\bar t g^*$ , with pair of $b$-jets coming from $Z$ and $g^*$,contribute to the irreducible backgrounds in the corresponding cases. Owing to the large cross section, the $t\bar t$ pair production could also give rise to the background, through event misconstuction.  

We will now turn to our numerical studies in the next section.

\section{Numerical Results}\label{numresults}

For our numerical study, we considered the event generator, WHIZARD~\cite{WHIZARD}, with the model files suitably modified\footnote{We refer to 
http://feynrules.irmp.ucl.ac.be/wiki/StandardModel in this regard.}  to accommodate the anomalous couplings being studied, viz, $f\bar f\Phi$ and
$VV\Phi$ couplings, where $f=t,~b,~\tau$ and $V=Z,~W$, parametrized through Eq.~\ref{eq:generic_coupl_t} and~\ref{eq:generic_coupl_b}.
We have cross checked the correctness of our implementation by verifying the 
results of Ref.~\cite{DDGMR} for the process being scrutinized.

To examine the effect of anomalous couplings on the process, and thus in deriving the Yukawa couplings, we consider the following values of the parameters in the two scenarios presented in the previous section.

\begin{itemize}
\item Model I: The only parameter in this scenario is denoted by $b$, which can assume any value in the range $0-1$. Specifically, we have considered $b=0, 0.1, 0.3, 0.5$ and 
$0.7$ for illustraton, where $b=0$ corrsponds to the SM. 

\item Model II:  As mentioned earlier, we consider a model independent approach in this scenario, with the parameters allowed to vary independent
of each other. While this is so, we have kept in mind  the most-likely-possibility with the physical Higgs particle being mostly CP-even, and therefore, $c$ is 
close to unity. At the same time, $a$ and $b$ can be quite different, and can be larger than one. Although we do not follow any particular model, 
in order to be close to realistic cases, we have chosen the first benchmark point from CP-violating MSSM~\cite{Ananthanarayan:2013cia}, and the second one 
from CP-violating 2HDM (type-II, without SUSY)~\cite{ZhangCPV}. 
Both P1 and P2 have the parameters $a$ and $c$ positive, while $b$ is negative. This seems to be the preferred direction in the case of the specific models considered. It is quite natural to ask what is the effect of signs on these parameters on the cross section, and other observables, and the conclusions drawn from those. To this effect, we consider a few other Bechmark Points (BP), somewhat arbitrarily, with different 
sign combinations. In Table~\ref{tab:2HDM_coupl} we present the values of the parameters corresponding to these BP's.

%\begin{table}
%\begin{tabular}{|c|c|c|c|c|c|r|r|r|r|}
%\hline
%Point&$\tan\beta$
%&${\cal O}_{11}$
%&${\cal O}_{21}$
%&${\cal O}_{31}$
%&$Z,~W$&\multicolumn{2}{c|}{top}&\multicolumn{2}{c|}{$b~/~\tau$}\\ \cline{6-10}
%&&&&&$c$&$a_{u}$&$b_{u}$&$a_{d}$&$b_{d}$\\ \cline{1-10}
%P1&2&0.22& 0.97&0.10 &  0.97 &   1.08 &  $-$0.05 & 0.50  & $-$0.20 \\\cline{1-10}
%P2&2&0.22& 0.92&0.32 &  0.92&    1.03&   $-$0.16&  0.50&   $-$0.64\\\cline{1-10}
%P3&2&0.22& 0.84&0.50&  0.85&    0.94&   $-$0.25&    0.50 &  $-$1.00\\\cline{1-10} 
%P4&0.8& 0.78&  0.51& 0.36&  0.93 &  0.82 &  $-$ 0.45  & 1.00 & $-$0.29\\\cline{1-10}
%\end{tabular}
%\caption{Couplings of the Higgs Bosons ($H_1$) in the CP-violating 2HDM, as defined in
%Eq.\,(\ref{eq:cpvmssm_coupl}), for different mixings given by $H_1={\cal O}_{11}~\phi_1+{\cal
%O}_{21}~\phi_2+{\cal O}_{31}~A$.
%}.
%\label{tab:2HDM_coupl}
%\end{table}

\begin{table}
\begin{tabular}{|c|r|r|r|r|r|}
\hline
Point
&$Z,~W$&\multicolumn{2}{c|}{top}&\multicolumn{2}{c|}{$b~/~\tau$}\\ \cline{2-6}
&$c$&$a_{u}$&$b_{u}$&$a_{d}$&$b_{d}$\\ \cline{1-6}
P1&0.97 &   1.08 &  $-$0.05 & 0.5  & $-$0.20 \\\cline{1-6}
P2& 0.93 &  0.82 &  $-$ 0.45  & 1.0 & $-$0.29\\\cline{1-6}
P3&0.93&    0.5&   $-$1.0&  0.5&   $-$1.0\\\cline{1-6}
P4&0.93&    $-$1.0&   1.0&    $-$1.0 &  1.0\\\cline{1-6} 
P5&0.93&    $-$1.0&   $-$1.0&    $-$1.0 &  $-$1.0\\\cline{1-6} 
P6&0.93&    1.0&   1.0&    1.0 &  1.0\\\cline{1-6} 
P7&-0.93&    1.0&   1.0&    1.0 &  1.0\\\cline{1-6} 
\end{tabular}
\caption{Benchmark points (BP) in the case of Model II.
}.
\label{tab:2HDM_coupl}
\end{table}
\end{itemize}

\begin{figure}[h]\centering
\begin{tabular}{c c} 
\hspace{-5mm}
\includegraphics[angle=0,width=110mm]{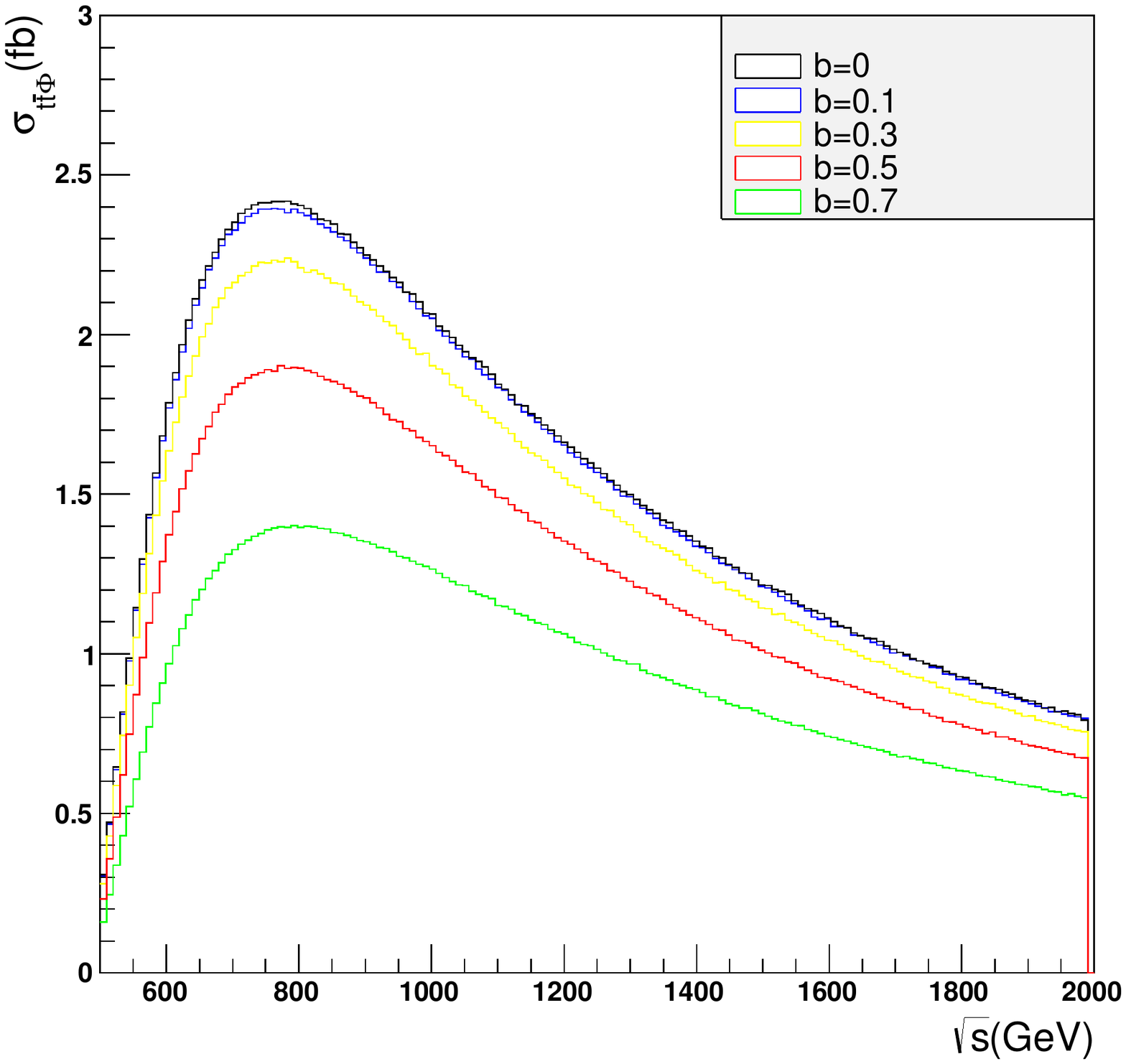} & \hspace{-5mm}
\includegraphics[angle=0,width=110mm]{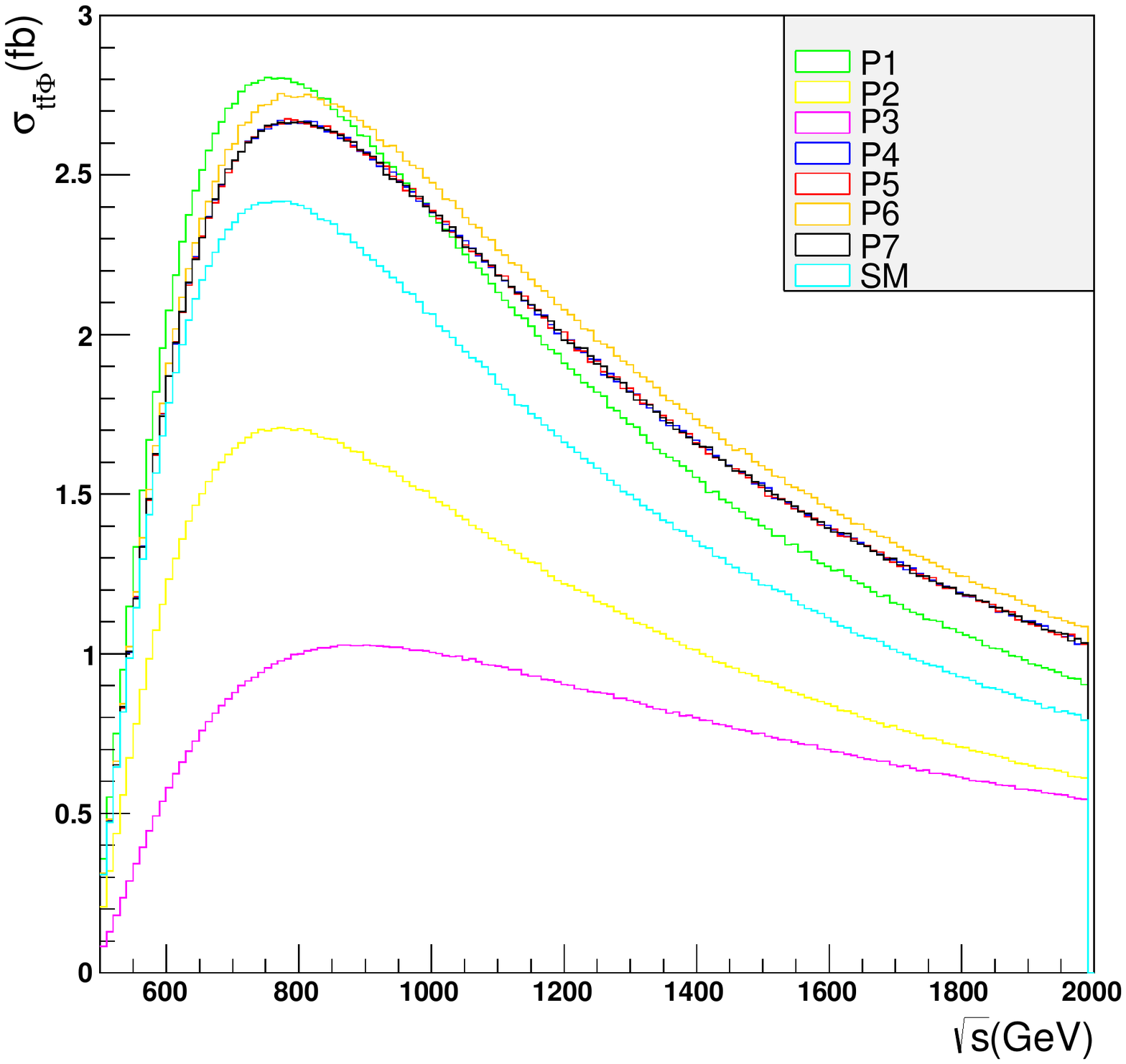}\\
\end{tabular}
%\vspace*{-7cm}
\caption{$\sigma_{t\bar{t}\Phi}$ vs $\sqrt{s}$ for different parameter values in Model I(left panel) and Model II(Right panel). }
\label{fig:cs}
\end{figure}

Considering the effect of the anomalous couplings in the total cross section, in Fig.~\ref{fig:cs} we present the cross section against the centre of mass energy for 
different values of $b$ in case of Model I, and for the BP's considered, in case of Model II. In the case of Model I, with 30\% pseudoscalar component ($b=0.3$), there is 
about 10\%  decrease in the cross section at the peak value corresponding to centre of mass energy of 800 GeV. This is increased to about 40\% with 70\% pseudoscalar 
component.  Coming to Model II, $a_d$ abd $b_d$ do not affect the production process. They only leave their signature in the decay of the Higgs boson. With P1 and P2, there is substantial difference in the cross section compared to the SM value, with the former having an enhanced effect, while the latter having a diminishing effect. This shows that the effect of the parameter $a$ is somewhat dominating compared to that of the $b$ parameter. Behaviour of the cases of P4 and P5 clearly indicate that the signs of $b$ has no perceivable effect in the total cross section. At the same time, a distinguishable effect of the sign of $a$ is visible comparing the points P4 and P6. Indeed, this is expected to be due to the interference between the diagram involving the $ZZ\Phi$ coupling with the others. In order to ascertain this, the signs of $a$ and $c$ are switched between P4 and P7, still keeping a relative negative sign.  A comparison of P1 and P3 also brings out the different $\sqrt{s}$ behaviour of the dependence 
of $a$ and $b$. While the effect of $a$ does not indicate any considerable change as the $\sqrt{s}$ is changed, in the case of $b$, the effects are substantially larger at larger $\sqrt{s}$. This advocates that the investigation of the process at a few chosen centre of mass energy values will be more enlightening compared to an analysis sitting only at one centre of mass energy. Between P4 and P7, the signs of $a$ and $c$ are switched, so as to keep a relative negative sign.

In order to understand how the above mentioned differences affect the Yukawa coupling measurement, we first focus on the case of ILC running at 
centre of mass energy of 800 GeV, as the cross section peaks around this value.  The cases of 500 GeV and 1000 GeV are also 
included so as to see the effect of the $\sqrt{s}$ dependence.  In the rest of this section we shall consider each of these cases separately, 
and discuss the effect of the anomalous couplings on the prefactor as well as the  signal significance. We would like to emphasize that, the 
purpose of this study is to bring home the issues to be addressed while performing Yukawa coupling measurements. We shall also present the decay 
distributions in order to assertain the need to adopt different strategies of background reduction.We do not attempt to present a full analysis 
to evaluate the best sensitivity of the measurements. With ILC, it is a general rule that the beam polarization help the phenomenology. In the 
present case, the initial state polarization does not play a direct role on the final states, except for a changed statistics.  As already mentioned, 
in this study we do not perform any analysis to enhance the signal significance. Rather, we rely on the procedure adopted in the earlier rigorous 
studies made at the respective centre of mass energies. With this limitation, we follow Ref.~\cite{AG2}, and consider unpolarized beam in the 
case of centre of mass energy of 800 GeV, while for 500 and 1000 GeV cases beam polarizations are considered as per Ref.~\cite{YITFKSY, LCtopyuk1000}. 
It is advisable to include a full detector simulation, exploring the advantages of beam polarization in the analysis to understand the complete picture,
which is not attempted in this report. 

\subsection{ILC with $\sqrt{s}=800$ GeV}

Top quark Yukawa coupling measurements at the ILC running with unpolarized beams at a centre of mass energy of 800 GeV were studied in great detail in Ref.~\cite{AG2} for 
$M_{\Phi}=$ 120 GeV within SM. While using the realistic detector simulations, they had neglected the effect of Higgs-strahlung off the Z boson, leading to a different 
extraction method compared to employing Eq.~\ref{eq:delta_g}, which is considered in more recent studies.

\begin{figure}[h]\centering
\begin{tabular}{c c} 
\hspace{-5mm}
\includegraphics[angle=0,width=80mm]{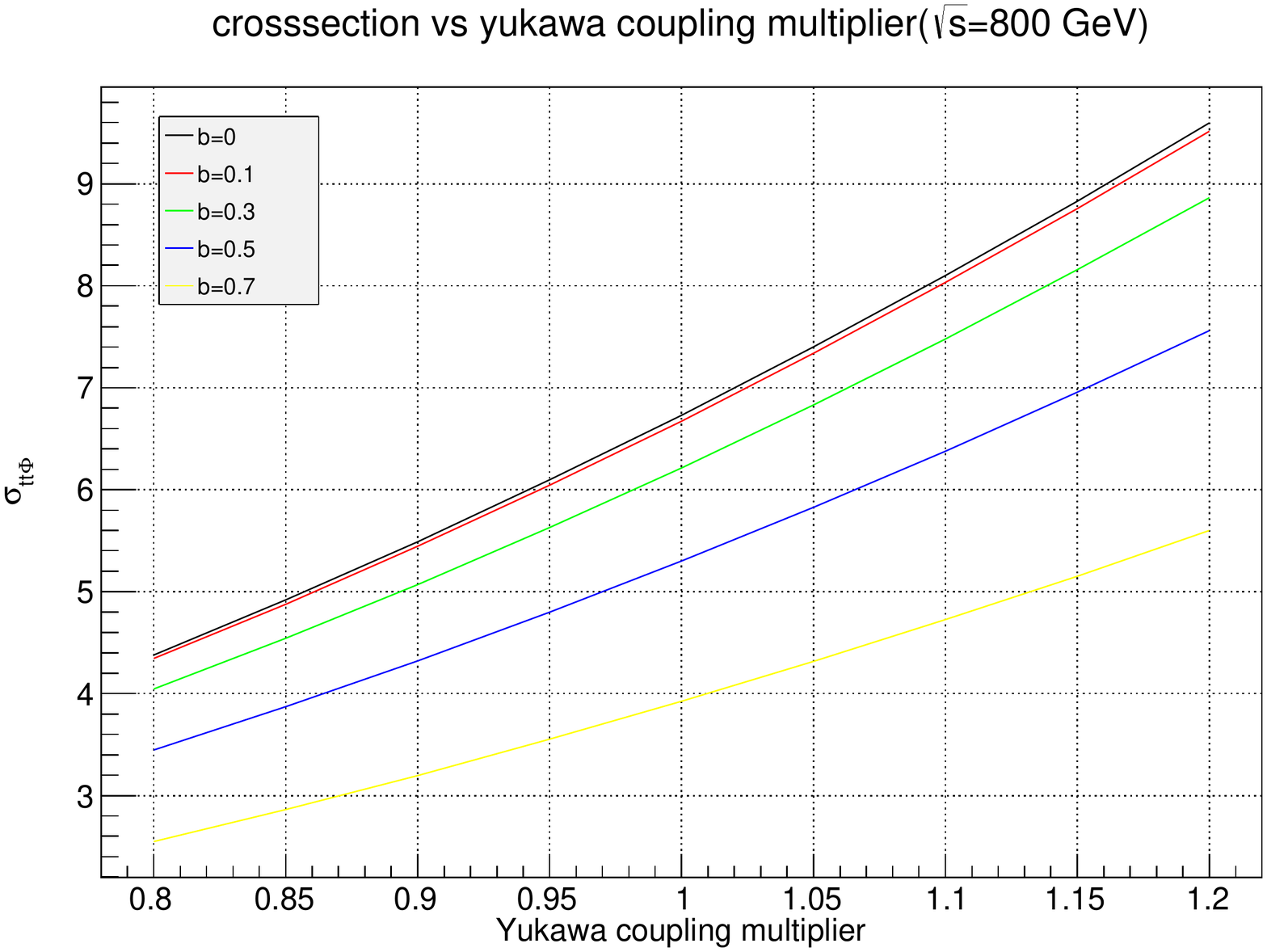} &
\includegraphics[angle=0,width=80mm]{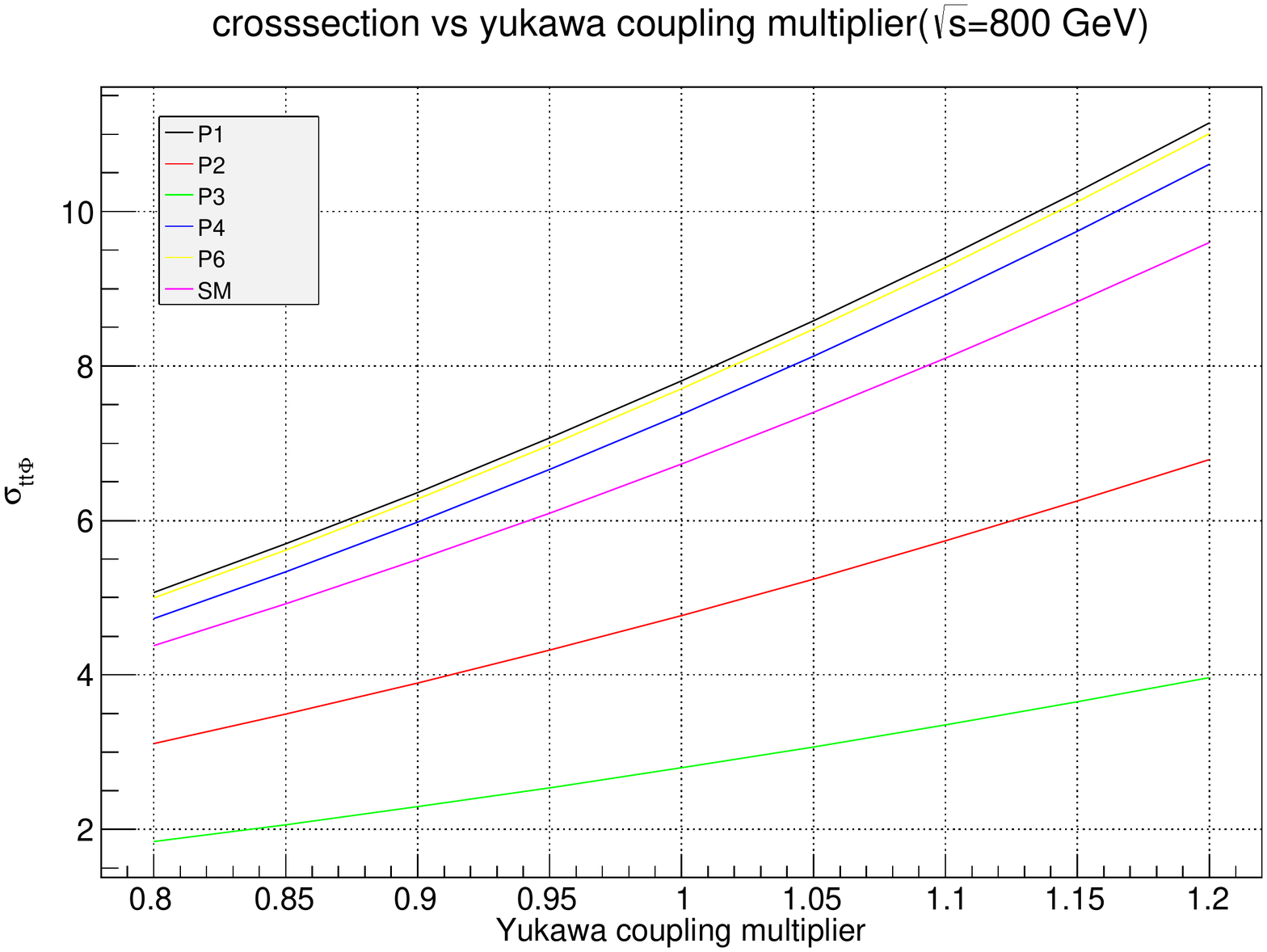}\\
\end{tabular}
%\vspace*{-7cm}
\caption{$\sqrt{s}=800$ GeV: $\sigma_{t\bar{t}\Phi}$ vs Yukawa multiplier for different parameter values in Model I(left fig.) and Model II(Right fig.). }
\label{fig1L.1}
\end{figure}

While considering scenarios involving more general Higgs sector with CP-mixed physical 
Higgs bosons, we obtained the prefactor for the illustrative parameter choices mentioned above for the two scenarios of Model I and Model II. The prefactor is obtained from the 
functional dependence of the cross section on the coupling.  In Fig.~\ref{fig1L.1}, we present cross section vs Yukawa coupling multiplier curves for Model I (left) and Model II
(right) separately. Here and thereafter we will not discuss P5 since it has same production cross section value as P4 and thus it will give no different results than P4. The behaviour of 
these curves is expected to fit a quadratic equation($A ~\lambda_t^2 + B~ \lambda_t+ C$), where we define a relative coupling, $\lambda_t=\frac{g_t}{g_t^M}$. For easy reference, we follow the representation of 
Ref.~\cite{LCtopyuk1000}, and present the equations of the curves corresponding to different parameter values in Eq.~\ref{eq:800mI} and~\ref{eq:800mII} below, after fitting the quadratic equation. 

\begin{eqnarray}
b=0.0: && \hspace{0.3cm}\sigma_{t\bar{t}\Phi} = 6.426 (\lambda_t-1)^2 + 13.040(\lambda_t-1)+ 6.731,\nonumber\\
b=0.1: && \hspace{0.3cm}\sigma_{t\bar{t}\Phi} = 6.404 (\lambda_t-1)^2 + 12.933(\lambda_t-1)+ 6.673,\nonumber\\
b=0.3: && \hspace{0.3cm}\sigma_{t\bar{t}\Phi} = 5.988 (\lambda_t-1)^2 + 12.049(\lambda_t-1)+ 6.215,\nonumber\\
b=0.5: && \hspace{0.3cm}\sigma_{t\bar{t}\Phi} = 5.090 (\lambda_t-1)^2 + 10.280(\lambda_t-1)+ 5.298,\nonumber\\
b=0.7: && \hspace{0.3cm}\sigma_{t\bar{t}\Phi} = 3.776 (\lambda_t-1)^2 + 7.632(\lambda_t-1)+ 3.923
\label{eq:800mI}
\end{eqnarray}

and

\begin{eqnarray}
P1: && \hspace{0.3cm}\sigma_{t\bar{t}\Phi} = 7.470 (\lambda_t-1)^2 + 15.194(\lambda_t-1)+ 7.805,\nonumber\\
P2: && \hspace{0.3cm}\sigma_{t\bar{t}\Phi} = 4.530 (\lambda_t-1)^2 + 9.189(\lambda_t-1)+ 4.768,\nonumber\\
P3: && \hspace{0.3cm}\sigma_{t\bar{t}\Phi} = 2.594 (\lambda_t-1)^2 + 5.300 (\lambda_t-1)+ 2.796, \nonumber\\
P4: && \hspace{0.3cm}\sigma_{t\bar{t}\Phi} =  7.367 (\lambda_t-1)^2 + 14.701 (\lambda_t-1)+ 7.373,\nonumber\\
P6: && \hspace{0.3cm}\sigma_{t\bar{t}\Phi} =  7.465(\lambda_t-1)^2 + 15.031 (\lambda_t-1)+ 7.703 
\label{eq:800mII}
\end{eqnarray}

As the equation is an exact quadratic equation, the fit is quite accurate, and the errors on the coefficiencts can be neglected. We have generically considered them to be below the  permille level. Please note that P4 and P5 fit to the same equation, as they differ by a sign of the parameter $b$, which has negligible effect. Therefore, we have not presentend the case of P5 in the following. Similarly the case of P7 is identical to that of P3, and therefore not presented here separately.
The value of prefactor for Model I obtained from the above turns out to be 0.516  for the SM value, and varies very slowly with change in $b$ to be 0.514 for $b=0.7$.  In the case of Model II, for P1, P2 and P6 as well the values of the prefactor remain close to the SM value,  giving 0.514, 0.519  and 0.512, respectively. The other points, P3 and P4 show slight deviation from the SM value, leading to 0.528 and 0.502, respectively. The case of P7 is identical to that of P3, again showing that it is the relative sign between $a$ and $c$, that matters, arising through the interference between the two relevant diagrams. 
By definition, the deviation of the prefactor from the value of $\frac{1}{2}$ indicated the influence of the $ZZ\Phi$ coupling through 
the Higgs-strahlung contribution to the cross section. As is clear from the parameter values considered, this influence is due to changed values of $a$ and $b$, rather than the change in $c$. 

The other factor entering $\Delta g_t$ is the signal sensitivity factor, $\Delta \sigma/\sigma=\sqrt{S+B}/S$. The extraction of this needs signal ($S$) as well as
background ($B$) events. In order to minimize the background, so as to get the best sensitivity, one employs suitable kinematical cuts, and other procedures. The sensitivity 
also depends on the efficiency of identification of the relevant final states, as well as the efficiency with which events could be reconstructed.  
Keeping to the limited scope of this study, we will assume the same machine efficiencies, and background reduction procedures followed by Ref.~\cite{Juste} in obtaining 
the sensitivity. The procedure we adopted in our analysis is described below.
\begin{table}[h]
\begin{center}
\begin{tabular}{ccc|ccc||ccc|ccc}

\hline

\hline
\multicolumn{6}{c||}{No. of events in Hadronic
case}&\multicolumn{6}{c}{No. of events in Smileptonic case:}\\
\cline{1-12}
\multicolumn{3}{c|}{Model I}&\multicolumn{3}{c||}{Model II}
&\multicolumn{3}{c|}{Model I}&\multicolumn{3}{c}{Model II}\\
\cline{1-12}
$b$ & Total&After cuts &Points & Total&After cuts  &$b$ & Total&After
cuts &Points & Total&After cuts\\
\hline%---------------------------------------------------------------------

0. & 399.0 & 93.8 & P1 & 243.7 &57.3 & 0. & 375.4 & 88.1 & P1 & 229.2 & 53.8\\

\hline%-------------------------------------------------------

0.1 & 396.0 & 93.1 & P2 &296.5 & 69.7& 0.1 & 372.8 & 87.5 & P2& 279.1 & 65.5\\

\hline%---------------------------------------------------------------------

0.3  & 373.5 & 87.8 & P3 &163.3& 38.4 & 0.3 & 351.5 & 82.5 & P3 & 153.8 & 36.1\\

\hline%---------------------------------------------------------------------

0.5  & 324.6 & 76.3  & P4 &488.4& 114.8 &0.5 & 305.5 & 71.7 & P4 &
459.8 & 107.9\\

\hline%---------------------------------------------------------------------

0.7  & 247.1 & 58.1  & P6 &505.4& 118.8 & 0.7 & 232.6 & 54.6 & P6 &
476.0 & 111.7\\

\hline%---------------------------------------------------------------------

\end{tabular}
\end{center}

\caption{Total number of events corresponding to $\sqrt{s}=800$ GeV
for an integrated luminosity of 1000 fb$^{-1}$ before and after the
kinematical cuts.}

\label{800_events}
\end{table}

 The signal events are obtained from $t\bar{t}\Phi$ production cross section by considering the braching ratios (BR's) appropriate to the specific channels.  That is, 
\begin{equation}
 \sigma^{\rm total}_{\rm signal} = \sigma_{t\bar{t}\Phi} \times BR(t\bar{t} \rightarrow X) \times BR(\Phi \rightarrow b\bar{b}),
\end{equation}
where $X$ denotes the specific final state corresponding to the hadronic or semileptonic channels, whichever is applicable. We note that, with CP-indefinite Higgs boson, the decay 
widths of $\Phi \rightarrow b\bar b,~\tau^+ \tau^-, ~\gamma\gamma,~gg,~WW^*, ~ZZ^{*}$ are all affected, as the $\Phi b \bar b$ and $\Phi VV$ vertices are all affected. We have 
taken care of this in obtaining the BR($\Phi \rightarrow b\bar b$). On the other hand,  the background remains the same in all cases, as the anomalous vertices do not appear 
in background processes. We present our results for an assumed integrated luminosity of 1000 fb$^{-1}$, corresponding to which, the total number of signal events in the hadronic and semilepton channels are presented in Table~\ref{800_events}. 
Following Ref.~\cite{Juste} in enhancing the signal over the background, the final number of events are also given in Table~\ref{800_events}, with  corresponding reduction 
factors of 0.234 and 0.235  for semileptonic
and hadronic cases, respectively. 
%The corresponding signal events are given in Table~\ref{800gtM1} and Table~\ref{800gtM2} for Model I and Model II respectively.

\begin{table} 
\begin{center}
\begin{tabular}{lccccccc}

\hline

\hline

$b$ & Prefactor &Signal($S_1$) & $\frac{\Delta\sigma}{\sigma}$ & $\frac{\Delta g_t}{g_t}$& Signal($S_2$) & $\frac{\Delta\sigma}{\sigma}$ & $\frac{\Delta g_t}{g_t}$ \\

\hline%---------------------------------------------------------------------

0. & 0.516 & 93.8 & 0.17 & 0.087 & 88.1 & 0.18 & 0.096 \\

\hline%-------------------------------------------------------

0.1 & 0.515 & 93.1 & 0.17 & 0.088 & 87.5 & 0.19 & 0.097\\

\hline%---------------------------------------------------------------------

0.3  & 0.515 & 87.8 & 0.18 & 0.092   & 82.5 & 0.20& 0.102\\

\hline%---------------------------------------------------------------------

0.5  & 0.515 & 76.3  & 0.20 & 0.104   & 71.7 & 0.22 & 0.115\\

\hline%---------------------------------------------------------------------

0.7  & 0.514 & 58.1  & 0.25 & 0.131   & 54.6 & 0.28& 0.146\\

\hline%---------------------------------------------------------------------

\end{tabular}
\end{center}

\caption{Model I: Yukawa coupling sensitivity for different parameters at $\sqrt{s}=800$ GeV. $S_1$ and $S_2$ are
          signal events in hadronic and semileptonic mode after kinematical cuts.}

\label{800gtM1} 
\end{table}

\begin{table} 
\begin{center}
\begin{tabular}{lccccccc}

\hline

\hline

Parameter & Prefactor & Signal($S_1$) & $\frac{\Delta\sigma}{\sigma}$ & $\frac{\Delta g_t}{g_t}$& Signal($S_2$) & $\frac{\Delta\sigma}{\sigma}$ & $\frac{\Delta g_t}{g_t}$ \\

\hline%---------------------------------------------------------------------

P1 & 0.513 & 57.3 & 0.26 & 0.133 & 53.8 & 0.29 & 0.148\\

\hline%-------------------------------------------------------

P2  & 0.518 & 69.7  & 0.22 & 0.112  & 65.5 & 0.24 & 0.124\\

\hline%---------------------------------------------------------------------

P3 & 0.527 & 38.4 & 0.37 &  0.196  & 36.1 & 0.41 & 0.220\\

\hline%---------------------------------------------------------------------

P4  & 0.501 & 114.8 & 0.14 & 0.073  & 107.9 & 0.16 & 0.080\\

\hline%---------------------------------------------------------------------

P6  & 0.512 & 118.8 & 0.14 & 0.072  & 111.7 & 0.15 & 0.079\\

\hline%---------------------------------------------------------------------

\end{tabular}
\end{center}

\caption{Model II: Yukawa coupling sensitivity for different parameters at $\sqrt{s}=800$ GeV. $S_1$ and $S_2$ are
          signal events in hadronic and semileptonic mode after kinematical cuts.}

\label{800gtM2} 
\end{table}

The signal sensitivity, and the top quark Yukawa coupling sensitivity for the parameter points of Model I and Model II, are presented in 
Tables~\ref{800gtM1} and \ref{800gtM2}, respectively. 
As can be seen, the dominating factor in the sensitivity of Yukawa coupling measurement is the signal significance, as the prefactor is close to its SM value. In the case of Model 
I, the increasing value of $b$, which corresponds to the increasing pseudoscalar composition in the Higgs, result in the worsening of the sensitivity monotonously.
While for the SM case (Model I with $b=0$), the sensitivity is about 8.7\% and 9.6\% for hadronic and semileptonic channels, respectively,  it is between 13\% and 15\% for a 
pseudoscalar composition of 70\%. In the case of Model II,  P2 and P3 cases in which $a$ is smaller than unity, the sensitivity is worsened, depending on how small the value 
of the parameter is. For example, with $a=0.5$, the sensitivity has gone down to about 20 - 22 \%. On the other hand, the case of P4 and P6 with $|a|=|b|$ can measure the 
coupling with sensitivity of about 7\%, which better than that of the SM case. Whereas, in the case of P1, although $a_u=1.08$, the signal events is small compared to the SM case, because of the reduced 
BR($\Phi \rightarrow b\bar b$), due to small value of $a_d$. As mentioned earlier, these variations are mainly due to the changes in the total cross section itself. 

In the above, we had assumed that the kinematical cuts that are employed in the case of SM are applicable, in the case with anomalous couplings as well. In reality, 
the kinematics of the final states could be different in these cases, and therefore, different strategy may be needed to study the machine capability and signal sensitivity. 
While it is beyond the scope of this study to present an exhaustive analysis in this regard, we shall present some of the simple kinematic distributions in the cases considered. 

Assuming partial reconstruction, we consider two different sets of final states to present the distributions. 
\begin{enumerate}
\item Higgs is fully reconstructed:  \(
 e^- e^+ \rightarrow t\bar{t}\Phi \rightarrow W^+ W^- b\bar{b}\Phi  \)
\item top quarks are fully reconstructed: \(
e^- e^+ \rightarrow t\bar{t}\Phi \rightarrow t\bar t b\bar{b}   \)
\end{enumerate}

In order to study the effect of these vertices on kinematical distributions related to signal process, we present 
various distributions corresponding to our chosen parameter points for Model I and Model II at 800 GeV. To reduce complexity, we consider the decay of top and Higgs
separately. Thus we use the following  signal prcoesses for the distributions
\begin{equation}
 e^- e^+ \rightarrow t\bar{t}\Phi \rightarrow W^+ W^- b\bar{b}\Phi;~ t\bar{t} b\bar{b}
\end{equation}

We consider the energy, angle and transverse momentum distributions of the final state particles in the two models, and compare those with the case of the SM. The case of 
Model I is found to have similar behaviour for most of the distributions, but with varying total number of events, as presented earlier. Whereas, the case of Model II some of 
the BP's have distinctly different functional dependence than others. To illustrate this, we consider the normalized distributions corresponding to Model I and Model II, 
and compare with the respective SM  distributions.

\begin{figure}[!t]\centering
\begin{tabular}{c c c  } 
\includegraphics[angle=0,width=60mm]{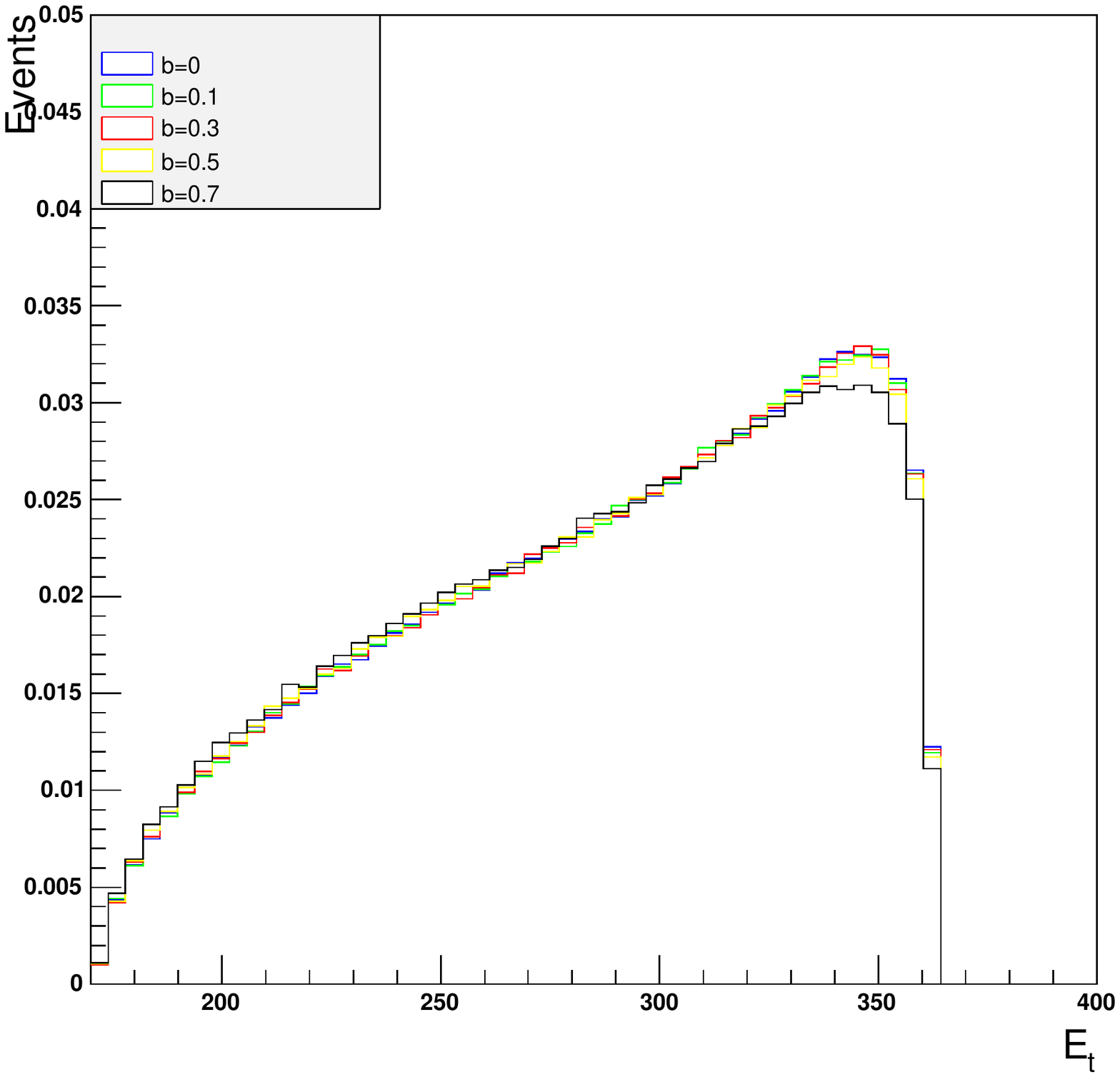} &
\includegraphics[angle=0,width=60mm]{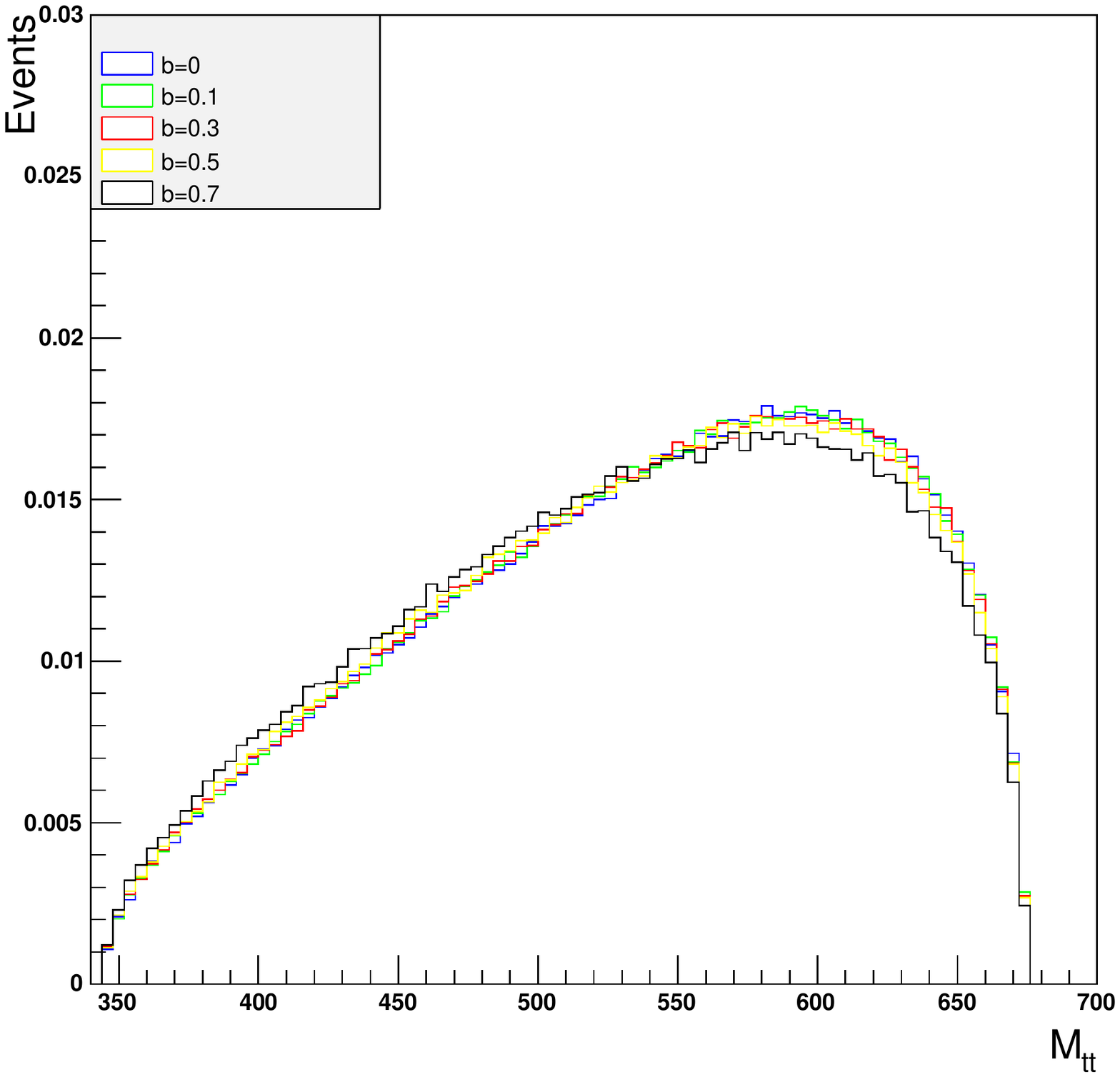}&
\includegraphics[angle=0,width=60mm]{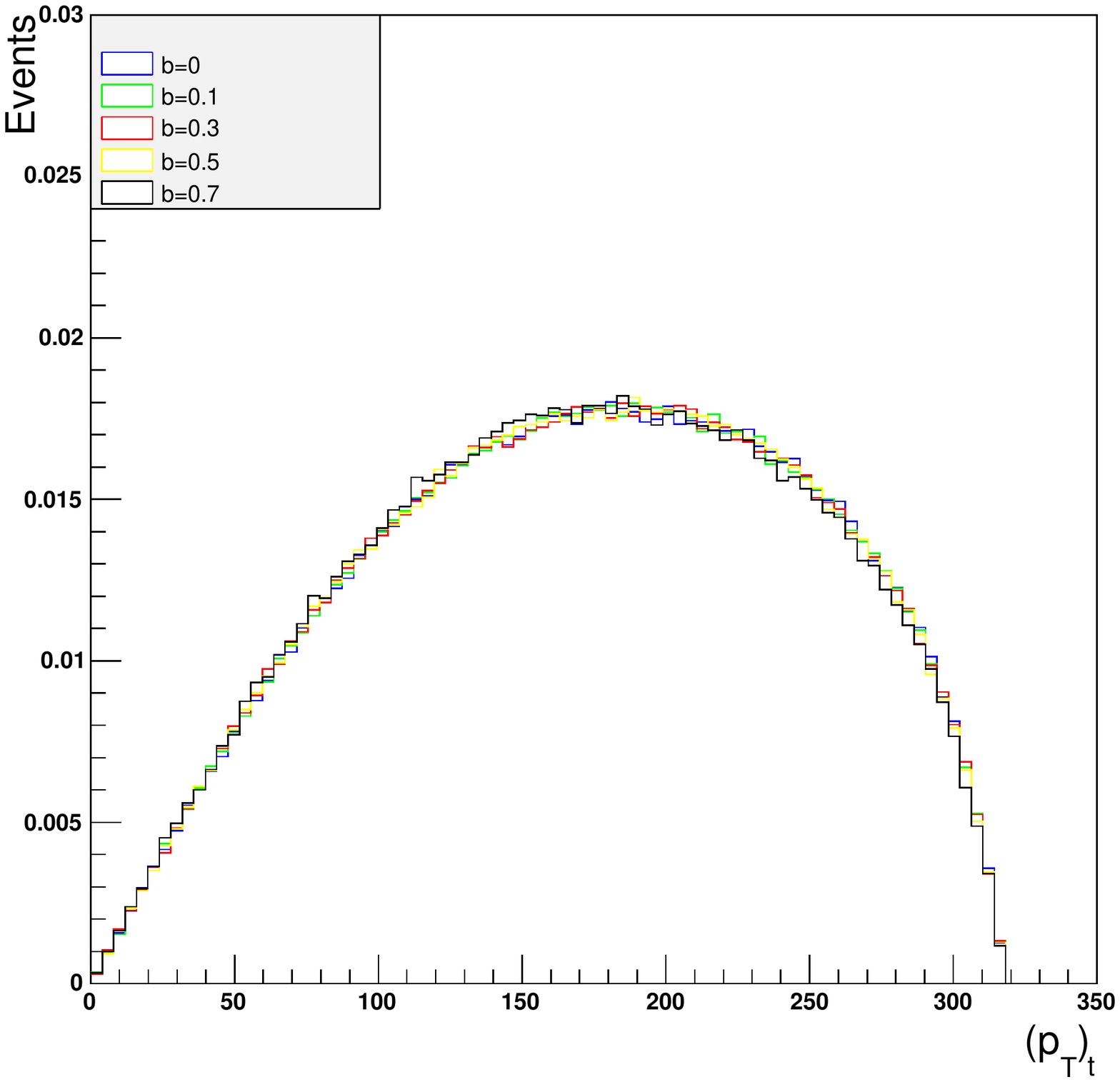} \\
\includegraphics[angle=0,width=60mm]{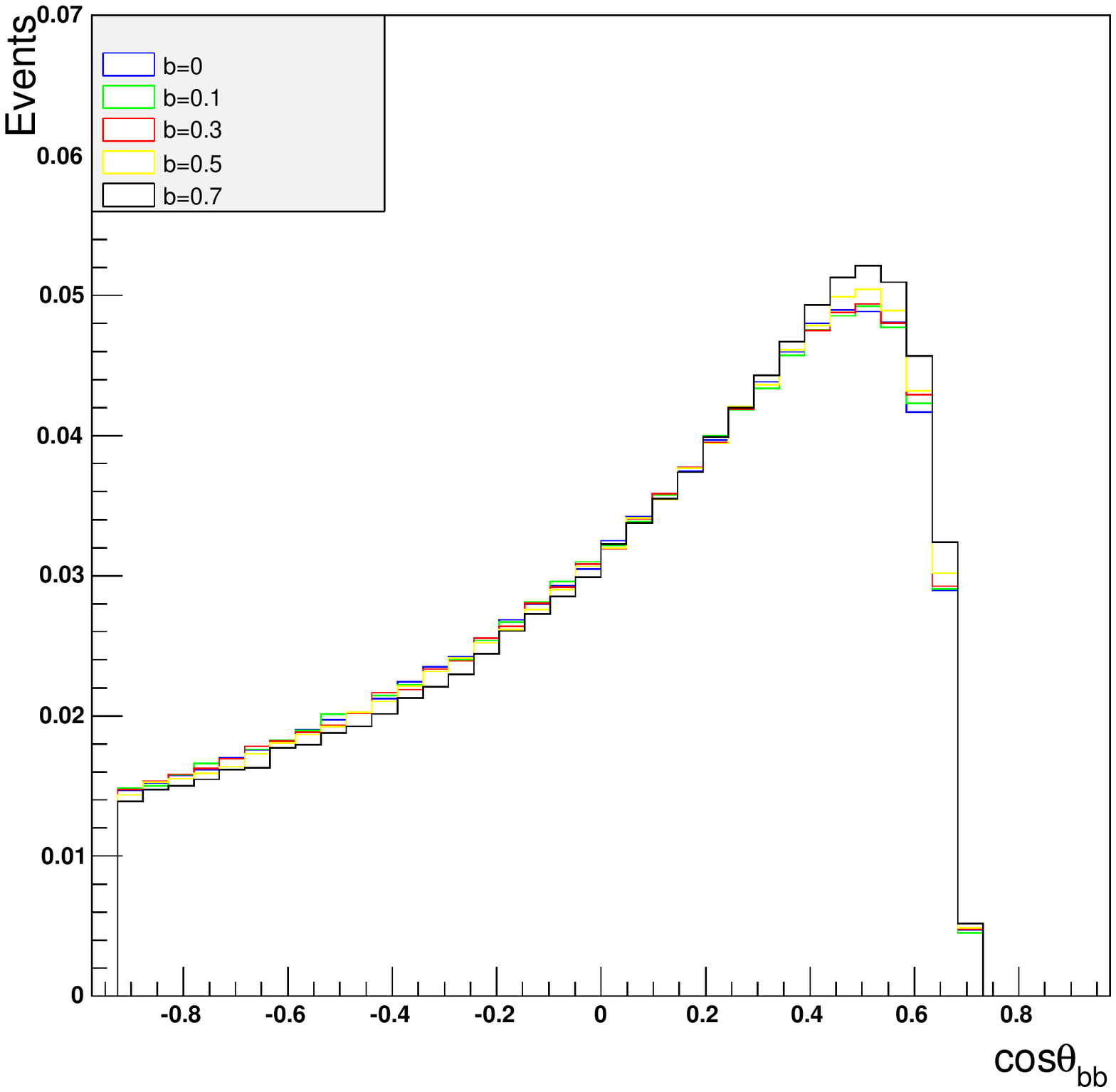}&
\includegraphics[angle=0,width=60mm]{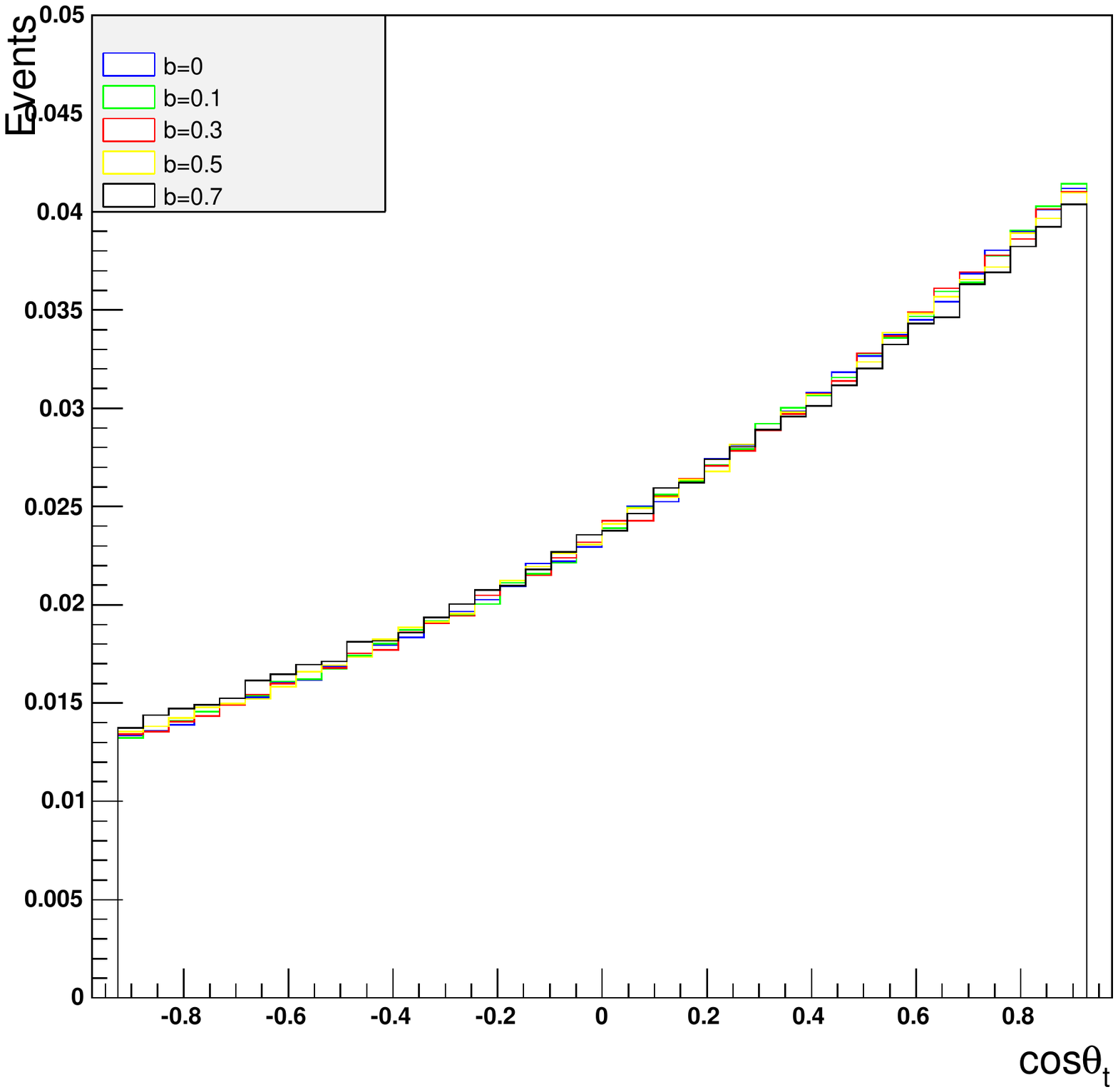} &
\includegraphics[angle=0,width=60mm]{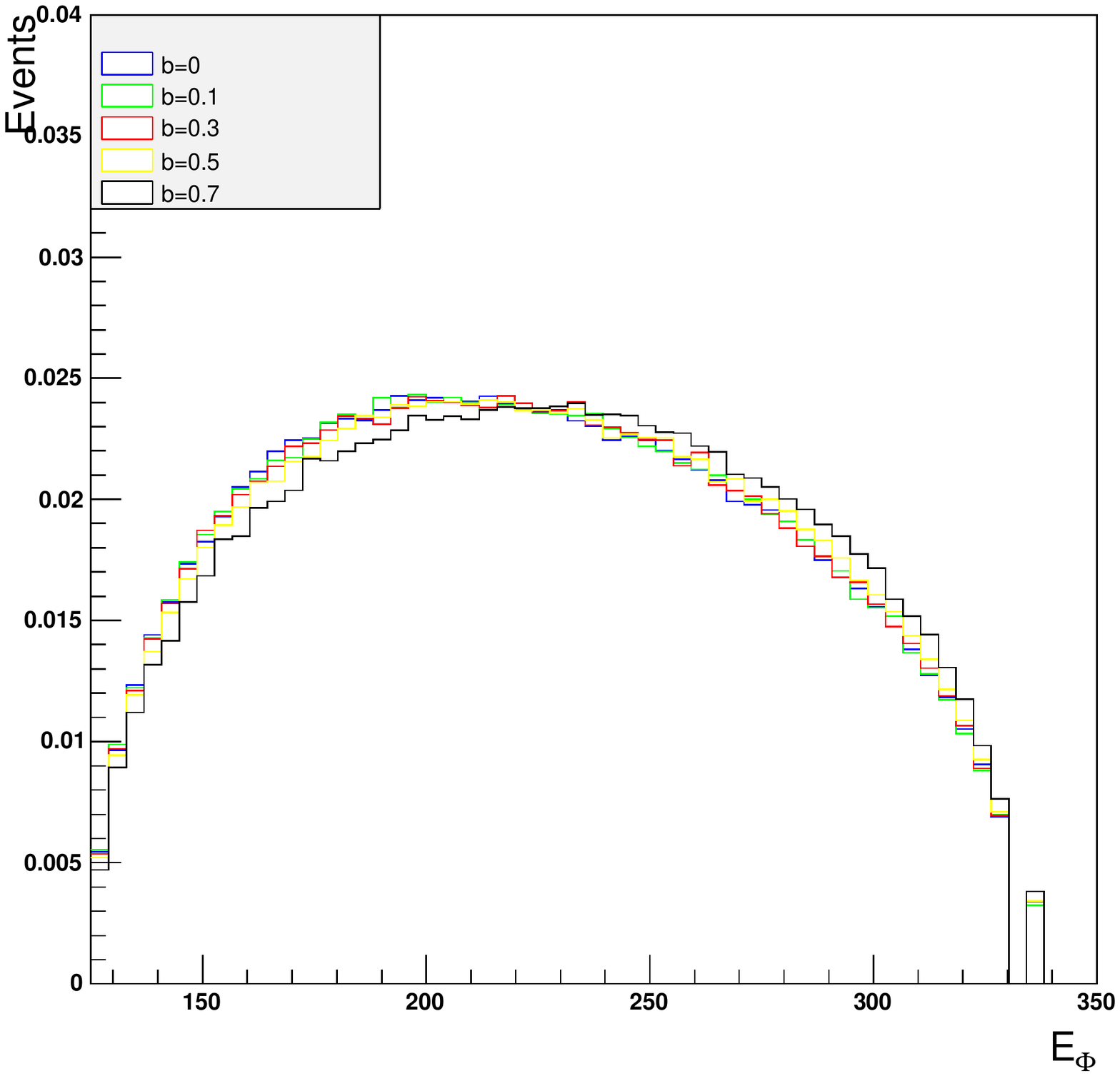}\\
\includegraphics[angle=0,width=60mm]{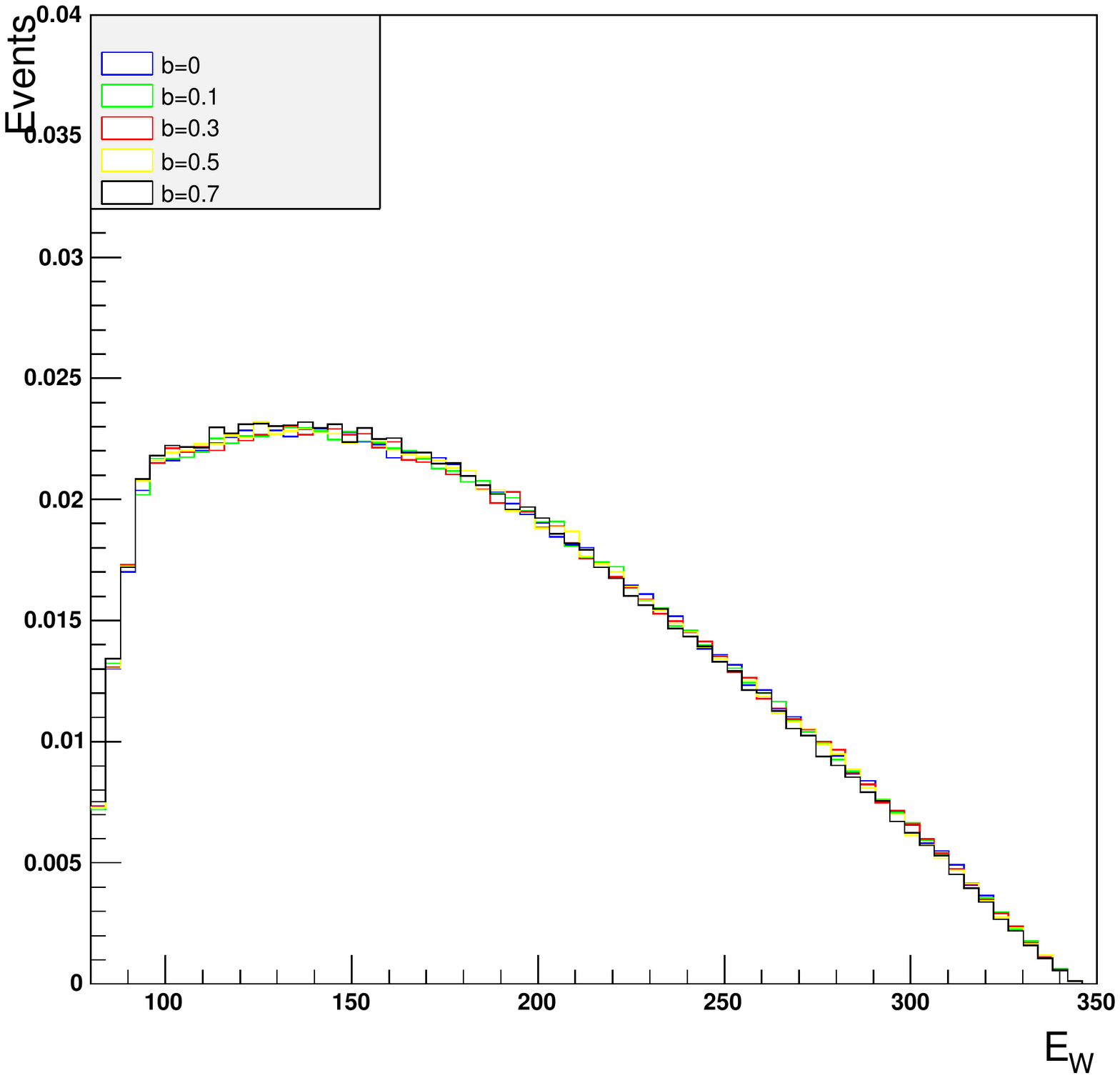} &
\includegraphics[angle=0,width=60mm]{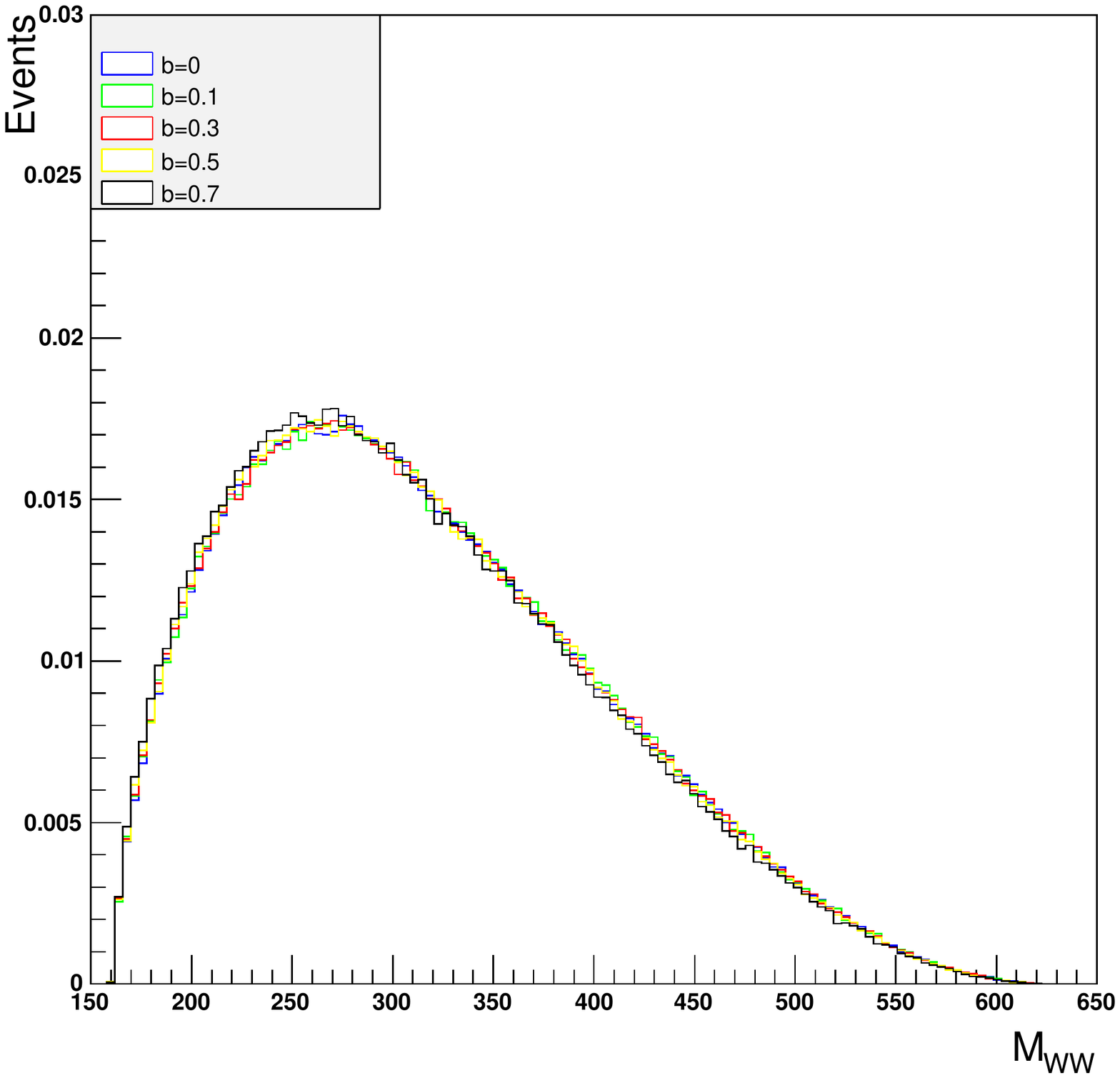}&
\includegraphics[angle=0,width=60mm]{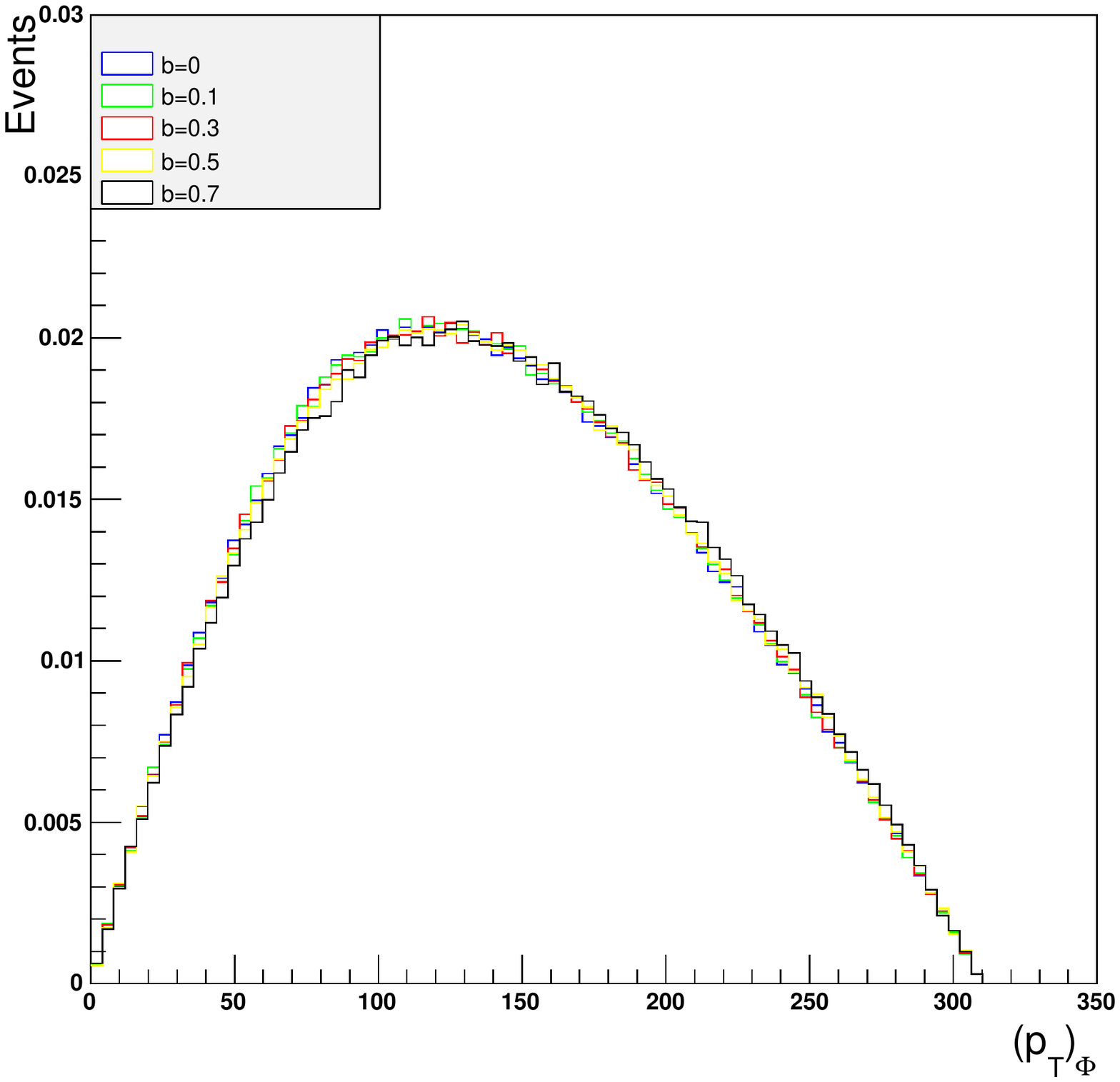} \\
\includegraphics[angle=0,width=60mm]{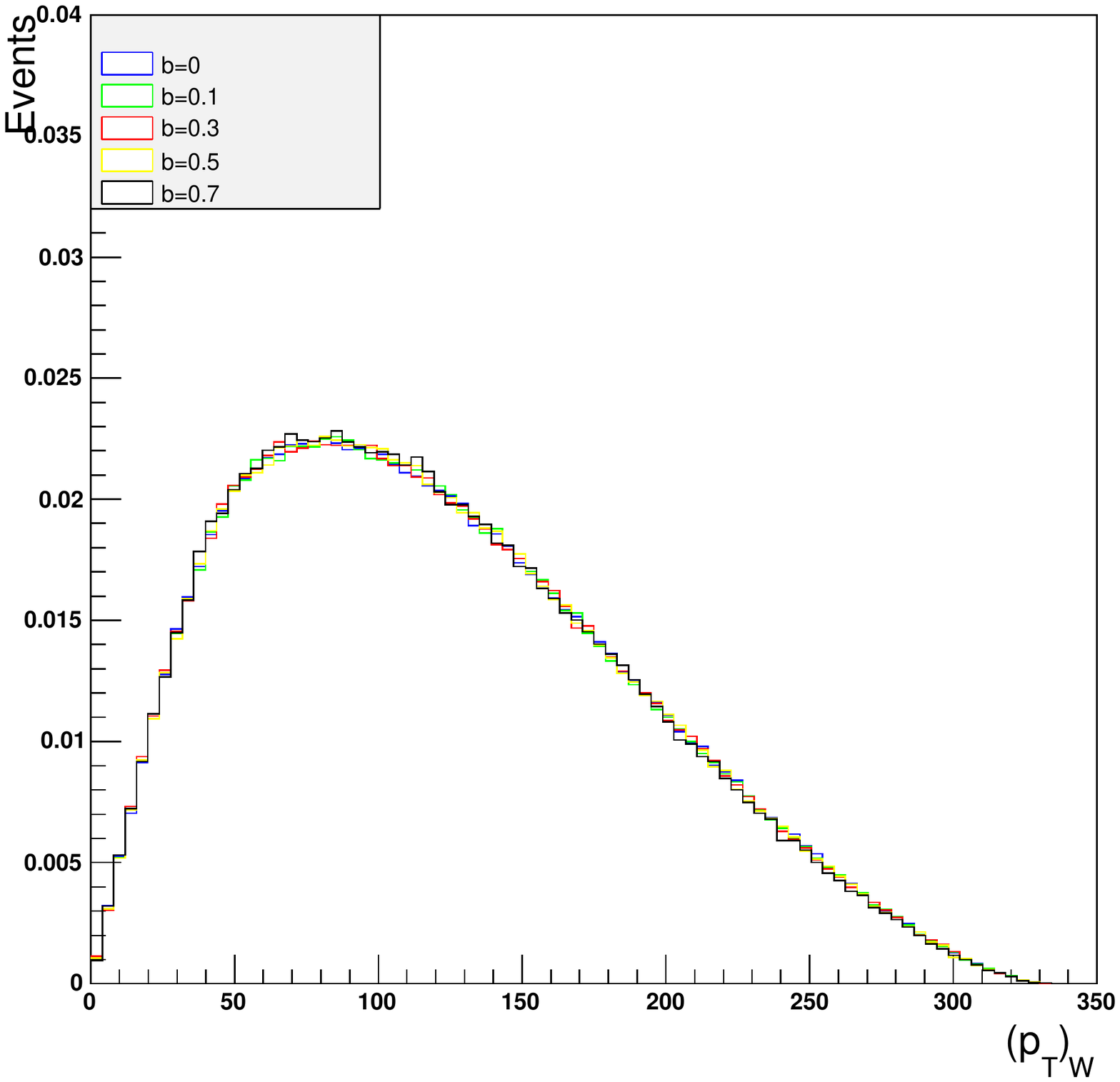}&
\includegraphics[angle=0,width=60mm]{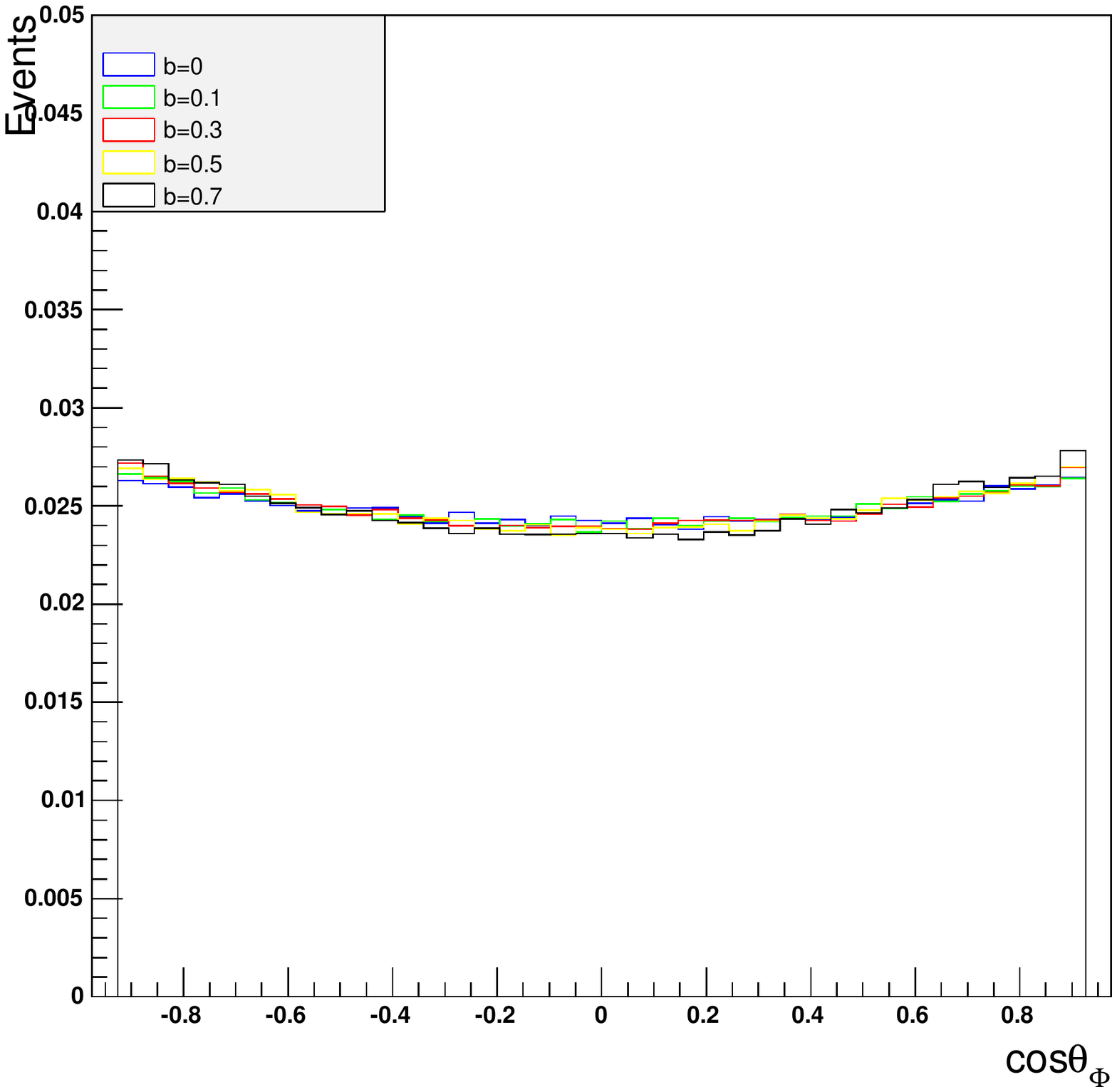}&
\includegraphics[angle=0,width=60mm]{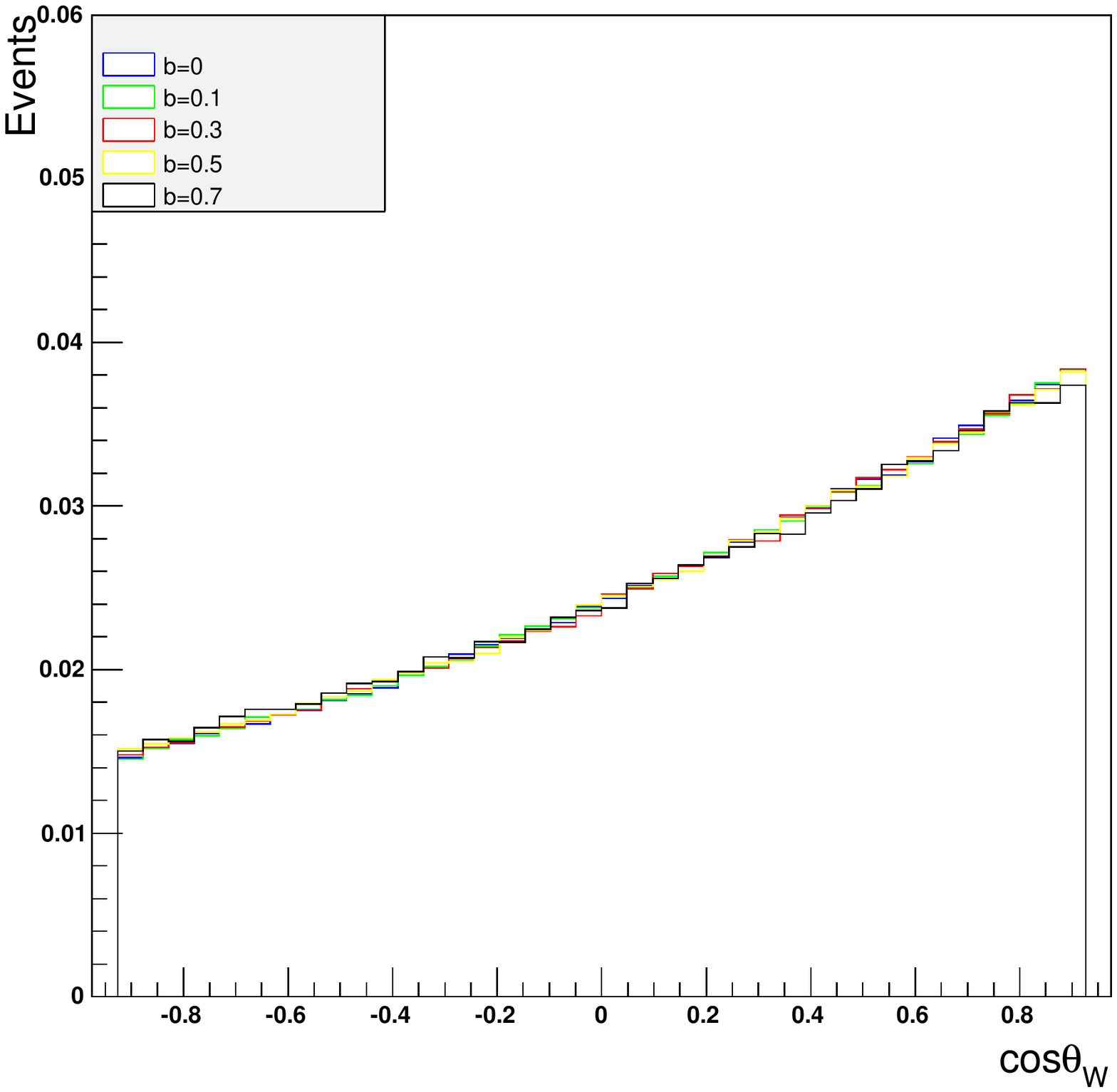}
\end{tabular}
%\vspace*{-2cm}
\caption{Model I: Normalized distributions of top quark energy, $E_t$, $\cos\theta_t$, invariant mass of $t\bar t$,  $M_{t\bar{t}}$, transverse momentum of $t$, $(p_T)_{t}$, and 
the cosine of angle between $b$ abd $\bar b$ in $e^+e^-\rightarrow t\bar t \Phi\rightarrow b\bar b t \bar{t}$; and energy , $E_\Phi$, transverse momentum, $(p_T)_H$ and
$\cos\theta_H$ of the Higgs boson, Energy and transverse momentum of $W$, $E_{\Phi}, (p_T)_W$, 
and $\cos\theta_W$, and the invariant mass of $WW$, $M_{WW}$ in $e^+e^-\rightarrow t\bar t \Phi\rightarrow b\bar b W^+W^-\Phi$. The centre of mass energy considered is, $\sqrt{s}=800$\,GeV, and an integrated
luminosity of $1000~{\rm fb}^{-1}$ is used. }
\label{fig:M1ndist800}
\end{figure}

\begin{figure}[!t]\centering
\begin{tabular}{c c c  } 
\includegraphics[angle=0,width=60mm]{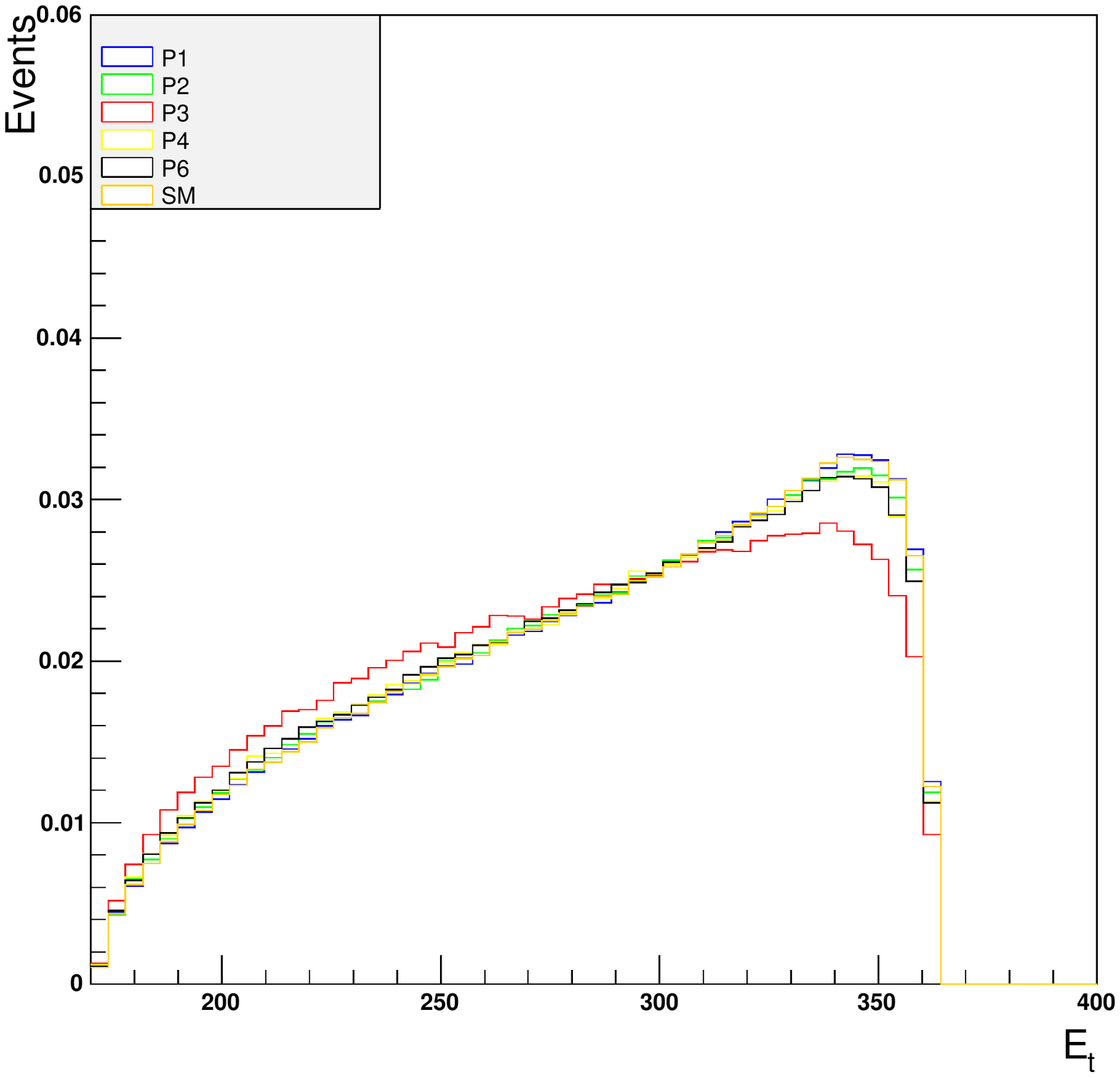} &
\includegraphics[angle=0,width=60mm]{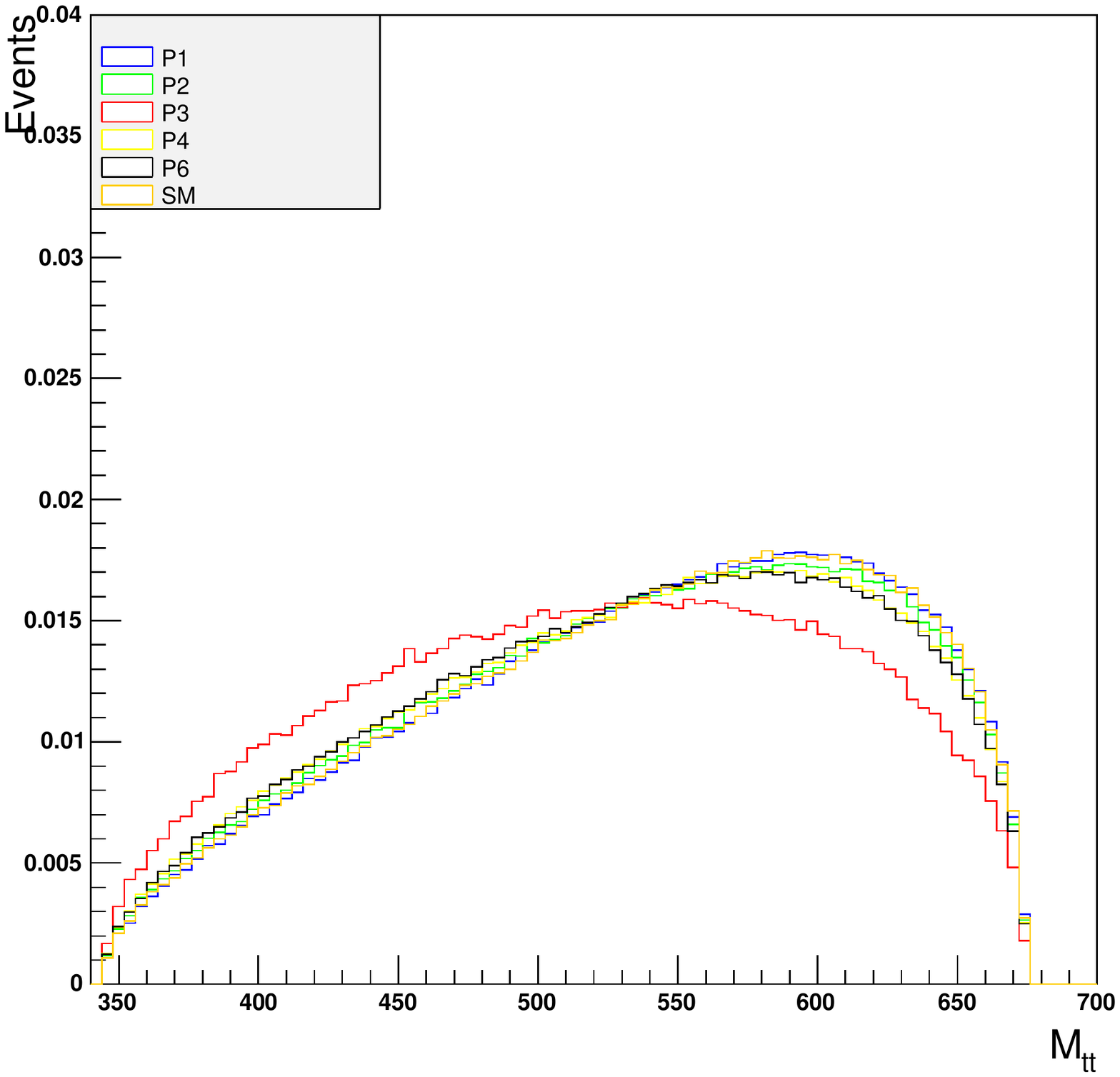}&
\includegraphics[angle=0,width=60mm]{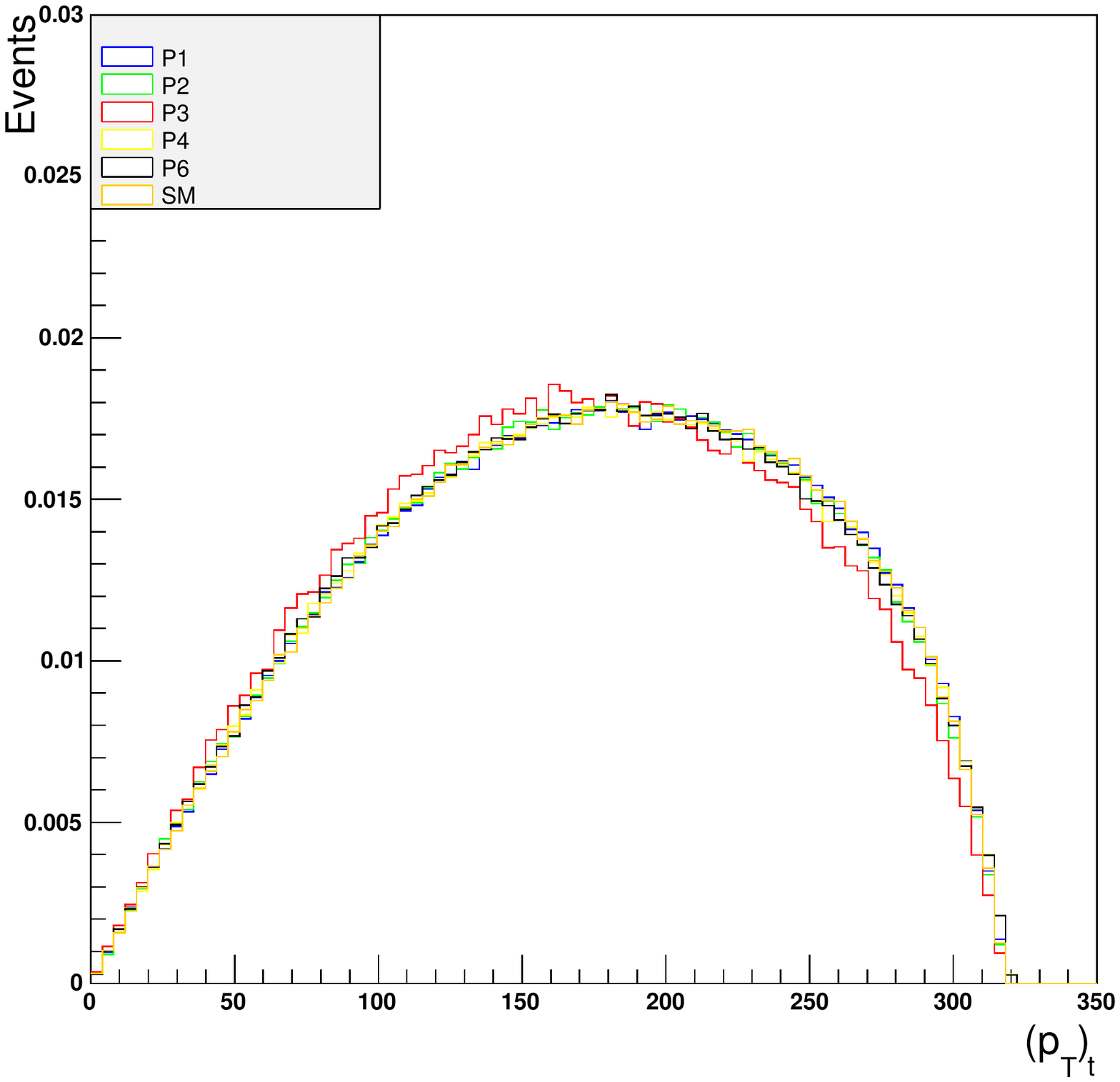} \\
\includegraphics[angle=0,width=60mm]{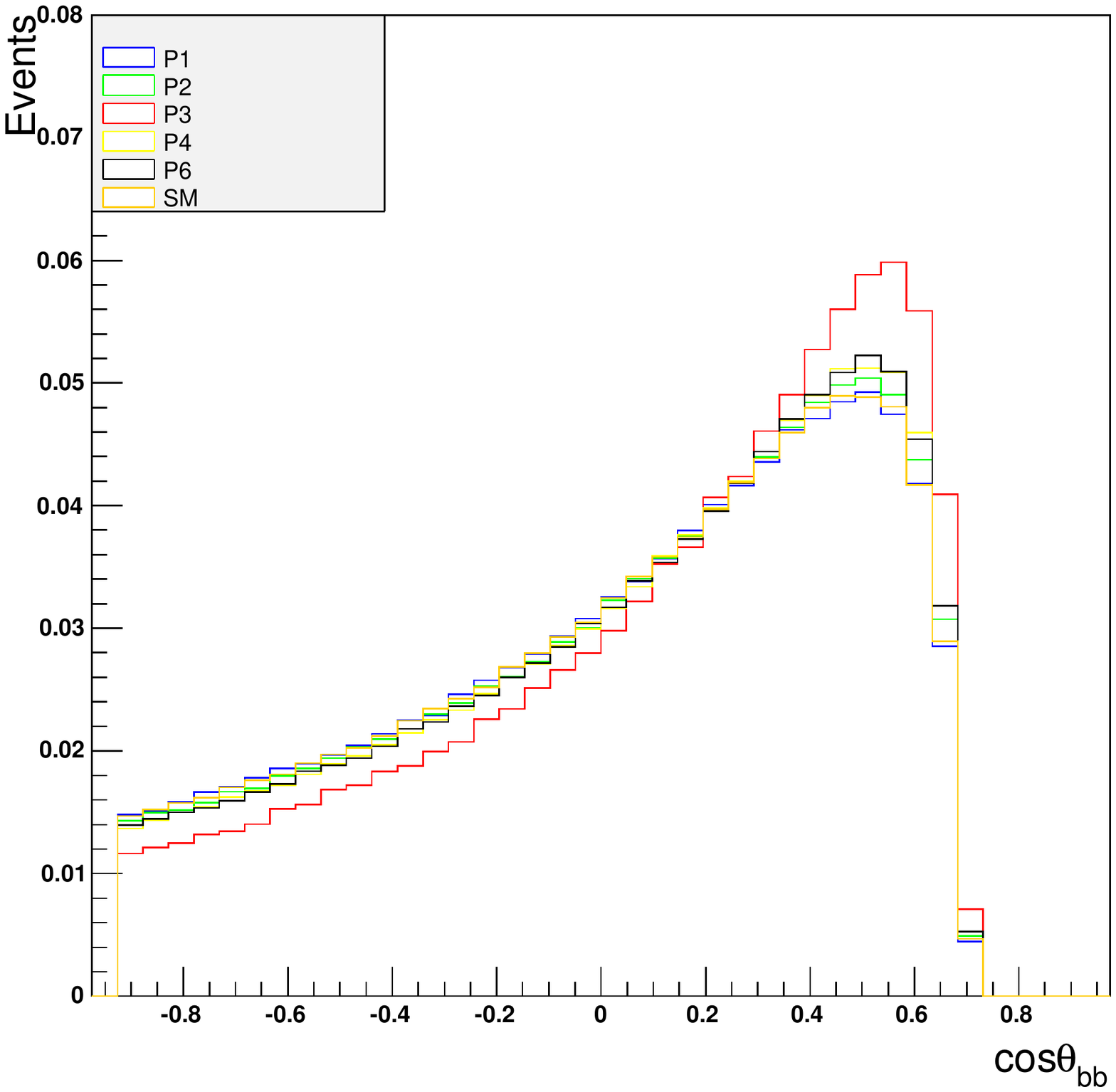}&
\includegraphics[angle=0,width=60mm]{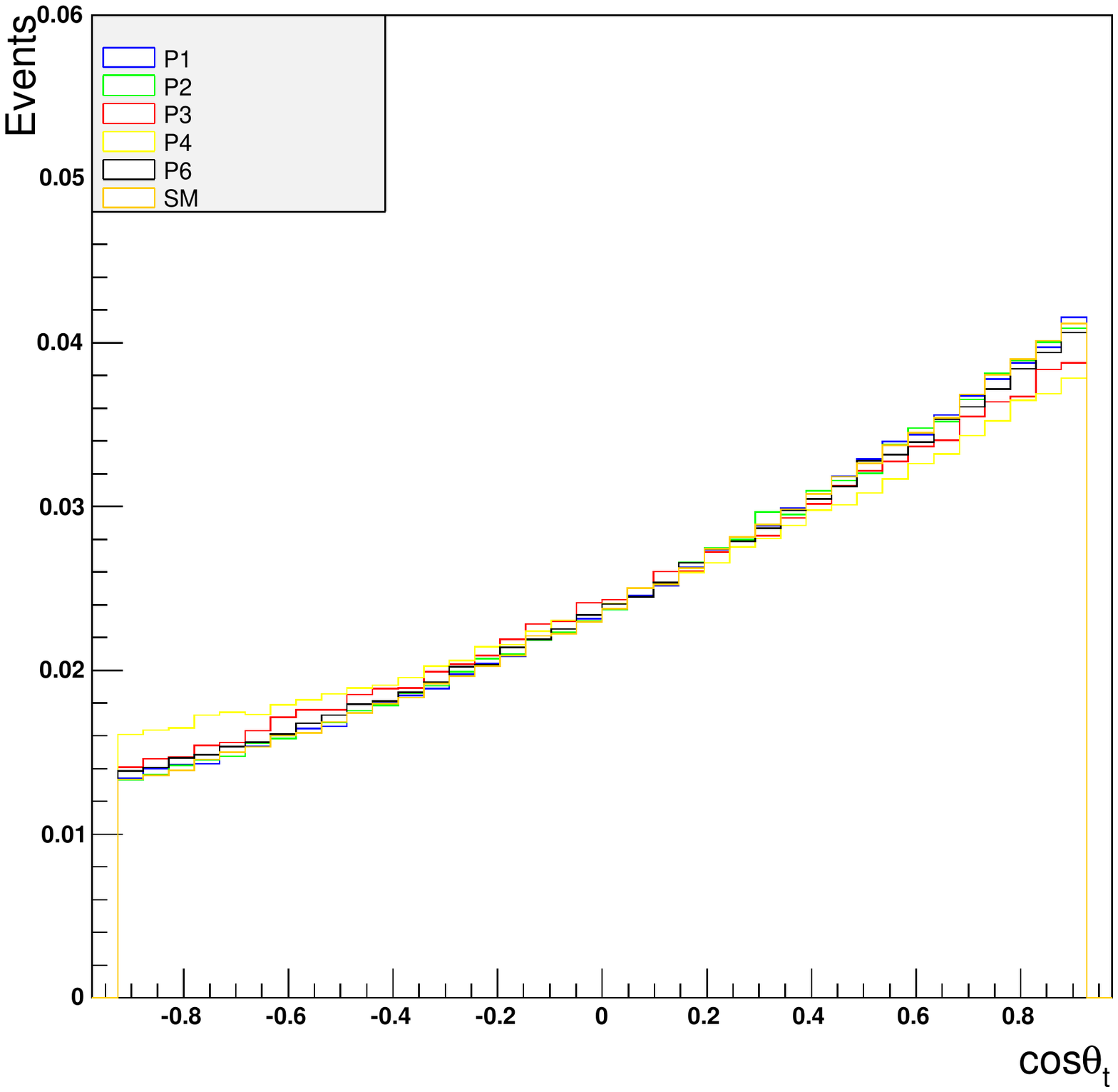} &
\includegraphics[angle=0,width=60mm]{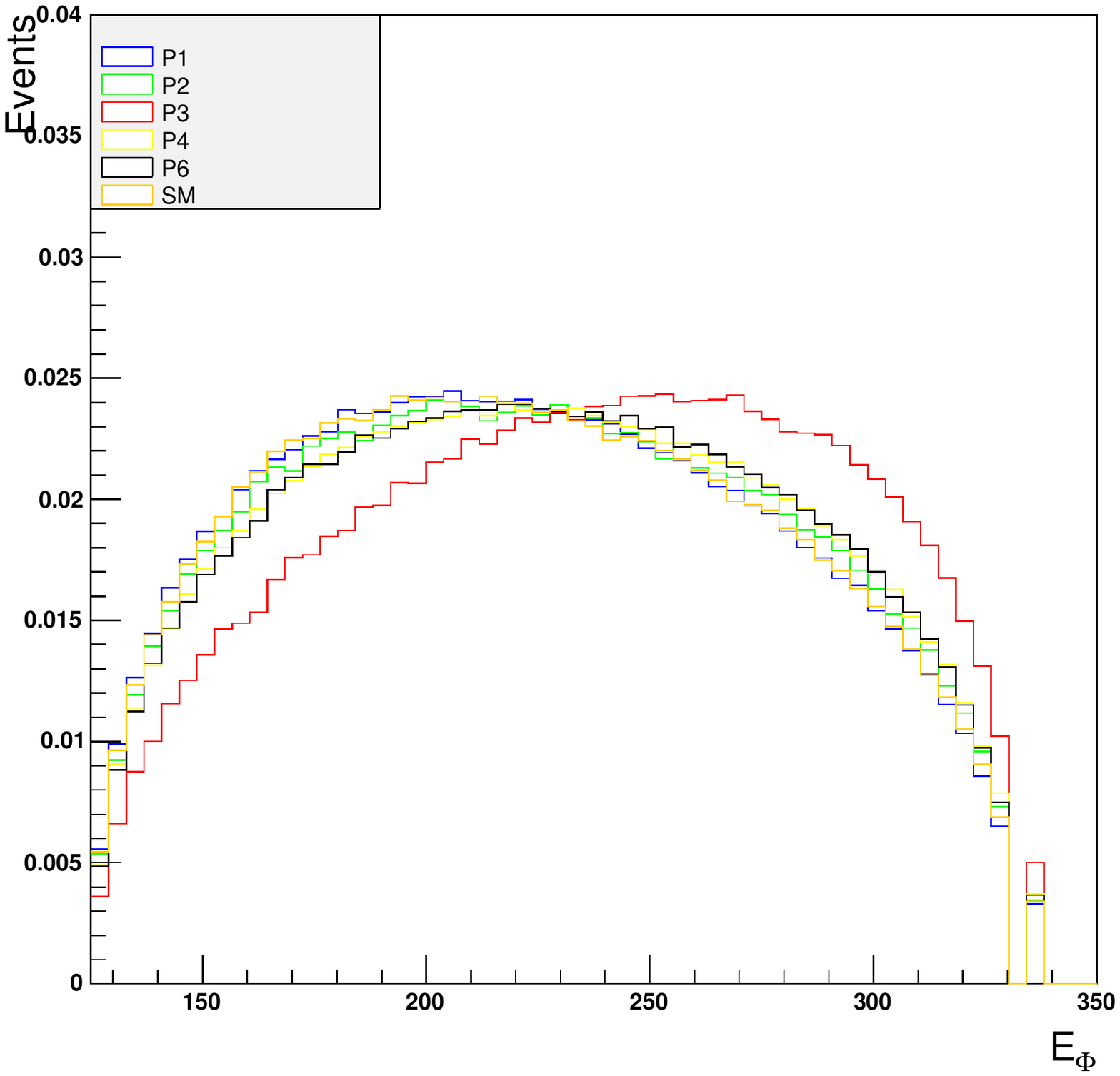}\\
\includegraphics[angle=0,width=60mm]{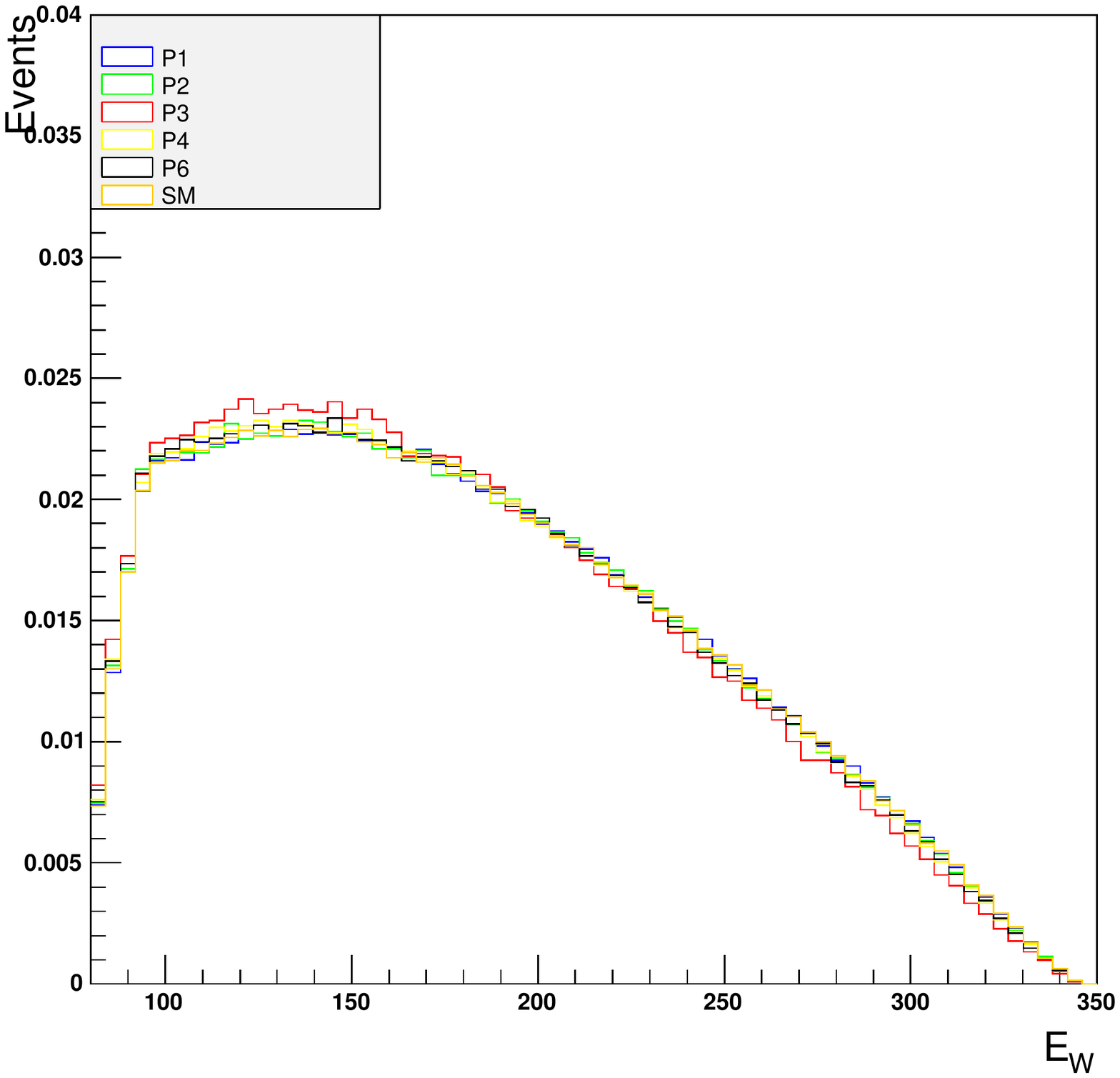} &
\includegraphics[angle=0,width=60mm]{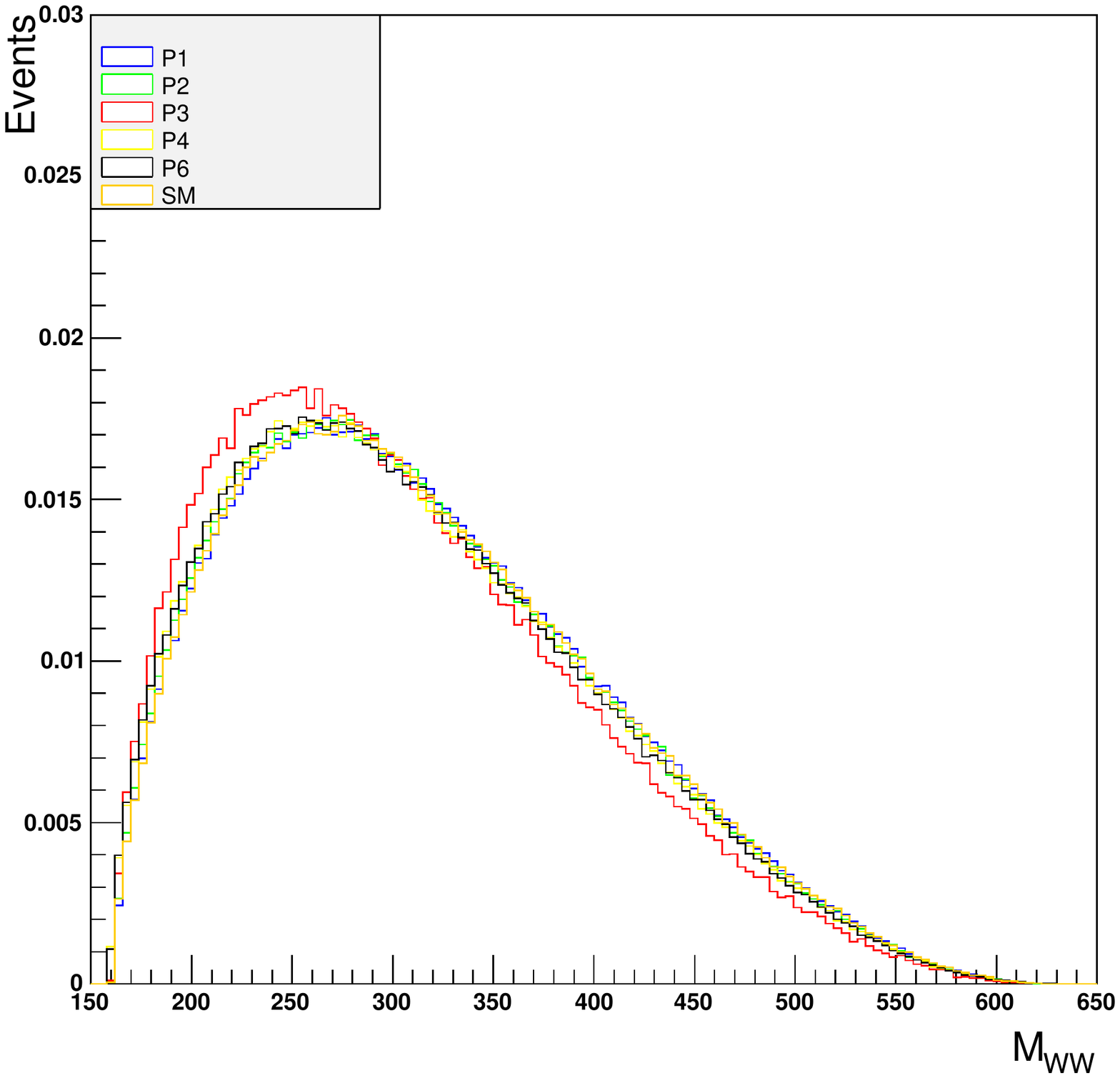}&
\includegraphics[angle=0,width=60mm]{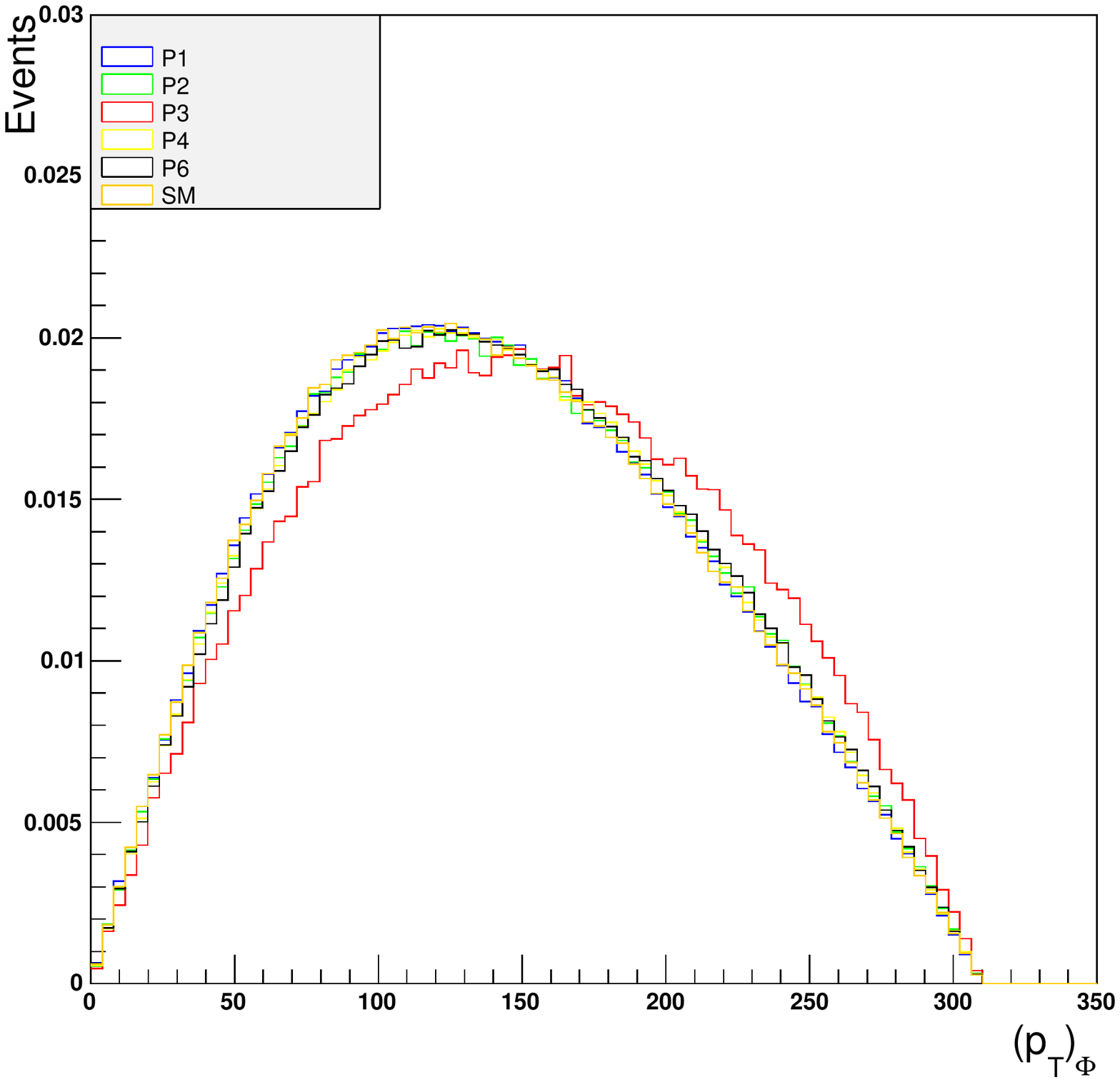} \\
\includegraphics[angle=0,width=60mm]{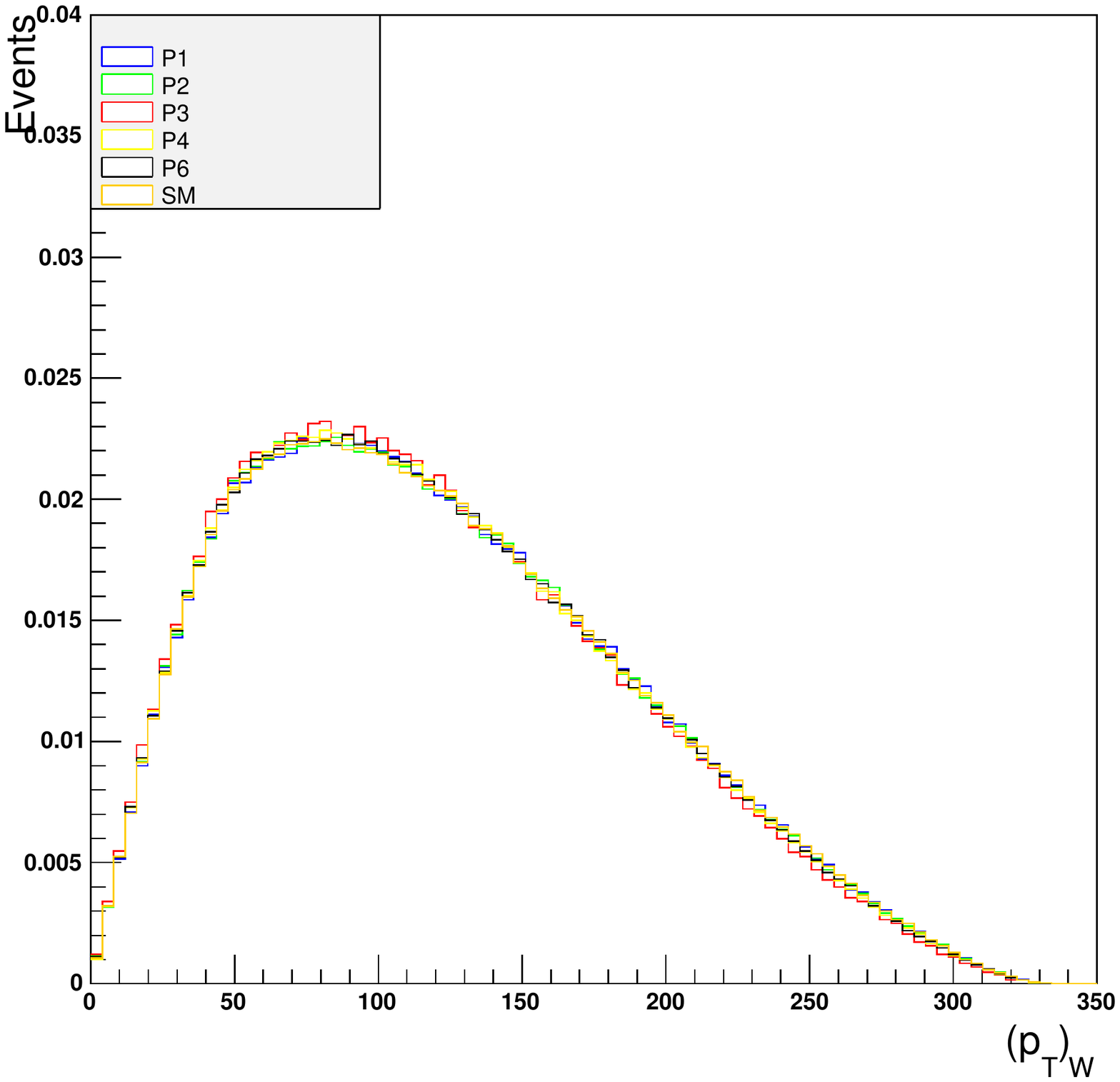}&
\includegraphics[angle=0,width=60mm]{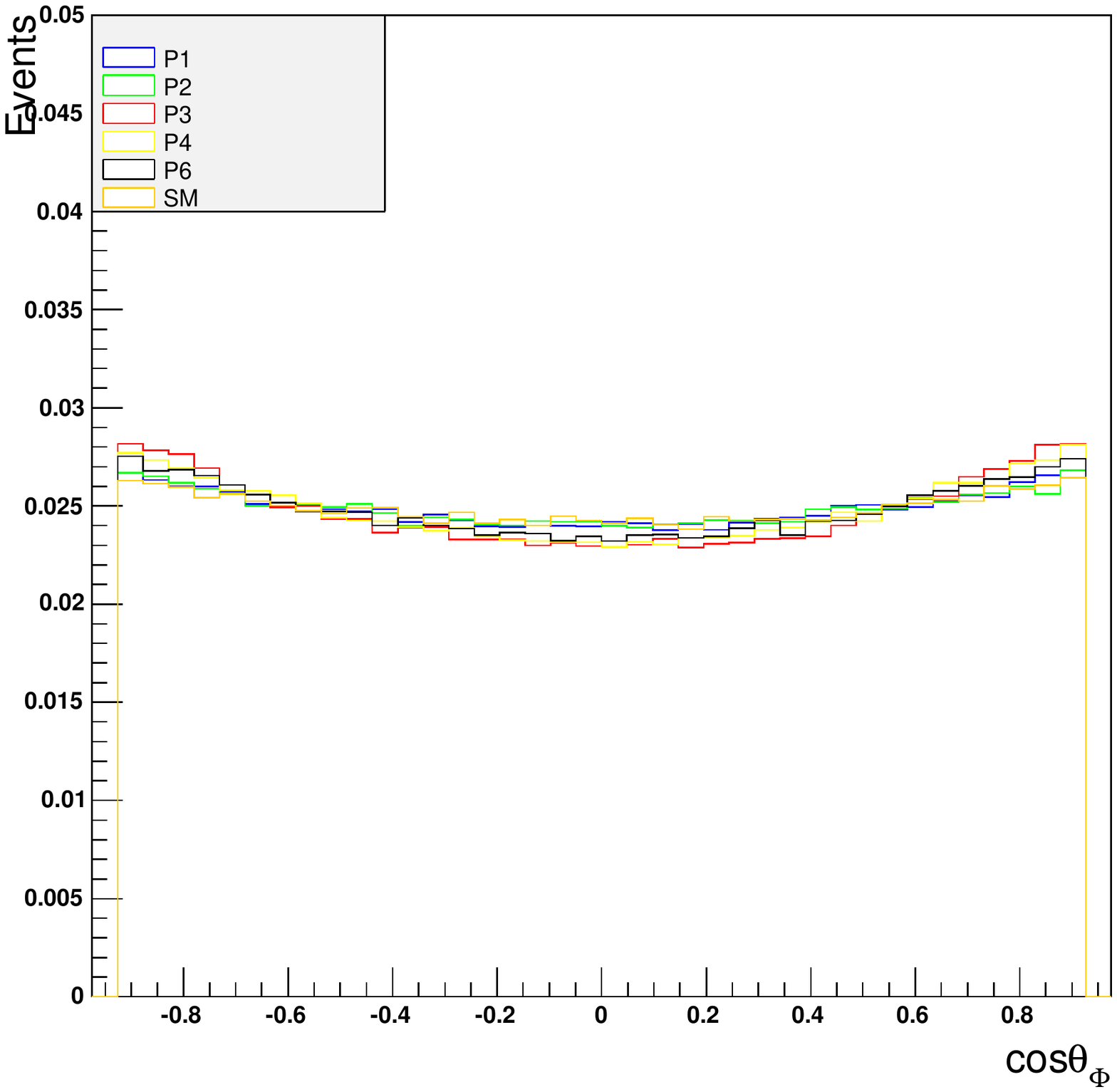}&
\includegraphics[angle=0,width=60mm]{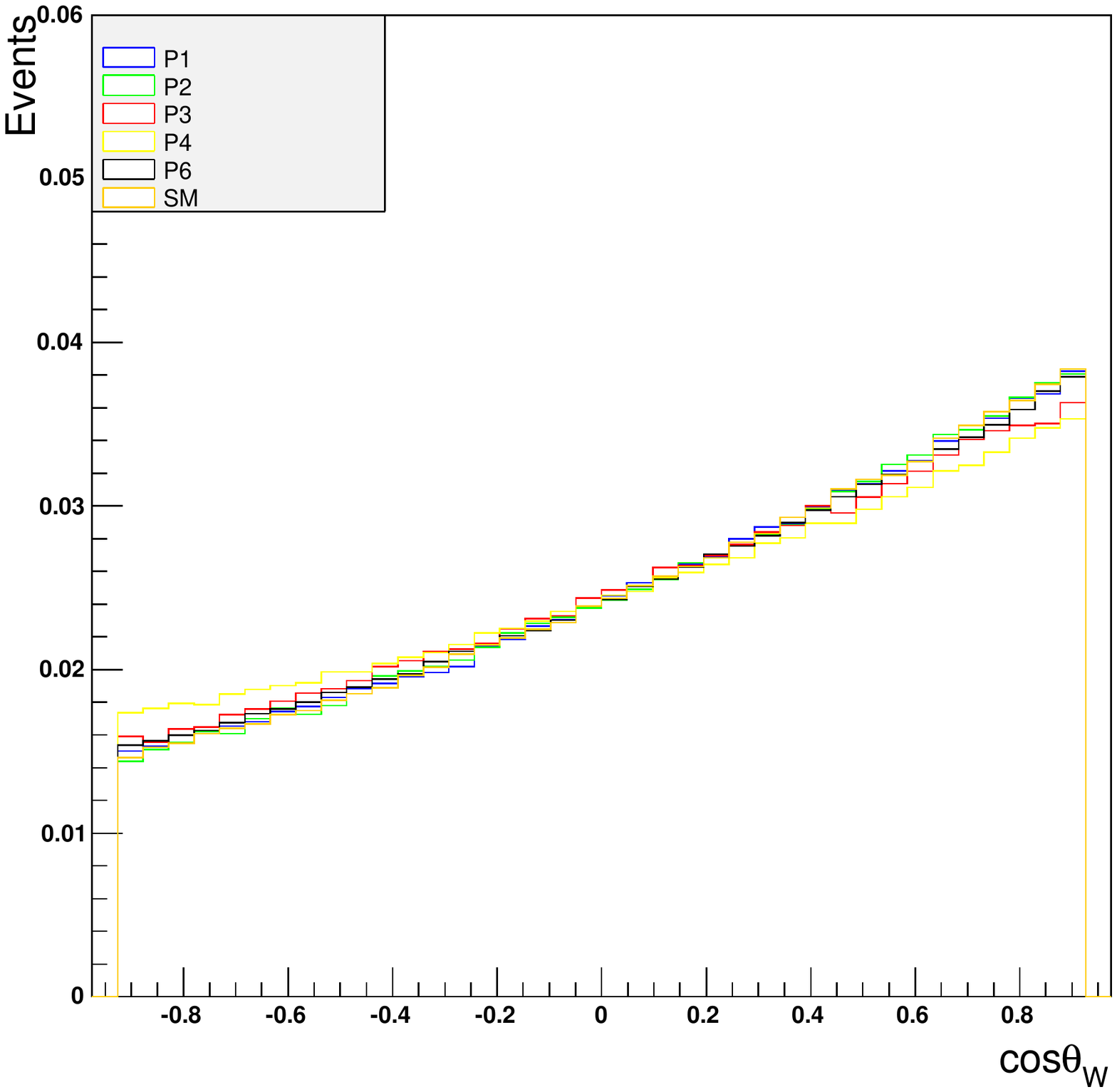}
\end{tabular}
%\vspace*{-2cm}
\caption{Model II: Normalized distributions of top quark energy, $E_t$, $\cos\theta_t$, invariant mass of $t\bar t$,  $M_{t\bar{t}}$, transverse momentum of $t$, $(p_t)_{t}$, and 
the cosine of angle between $b$ abd $\bar b$ in $e^+e^-\rightarrow t\bar t \Phi\rightarrow b\bar b t \bar{t}$; and energy , $E_\Phi$, transverse momentum, $(p_T)_H$ and
$\cos\theta_H$ of the Higgs boson, Energy and transverse momentum of $W$, $E_{\Phi}, (p_T)_W$, 
and $\cos\theta_W$, and the invariant mass of $WW$, $M_{WW}$ in $e^+e^-\rightarrow t\bar t \Phi\rightarrow b\bar b W^+W^-\Phi$. The centre of mass energy considered is, 
$\sqrt{s}=800$\,GeV, and an integrated luminosity of $1000~{\rm fb}^{-1}$ is used. }
\label{fig:M2ndist800}
\end{figure}

In Fig.~\ref{fig:M1ndist800} and Fig.~\ref{fig:M2ndist800}, we present various distributions corresponding to $t$ and $\Phi$ decay for Model 
I and Model II resepectively. As visible from the figures of Model I, only $M_{t\bar{t}}$ and $E_{\Phi}$ show a very slight departure from the SM case
for a much larger $b$ value of 0.7. However for Model II, except for the benchmark point P3 and in some cases P4, the distributions follow the SM trend. Like in the case of total cross section, the sign of $b$ does not really show up in the distributions  as well. While P3 shows perceivable deviation from the rest of the cases in the $E_t$, $M_{tt}$, $\cos\theta_{bb}$, $E_\Phi$, $M_{WW}$ and $(P_T)_\Phi$ distributions, P4 differes from others in $\cos\theta_t$ and $\cos\theta_W$ distributions. 
Note that, P3 corresponds to the case with small value of $a$ and large value of $b$.  Whereas in P3 it is the differnece in the
magnitude of the parameters that play the role, in the case of P4 it is the sign of the parameter $a$ that seems to affect the distribution.  We must note here that it is in fact the relative sign between $a$ and $c$ that matters, indicating that the interference term between the Higgs-stahlung process with the Higgs radiation off the top-quark is indeed the case of this, as expected. This was confirmed by comparing the results between P6 and P7, which concide with each other.
Thus, the conclusion that we may draw from these analyses is that, while for small deviations from the SM case the strategy used to obtain the signal sensitivity may be followed, 
one need to be cautious in general. We would like to alert the reader, that in general, the values of the parameters $a$ and $b$ do not directly relate to the scalar-psuedoscalar mixing. Rather, these parameters could be more complicated functions involving the mixing angle along with other parameters of the model. It is therefore, in principle possible to have deviations of $a$ and $b$ from the SM values of unity and zero, respectively, even if the Higgs resonance is mostly CP-even.

In the rest of this section we shall consider the cases with centre of mass energies of  $500$ GeV and $1000$ GeV.

\subsection{$\sqrt{s}=500$ GeV}

This case for SM was studied in Ref.~\cite{YITFKSY} using initial beam polarization with $(P_{e^-}, P_{e^+})=(-0.8, +0.3)$ and an integrated luminosity
of 1000 fb$^{-1}$ for a Higgs mass of $M_{\Phi}=120$ GeV. Here we obtained our signal events from WHIZARD for $M_{\Phi}=125$ GeV by factorizing the 
process into its production and decay part as explained previously. We follow Ref.~\cite{YITFKSY}, and work with polarized beams as specified above. This deviation in the treatment compared to the $\sqrt{s}=800$ GeV case is purely due to the limited scope adopted in this study. As made clear in the earlier sections, our main goal is to illustrate the complications that might arise in the case of a general $t\bar t\Phi$ coupling, which is more apt to perform a model independent investigation. To this effect, we would like to make direct comparisons with the existing detailed study performed strictly within the framework of the SM. A more complete analysis including advantages of beam polarization over the case of unpolarized beams, with more realistic detector simulations, and adopting strategies independent of the ones considered for the SM case is beyond the scope of this study.

Being close to the threshold, unlike the case of $\sqrt{s}=800$ GeV,  in the present case of  $\sqrt{s}=500$ GeV, $t\bar{t}$  bound state effects play significant role, which should be taken care of properly in the signal and background processes.Due to this bound state effects, the
tree level amplitudes change by a factor ~\cite{YITFKSY, Kolodziej:2006nr}
\begin{equation}
 R_{t\bar{t}} = \frac{A_{t\bar{t}}(i\rightarrow f)}{[A_{t\bar{t}}(i\rightarrow f)]_{tree}} = \sqrt{K_{i\rightarrow f}}\times F(\hat{s}_{t\bar{t}},
 \overrightarrow{p}; m_t, \Gamma_t, \alpha_s)
\end{equation}

Here $F$ encodes the process independent bound state effects, which is a function of the centre of mass energy of $t\bar t$, $\sqrt{\hat{s}_{t\bar{t}}}$, 
the three momentum of $t$ in the CM frame of $t\bar t$, $\overrightarrow{p}$,  the pole mass ($m_t$) and the width ($\Gamma_t$) of the top quark. The factor $K_{i\rightarrow f}$ is the hard vertex correction
factor, which is taken to be $0.843$ for the signal, whereas considered to be unity for background processes, giving an overall enhancement factor($R_{t\bar{t}}$) of  1.28 in signal events \cite{YITFKSY}. Since these effects are independent of the CP parameters, we will use the same factor
in our study. \\

\begin{figure}[!t]\centering
\begin{tabular}{c c} 
\hspace{-5mm}
\includegraphics[angle=0,width=80mm]{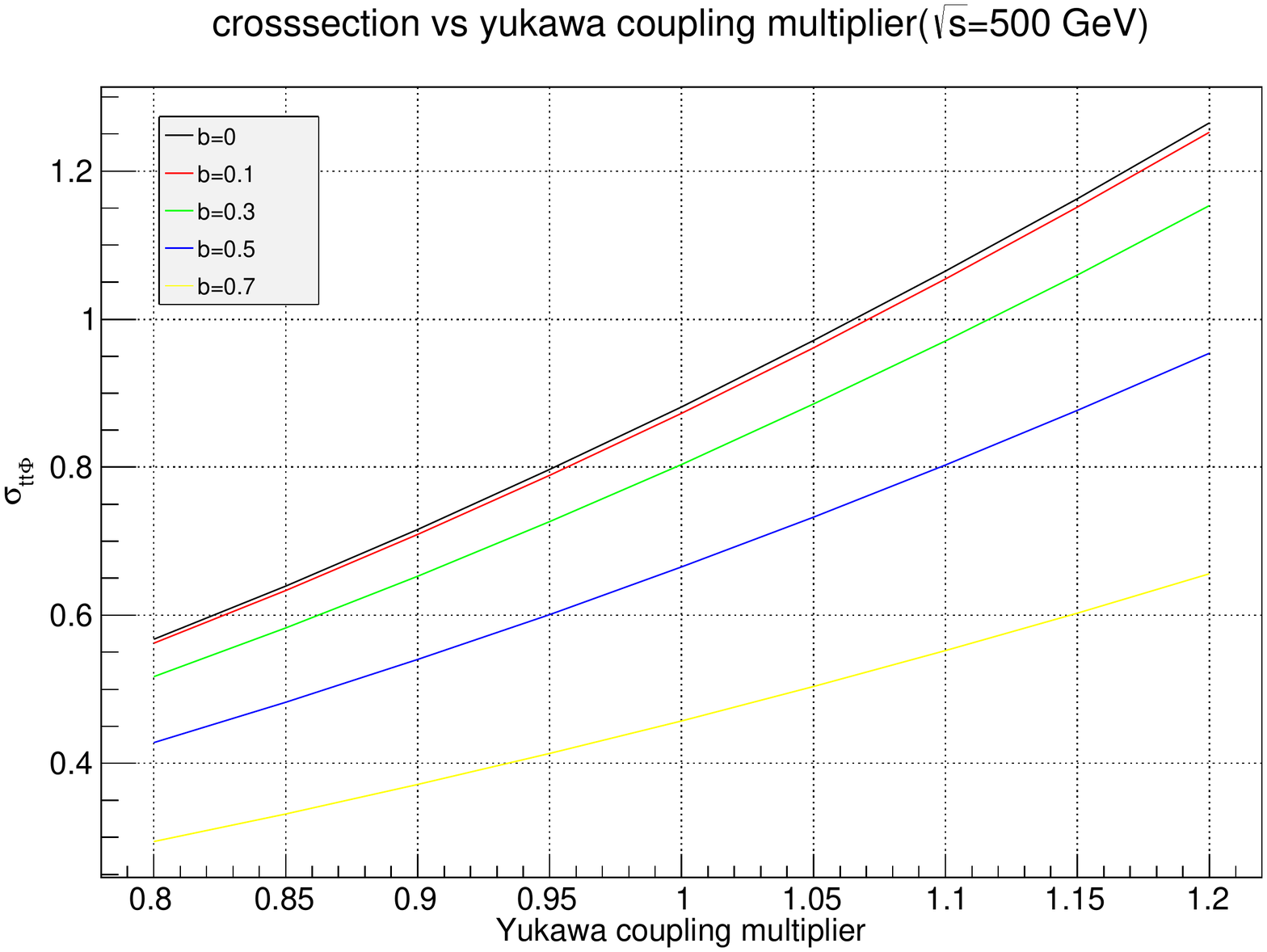} &
\includegraphics[angle=0,width=80mm]{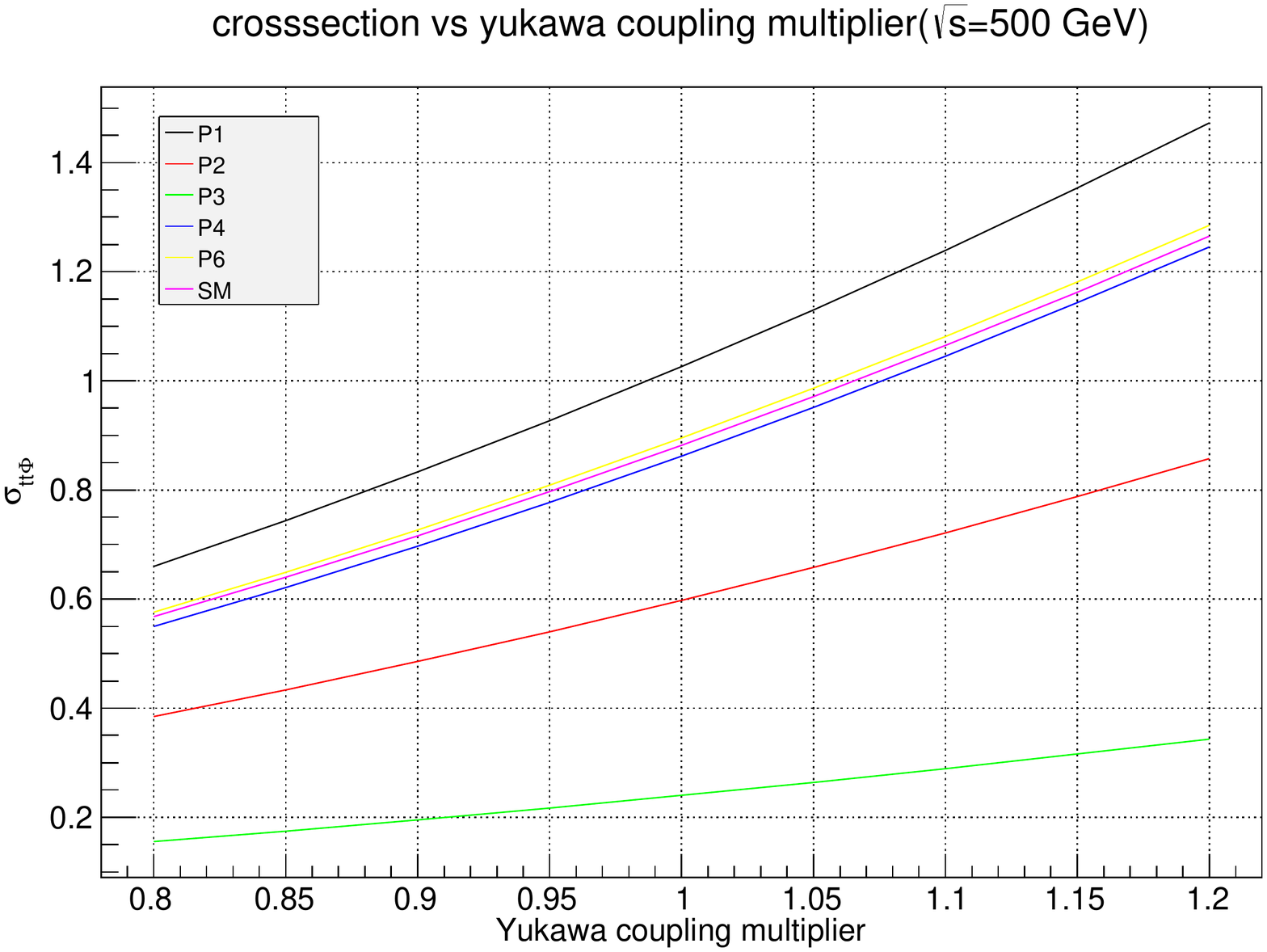}\\
\end{tabular}
%\vspace*{-7cm}
\caption{$\sqrt{s}=500$ GeV: $\sigma_{t\bar{t}\Phi}$ vs Yukawa multiplier for different parameter values in Model I(left fig.) and Model II(Right fig.). }
\label{fig1L.3}
\end{figure}

To determine the prefactor, we consider the dependence of the cross section on the Yukawa coupling. In Fig.~\ref{fig1L.3} we present the variation of the cross section with the Yukawa coupling multiplier for our Model points. Fitting to the quadratic polynomial leads to the following equations for the respective cases.

\begin{eqnarray}
b=0.0: && \hspace{0.3cm}\sigma_{t\bar{t}\Phi} = 0.858 (\lambda_t-1)^2 + 1.743(\lambda_t-1)+ 0.881,\nonumber\\
b=0.1: && \hspace{0.3cm}\sigma_{t\bar{t}\Phi} = 0.858 (\lambda_t-1)^2 + 1.726(\lambda_t-1)+ 0.872,\nonumber\\
b=0.3: && \hspace{0.3cm}\sigma_{t\bar{t}\Phi} = 0.776 (\lambda_t-1)^2 + 1.589(\lambda_t-1)+ 0.803,\nonumber\\
b=0.5: && \hspace{0.3cm}\sigma_{t\bar{t}\Phi} = 0.656 (\lambda_t-1)^2 + 1.315(\lambda_t-1)+ 0.665,\nonumber\\
b=0.7: && \hspace{0.3cm}\sigma_{t\bar{t}\Phi} = 0.446 (\lambda_t-1)^2 + 0.904(\lambda_t-1)+ 0.457
\end{eqnarray}

and

\begin{eqnarray}
P1: && \hspace{0.3cm}\sigma_{t\bar{t}\Phi} = 1.018 (\lambda_t-1)^2 + 2.031(\lambda_t-1)+ 1.026,\nonumber\\
P2: && \hspace{0.3cm}\sigma_{t\bar{t}\Phi} = 0.584 (\lambda_t-1)^2 + 1.180(\lambda_t-1)+ 0.597,\nonumber\\
P3: && \hspace{0.3cm}\sigma_{t\bar{t}\Phi} = 0.226 (\lambda_t-1)^2 + 0.470 (\lambda_t-1)+ 0.240, \nonumber\\
P4: && \hspace{0.3cm}\sigma_{t\bar{t}\Phi} = 0.871(\lambda_t-1)^2 + 1.739 (\lambda_t-1)+ 0.862, \nonumber\\
P6: && \hspace{0.3cm}\sigma_{t\bar{t}\Phi} = 0.876 (\lambda_t-1)^2 + 1.772 (\lambda_t-1)+ 0.895
\end{eqnarray}

In the SM, the prefactor is 0.50 which means the contribution of the 3rd diagram is negligible. For Model I, the prefactor
value does not change at all for all considered points. The similar situation persists in Model II except for the points P3 and P4 for which 
its value is 0.51 and 0.49, respectively. Thus at this centre of mass energy the sensitivity of the Yukawa coupling will be governed purely
by $\Delta\sigma/\sigma=\sqrt{S+B}/S$ factor.

As far as the determination of $\Delta\sigma/\sigma$ is concerned, the introduction of various kinematical cuts~\cite{YITFKSY} like mass, 
b-tagging, thrust etc leads to the depletion of signal events by a factor of 0.173 and 0.139 in the hadronic and semileptonic mode respectively. 
As emphasized earlier we will use the same reduction factors for the representative points in our scenarios. The corresponding top Yukawa
sensitivity($g_t$) with associated signal significance is given in Table~\ref{500gtM1} and Table~\ref{500gtM2} for Model I and Model II respectively.
This case gives sensitivity of around 21.4\% and 20.7\% for hadronic and semileptonic mode in SM. The sensitivity is further dropped to 37.3\% and
35.5\% for the largest considered mixing $b=0.7$ in Model I. For Model II, the best sensitivity of around 19\% is obtained in the semileptonic mode
for P6 and the worst scenario is for P3 where it reaches 76.2\% in the hadronic mode.

\begin{table} 
\begin{center}
\begin{tabular}{lccccccc}

\hline

\hline

$b$ & Prefactor & Signal($S_1$) & $\frac{\Delta\sigma}{\sigma}$ & $\frac{\Delta g_t}{g_t}$& Signal($S_2$) & $\frac{\Delta\sigma}{\sigma}$ & $\frac{\Delta g_t}{g_t}$ \\

\hline%---------------------------------------------------------------------

0. & 0.505 & 30.1 & 0.42 & 0.214 & 23.4 & 0.41 & 0.207 \\

\hline%-------------------------------------------------------

0.1 & 0.505 & 29.8 & 0.43 & 0.216 & 23.2& 0.41 & 0.209\\

\hline%---------------------------------------------------------------------

0.3  & 0.505 & 27.7 & 0.46 & 0.231   & 21.6& 0.44& 0.223\\

\hline%---------------------------------------------------------------------

0.5  & 0.505 & 23.4  & 0.54 & 0.270   & 18.2& 0.51 & 0.259\\

\hline%---------------------------------------------------------------------

0.7  & 0.505 & 16.6  & 0.74 & 0.373   & 12.9& 0.71& 0.355\\

\hline%---------------------------------------------------------------------

\end{tabular}
\end{center}

\caption{Model I: Yukawa coupling sensitivity for different parameters at $\sqrt{s}=500$ GeV. $S_1$ and $S_2$ are
          signal events in hadronic and semileptonic mode after kinematical cuts.}

\label{500gtM1} 
\end{table}

\begin{table} 
\begin{center}
\begin{tabular}{lccccccc}

\hline

\hline

Parameter & Prefactor & Signal($S_1$) & $\frac{\Delta\sigma}{\sigma}$ & $\frac{\Delta g_t}{g_t}$& Signal($S_2$) & $\frac{\Delta\sigma}{\sigma}$ & $\frac{\Delta g_t}{g_t}$ \\

\hline%---------------------------------------------------------------------

P1 & 0.505 & 18.4 & 0.67 & 0.339 & 14.3 & 0.64 & 0.323\\

\hline%-------------------------------------------------------

P2  & 0.505 & 21.4  & 0.58 & 0.294  & 16.6&  0.56& 0.281\\

\hline%---------------------------------------------------------------------

P3 & 0.510& 8.0& 1.49 & 0.762  & 6.2 & 1.40 & 0.716\\

\hline%---------------------------------------------------------------------

P4  & 0.495& 32.5 & 0.40 & 0.198  & 25.3&  0.38& 0.192\\

\hline%---------------------------------------------------------------------

P6  & 0.505& 33.7 & 0.38 & 0.195  & 26.2& 0.37 & 0.190\\

\hline%---------------------------------------------------------------------

\end{tabular}
\end{center}

\caption{Model II: Yukawa coupling sensitivity for different parameters at $\sqrt{s}=500$ GeV. $S_1$ and $S_2$ are
          signal events in hadronic and semileptonic mode after kinematical cuts.}

\label{500gtM2} 

\end{table}

\subsection{$\sqrt{s}=1000$ GeV}

This case for SM was considered~\cite{LCtopyuk1000} with a luminosity of $1000$ fb$^{-1}$ split equally between two polarization
states, $(e^+, e^{-})=(\pm 0.2, \mp 0.8)$. The analysis was performed using a cut based and a mutlivariate approach. The background
events were reduced using the number of selection variables like the number of isolated events in the sample, jet clustering algorithm, flavour
tagging, the helicity of the $b\bar{b}$ pair associated with Higgs boson etc. This case gives a signal significance of 7.0(5.2) and a statistical
uncertainty of $7.4\%(9.9\%)$ for hadronic(semileptonic) case on the value of $g_t$.

\begin{figure}[!t]\centering
\begin{tabular}{c c} 
\hspace{-5mm}
\includegraphics[angle=0,width=80mm]{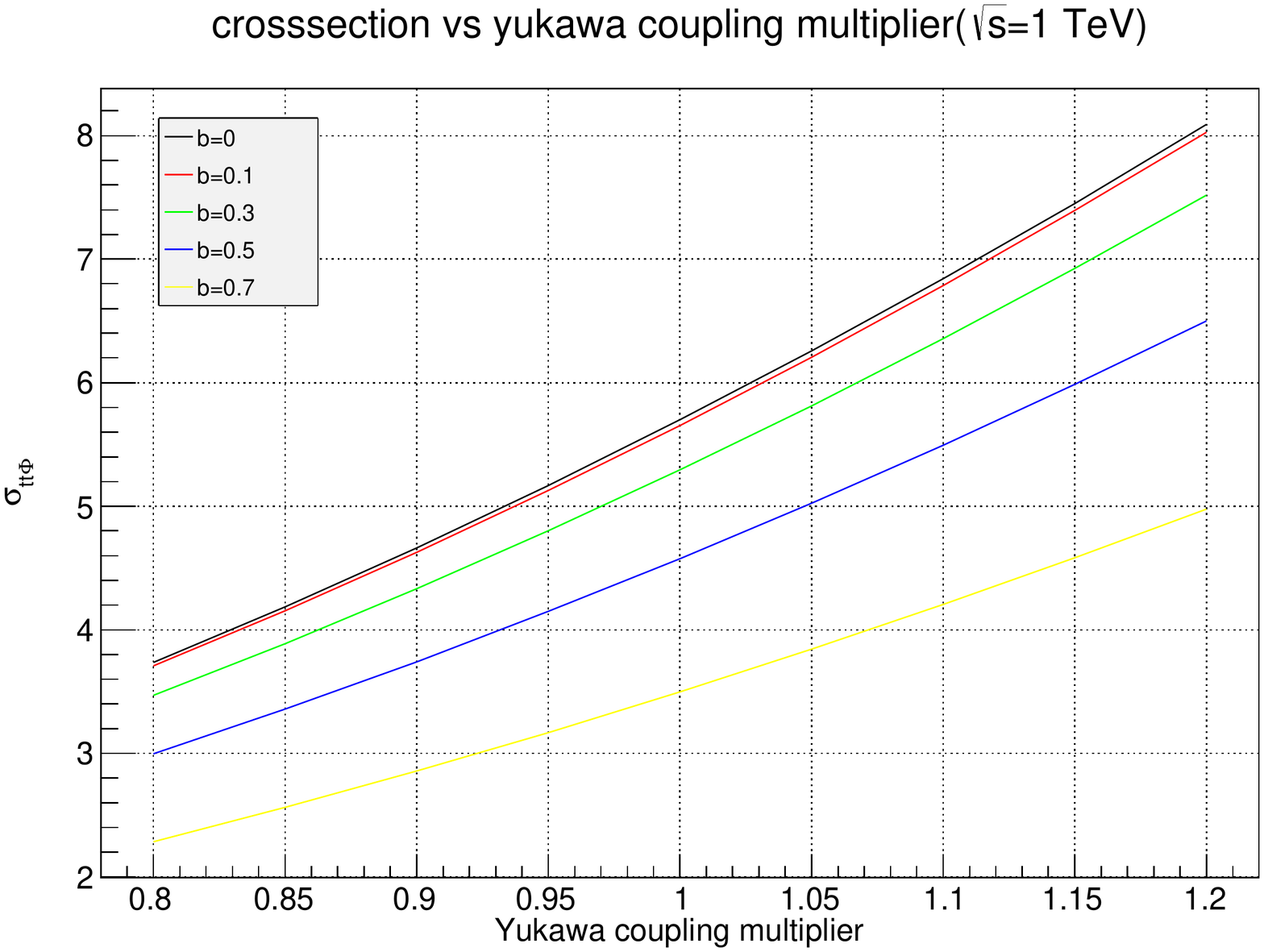} &
\includegraphics[angle=0,width=80mm]{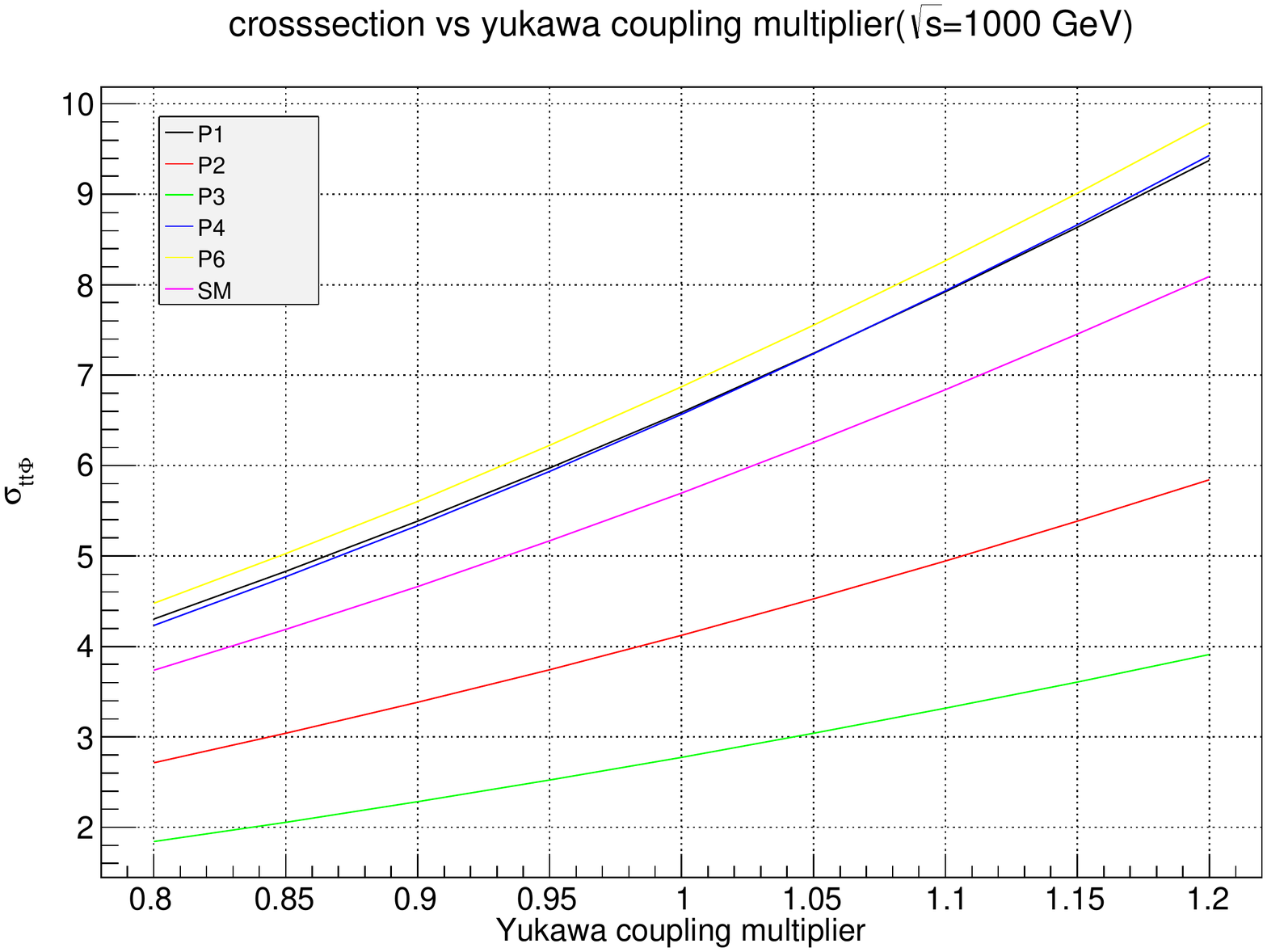}\\
\end{tabular}
%\vspace*{-7cm}
\caption{$\sqrt{s}=1000$ GeV: $\sigma_{t\bar{t}\Phi}$ vs Yukawa multiplier for different parameter values in Model I(left fig.) and Model II(Right fig.). }
\label{fig1L.2}
\end{figure}

Like the previous case, in Fig.~\ref{fig1L.2} we present $\sigma_{t\bar{t}\Phi}$ vs Yukawa coupling multiplier
plots for different considered parameter points. The corresponding quadratic equations are given below:

\begin{eqnarray}
b=0.0: && \hspace{0.3cm}\sigma_{t\bar{t}\Phi} = 5.308 (\lambda_t-1)^2 + 10.884(\lambda_t-1)+ 5.698,\nonumber\\
b=0.1: && \hspace{0.3cm}\sigma_{t\bar{t}\Phi} = 5.370 (\lambda_t-1)^2 + 10.800(\lambda_t-1)+ 5.653,\nonumber\\
b=0.3: && \hspace{0.3cm}\sigma_{t\bar{t}\Phi} = 4.958 (\lambda_t-1)^2 + 10.123(\lambda_t-1)+ 5.293,\nonumber\\
b=0.5: && \hspace{0.3cm}\sigma_{t\bar{t}\Phi} = 4.346 (\lambda_t-1)^2 + 8.767(\lambda_t-1)+ 4.575,\nonumber\\
b=0.7: && \hspace{0.3cm}\sigma_{t\bar{t}\Phi} = 3.344 (\lambda_t-1)^2 + 6.733(\lambda_t-1)+ 3.496
\end{eqnarray}

and

\begin{eqnarray}
P1: && \hspace{0.3cm}\sigma_{t\bar{t}\Phi} = 6.250 (\lambda_t-1)^2 + 12.680(\lambda_t-1)+ 6.590,\nonumber\\
P2: && \hspace{0.3cm}\sigma_{t\bar{t}\Phi} = 3.836 (\lambda_t-1)^2 + 7.822(\lambda_t-1)+ 4.123,\nonumber\\
P3: && \hspace{0.3cm}\sigma_{t\bar{t}\Phi} = 2.574 (\lambda_t-1)^2 + 5.167(\lambda_t-1)+ 2.772,\nonumber\\
P4: && \hspace{0.3cm}\sigma_{t\bar{t}\Phi} = 6.637 (\lambda_t-1)^2 + 12.986(\lambda_t-1)+ 6.568,\nonumber\\
P6: && \hspace{0.3cm}\sigma_{t\bar{t}\Phi} = 6.589 (\lambda_t-1)^2 + 13.286(\lambda_t-1)+ 6.869
\end{eqnarray}

Here the SM value of prefactor is 0.52 which means the 3rd diagram contributes  4\% in the total cross section. 
In Model I, the prefactor decreases very slightly with the increase of CP mixing and its value becomes 0.519 for $b=0.7$.
However the prefactor changes significantly in Model II and its value is 0.53, 0.50 and 0.51 for P3, P4 and P6
respectively. For P1 and P2 it remains close to the SM value of 0.52.

The  introduction of various kinematical cuts~\cite{LCtopyuk1000} for minimizing the background results in the reduction of signal events by a 
factor of 0.391 and 0.151 in the hadronic and semileptonic mode respectively. 
The corresponding top Yukawa sensitivity($g_t$) with associated signal significance is given in Table~\ref{1000gtM1} and Table~\ref{1000gtM2} for
Model I and Model II respectively. As observed from these tables, this case provides the best sensitivity for measuring the top Yukawa coupling. It 
gives coupling sensitivity of around 7.4\% and 9.9\% for the hadronic and semileptonic mode in the SM. With non zero CP violation, the sensitivity is further dropped 
to 11.0\% and 14.4\% for largest mixing case($b=0.7$) in Model I. For Model II, the best sensitivity of around 5.7\% is obtained in the hadronic mode
for P6 while the worst case is for P3 where it reaches about 19.9\% in the semileptonic mode.

\begin{table} 
\begin{center}
\begin{tabular}{lccccccc}

\hline

\hline

$b$ &Prefactor& Signal($S_1$) & $\frac{\Delta\sigma}{\sigma}$ & $\frac{\Delta g_t}{g_t}$& Signal($S_2$) & $\frac{\Delta\sigma}{\sigma}$ & $\frac{\Delta g_t}{g_t}$ \\

\hline%---------------------------------------------------------------------

0. &0.523& 245.4 & 0.14 & 0.074 & 91.3 & 0.19 & 0.099\\

\hline%-------------------------------------------------------

0.1 &0.523& 243.8 & 0.14 & 0.075 & 90.7& 0.19 & 0.100\\

\hline%---------------------------------------------------------------------

0.3  &0.522 & 231.2 & 0.15 & 0.079   & 86.0& 0.20& 0.104\\

\hline%---------------------------------------------------------------------

0.5  &0.521& 203.6  & 0.17 & 0.088   & 75.8& 0.22 & 0.117\\

\hline%---------------------------------------------------------------------

0.7  &0.519& 160.2  & 0.21 & 0.110   & 59.6 & 0.27& 0.144\\

\hline%---------------------------------------------------------------------

\end{tabular}
\end{center}

\caption{Model I: Yukawa coupling sensitivity for different parameters at $\sqrt{s}=1000$ GeV. $S_1$ and $S_2$ are
          signal events in hadronic and semileptonic mode after kinematical cuts. }

\label{1000gtM1} 
\end{table}

\begin{table} 
\begin{center}
\begin{tabular}{lccccccc}

\hline

\hline

Parameter & Prefactor & Signal($S_1$) & $\frac{\Delta\sigma}{\sigma}$ & $\frac{\Delta g_t}{g_t}$& Signal($S_2$) & $\frac{\Delta\sigma}{\sigma}$ & $\frac{\Delta g_t}{g_t}$ \\

\hline%---------------------------------------------------------------------

P1 & 0.519 & 148.4 & 0.22 & 0.118 & 55.2 & 0.29 & 0.154\\

\hline%-------------------------------------------------------

P2  & 0.527 &186.8 & 0.18 & 0.096  &69.5 & 0.24 & 0.126\\

\hline%---------------------------------------------------------------------

P3 & 0.536 & 118.3 & 0.28 & 0.152   & 44.0 & 0.37 & 0.196\\

\hline%---------------------------------------------------------------------

P4  & 0.505 & 313.9 & 0.11 & 0.058  & 116.8& 0.15 & 0.078\\

\hline%---------------------------------------------------------------------

P6  & 0.517 & 326.2  & 0.11 & 0.057  & 121.4 & 0.15 & 0.078 \\

\hline%---------------------------------------------------------------------

\end{tabular}
\end{center}

\caption{Model II: Yukawa coupling sensitivity for different parameters at $\sqrt{s}=1000$ GeV. $S_1$ and $S_2$ are
          signal events in hadronic and semileptonic mode after kinematical cuts.}

\label{1000gtM2} 
\end{table}

\section{Summary and Conclusions}\label{discussion}
The discovery of the 125 GeV scalar resonance at the LHC opens the way for studying its properties in great detail.
Being a hadron collider, LHC might not fulfill this task completely which is likely to be followed by a linear collider.  On the other hand, the
ILC, is perceived as a precision machine, which will be crucial in establishing the properties of this new degree of freedom, at a very high precision. While the mass and branching ratio measurements of the new resonance, and the spin and parity studies so far, indicate an SM-like Higg boson, the verdict is yet to come in this regard. Moreover, the SM Higgs mechanism is marred with difficulties like the hierarchy problem, which require inputs from beyond the SM. Precise measurement of the Higgs couplings with all the particles will be the key to understand what kind of new physics is in store.  Top quark being the heaviest state, couples strongly to the Higgs boson, making a detailed and precise study of the $t\bar t\Phi$ coupling essential to help establish the electroweak sector of SM at the energies explored, and at the same time to provide hints of new physics beyond the electroweak energies. 

In this work, we have studied the measurement of a general $t\bar t\Phi$ coupling, including CP-violating anomalous effects. Such couplings naturally arise in many
extensions of the SM, like the 2HDM with CP violation, where Higgs can be a CP mixture of scalar and pseudoscalar components. We note that, although pure CP-odd sate is ruled out 
 by the LHC data, the possibility of a mixed CP state is still a viable option, despite perhaps being small.

The issue of the measurement of the Yukawa coupling and the sensitivity achievable at the ILC at design energies of 500, 800 and 1000 GeV have been the subject of recent studies.  In all these, 
only a SM Higgs was assumed and it was shown that with typical luminosities of 1000 fb$^{-1}$, it was possible to achieve sensitivities typically in 
the range of 10\%.  At 500 GeV the issue of bound state effects was also studied, since the available kinetic energy for the final state particles 
is very small.  In the present work, we have considered the implication of departing from the SM hypothesis for the Higgs boson.  
We have considerd a generalized coupling, and studied the effect on the Yukawa coupling measurements.

It is also important to ask 
whether the methodology is internally consistent or not. We considered distributions in detail in these models.  Our study is 
validated by the features of these distributions, which is that kinematical cuts affects signal events in the same way for different CP parameters which is clear 
from the distributions and we have used the same reduction factor as for the SM case which has been studied in the literature.

The main conclusion is that the sensitivity worsens as the departure from the SM grows, partly due to the falling signal cross section, and partly due to different functional dependence of the cross section on the coupling. Assuming the same strategy adopted in the previous studies of the SM case, 
we find that the measurement of the top Yukawa coupling is possible down to about 20\% for 500 GeV 
energy and 1000 fb$^{-1}$ integrated luminosity in the CP-violating scenario. The case of 800 GeV gives a picture of both improvement, as well as worsening of the sensitivity compared to the case of SM, depending on the values of the parameters $a$ and $b$, ranging between 7\% and 22\% for different benchmark points considered. The higher energies of about 1000 GeV, shows improved significance in general ranging between 5.7\% and about 20\%, where the worst case corresponding to the benchmark point P3, with smaller value of the parameter $a$. To conclude, this first study demonstrates the need for more detailed investigations on the impact of BSM physics on the measurements of the properties of particles at the ILC.

\noindent {\bf{Acknowledgements:}} \\
We thank the authors of {\sc WHIZARD} and the FeynRules interface, 
especially J. Reuter and C. Speckner, for very helpful discussions
regarding the implementation of our models. The work of SKG
and CSK is supported 
by the National Research Foundation of Korea(NRF) grant funded
by Korea government of the Ministry of Education, Science and Technology(MEST)
(Grant
No. 2011-0017430 and Grant No. 2011-0020333). PP acknowledges the support of BRNS, DAE, Government of India (Project No.:
2010/37P/49/BRNS/1446).


\bigskip


\begin{thebibliography}{99}

%\cite{:2012gu}
\bibitem{cms} 
  S.~Chatrchyan {\it et al.}  [CMS Collaboration],
  %``Observation of a new Boson at a mass of 125 GeV with the 
%CMS experiment at the LHC,''
  Phys.\ Lett.\ B {\bf 716}, 30 (2012)
  [arXiv:1207.7235 [hep-ex]].
  %%CITATION = ARXIV:1207.7235;%%

\bibitem{atlas} 
  G.~Aad {\it et al.}  [ATLAS Collaboration],
  %``Observation of a new particle in the search for the 
%Standard Model Higgs Boson with the ATLAS detector at the LHC,''
  Phys.\ Lett.\ B {\bf 716}, 1 (2012)
  [arXiv:1207.7214 [hep-ex]].
  %%CITATION = ARXIV:1207.7214;%%
  

\bibitem{Moroind1}
The Atlas Collaboration, ATLAS-CONF-2013-029,\,
http://cds.cern.ch/record/1527124/files/ATLAS-CONF-2013-029.pdf

\bibitem{Moroind2}
The Atlas Collaboration, ATLAS-CONF-2013-013,\,
http://cds.cern.ch/record/1523699/files/ATLAS-CONF-2013-013.pdf


\bibitem{Moroind3}
The Atlas Collaboration, ATLAS-CONF-2013-031,\,
http://cds.cern.ch/record/1527127/files/ATLAS-CONF-2013-031.pdf

\bibitem{Moroind4}
The CMS Collaboration, HIG-13-002-pas,\,
http://cds.cern.ch/record/1523767/files/HIG-13-002-pas.pdf

\bibitem{Moroind5}
The CMS Collaboration, HIG-13-003-pas,\,
http://cds.cern.ch/record/1523673/files/HIG-13-003-pas.pdf   



%\cite{Maltoni:2002jr}
\bibitem{Maltoni:2002jr} 
  F.~Maltoni, D.~L.~Rainwater and S.~Willenbrock,
  %``Measuring the top quark Yukawa coupling at hadron colliders via $t\bar{t}H,H\to W^+W^-$,''
  Phys.\ Rev.\ D {\bf 66}, 034022 (2002)
  [hep-ph/0202205].
  %%CITATION = HEP-PH/0202205;%%
  %54 citations counted in INSPIRE as of 29 Jul 2013

%\cite{Ellis:2013yxa}
\bibitem{Ellis:2013yxa}
  J.~Ellis, D.~S.~Hwang, K.~Sakurai and M.~Takeuchi,
  %``Disentangling Higgs-Top Couplings in Associated Production,''
  JHEP {\bf 1404} (2014) 004
  [arXiv:1312.5736 [hep-ph]].
  %%CITATION = ARXIV:1312.5736;%%
  %9 citations counted in INSPIRE as of 20 May 2014

%\cite{Chang:2014rfa}
\bibitem{Chang:2014rfa}
  J.~Chang, K.~Cheung, J.~S.~Lee and C.~-T.~Lu,
  %``Probing the Top-Yukawa Coupling in Associated Higgs production with a Single Top Quark,''
  arXiv:1403.2053 [hep-ph].
  %%CITATION = ARXIV:1403.2053;%%
  %1 citations counted in INSPIRE as of 20 May 2014

 %\cite{Nishiwaki:2013cma}
\bibitem{Nishiwaki:2013cma}
  K.~Nishiwaki, S.~Niyogi and A.~Shivaji,
  %``$ttH$ Anomalous Coupling in Double Higgs Production,''
  JHEP {\bf 1404} (2014) 011
  [arXiv:1309.6907 [hep-ph]].
  %%CITATION = ARXIV:1309.6907;%%
  %4 citations counted in INSPIRE as of 20 May 2014

%\cite{BrauJames:2007aa}
\bibitem{ILC1} 
  J.~Brau, (Ed.) {\it et al.}  [ILC Collaboration],
  %``ILC Reference Design Report: ILC Global Design Effort and 
%World Wide Study,''
  arXiv:0712.1950 [physics.acc-ph].
  %%CITATION = ARXIV:0712.1950;%%

%\cite{Djouadi:2007ik}
\bibitem{ILC2} 
  G.~Aarons {\it et al.}  [ILC Collaboration],
  %``International Linear Collider Reference Design Report Volume 2: 
%Physics At The Ilc,''
  arXiv:0709.1893 [hep-ph].
  %%CITATION = ARXIV:0709.1893;%%
  


%\cite{MoortgatPick:2005cw}
\bibitem{polarizationreview} 
  G.~Moortgat-Pick, T.~Abe, G.~Alexander, B.~Ananthanarayan, 
A.~A.~Babich, V.~Bharadwaj, D.~Barber and A.~Bartl {\it et al.},
  %``The Role of polarized positrons and electrons in revealing fundamental 
%interactions at the linear collider,''
  Phys.\ Rept.\  {\bf 460}, 131 (2008)
  [hep-ph/0507011].
  %%CITATION = HEP-PH/0507011;%%
  
  
  %\cite{Baer:1999ge}
\bibitem{Baer:1999ge} 
  H.~Baer, S.~Dawson and L.~Reina,
  %``Measuring the top quark Yukawa coupling at a linear e+ e- collider,''
  Phys.\ Rev.\ D {\bf 61}, 013002 (2000)
  [hep-ph/9906419].
  %%CITATION = HEP-PH/9906419;%%
  %75 citations counted in INSPIRE as of 04 Apr 2014
  
      
\bibitem{AG1}
  A.~Gay,
  ``Measurement of the top-Higgs Yukawa coupling at a Linear e+ e- Collider,''
LC-PHSM-2006-002 (Linear Collider Note)

%\cite{Gay:2006vs}
\bibitem{AG2}
  A.~Gay,
  %``Measurement of the top-Higgs Yukawa coupling at a Linear e+ e- Collider,''
  Eur.\ Phys.\ J.\ C {\bf 49}, 489 (2007)
  [hep-ph/0604034].
  %%CITATION = HEP-PH/0604034;%%
  
  
%\cite{Kolodziej:2009cx}
\bibitem{KS1} 
  K.~Kolodziej and S.~Szczypinski,
  %``e+ e- ---> t anti-t H including decays: 
%On the size of background contributions,''
  Eur.\ Phys.\ J.\ C {\bf 64}, 645 (2009)
  [arXiv:0903.4606 [hep-ph]].
  %%CITATION = ARXIV:0903.4606;%%

%\cite{Kolodziej:2009bj}
\bibitem{KS2} 
  K.~Kolodziej and S.~Szczypinski,
  %``Signal and Background in e+ e- ---> t anti-t H,''
  Acta Phys.\ Polon.\ B {\bf 40}, 3015 (2009)
  [arXiv:0911.1085 [hep-ph]].
  %%CITATION = ARXIV:0911.1085;%%

%\cite{Yonamine:2010su}
\bibitem{YIUF} 
  R.~Yonamine, K.~Ikematsu, S.~Uozumi and K.~Fujii,
  %``A Study of top-quark Yukawa coupling measurement in $e^+e^- -> t \bar{t} H$ at sqrt(s) = 500 GeV,''
  arXiv:1008.1110 [hep-ex].
  %%CITATION = ARXIV:1008.1110;%%

%\cite{Yonamine:2011jg}
\bibitem{YITFKSY} 
  R.~Yonamine, K.~Ikematsu, T.~Tanabe, K.~Fujii, Y.~Kiyo, Y.~Sumino and H.~Yokoya,
  %``Measuring the top Yukawa coupling at the ILC at sqrt(s) = 500 GeV,''
  Phys.\ Rev.\ D {\bf 84}, 014033 (2011)
  [arXiv:1104.5132 [hep-ph]].
  %%CITATION = ARXIV:1104.5132;%% 

%\cite{Tabassam:2012it}
\bibitem{TM} 
  H.~Tabassam, V.~Martin,
  %``Top Higgs Yukawa Coupling Analysis from e+e- -> ttH -> bW bW bb,''
  arXiv:1202.6013 [hep-ex].
  %%CITATION = ARXIV:1202.6013;%%
 
    
%\cite{Gay:2006vs}
\bibitem{LCtopyuk1000}
  T.~Price, T.~Tanabe etal
  %``Measurement of the top-Higgs Yukawa coupling at a Linear e+ e- Collider,''
  LC-REP-2013-004.
  %%CITATION = HEP-PH/0604034;%%   


%\cite{Han:1999xd}
\bibitem{Han:1999xd} 
  T.~Han, T.~Huang, Z.~H.~Lin, J.~X.~Wang and X.~Zhang,
  %``e+ e- ---> t anti-t H with nonstandard Higgs boson couplings,''
  Phys.\ Rev.\ D {\bf 61}, 015006 (2000)
  [hep-ph/9908236].
  %%CITATION = HEP-PH/9908236;%%
  %22 citations counted in INSPIRE as of 29 Jul 2013
  
  
%\cite{Atwood:2000tu}
\bibitem{Atwood:2000tu} 
  D.~Atwood, S.~Bar-Shalom, G.~Eilam and A.~Soni,
  %``CP violation in top physics,''
  Phys.\ Rept.\  {\bf 347}, 1 (2001)
  [hep-ph/0006032].
  %%CITATION = HEP-PH/0006032;%%
  %120 citations counted in INSPIRE as of 29 Jul 2013 

 

%\cite{El Kaffas:2006nt}
\bibitem{2HDMcpv} 
  A.~W.~El Kaffas, W.~Khater, O.~M.~Ogreid and P.~Osland,
  %``Consistency of the two Higgs doublet model and CP violation in top production at the LHC,''
  Nucl.\ Phys.\ B {\bf 775}, 45 (2007)
  [hep-ph/0605142];
  %%CITATION = HEP-PH/0605142;%%
  %32 citations counted in INSPIRE as of 25 Feb 2013
%\cite{Khater:2003wq}
%\bibitem{Khater:2003wq} 
  W.~Khater and P.~Osland,
  %``CP violation in top quark production at the LHC and two Higgs doublet models,''
  Nucl.\ Phys.\ B {\bf 661}, 209 (2003)
  [hep-ph/0302004];
  %%CITATION = HEP-PH/0302004;%%
  %47 citations counted in INSPIRE as of 25 Feb 2013
%\cite{Barroso:2012wz}
%\bibitem{Barroso:2012wz} 
  A.~Barroso, P.~M.~Ferreira, R.~Santos and J.~P.~Silva,
  %``Probing the scalar-pseudoscalar mixing in the 125 GeV Higgs particle with current data,''
  Phys.\ Rev.\ D {\bf 86}, 015022 (2012)
  [arXiv:1205.4247 [hep-ph]];
  %%CITATION = ARXIV:1205.4247;%%
  %17 citations counted in INSPIRE as of 25 Feb 2013
 %\cite{Ellis:2005ika}
%\bibitem{Ellis:2005ika} 
  J.~R.~Ellis, J.~S.~Lee and A.~Pilaftsis,
  %``Resonant CP violation in Higgs radiation at e+ e- linear collider,''
  Phys.\ Rev.\ D {\bf 72}, 095006 (2005)
  [hep-ph/0507046].
  %%CITATION = HEP-PH/0507046;%%
  %29 citations counted in INSPIRE as of 25 Feb 2013 
  
\bibitem{MSSMcpv} 
  S.~Y.~Choi and J.~S.~Lee,
  %``Decays of the MSSM Higgs bosons with explicit CP violation,''
  Phys.\ Rev.\ D {\bf 61}, 015003 (1999)
  [hep-ph/9907496];
  %%CITATION = HEP-PH/9907496;%%
  %72 citations counted in INSPIRE as of 12 Apr 2013
  S.~Hesselbach, S.~Moretti, S.~Munir and P.~Poulose,
  %``Explicit CP violation in the MSSM through gg --> H(1) --> gamma gamma,''
  Phys.\ Rev.\ D {\bf 82}, 074004 (2010)
  [arXiv:0903.0747 [hep-ph]];
  %%CITATION = ARXIV:0903.0747;%%
  %5 citations counted in INSPIRE as of 12 Apr 2013   
  S.~Hesselbach, S.~Moretti, S.~Munir and P.~Poulose,
  %``Explicit CP violation in the MSSM Higgs sector,''
  AIP Conf.\ Proc.\  {\bf 1200}, 498 (2010)
  [arXiv:0910.0230 [hep-ph]].
  %%CITATION = ARXIV:0910.0230;%%
  %4 citations counted in INSPIRE as of 12 Apr 2013 
  

\bibitem{MSSMcpvPP} 
  A.~Chakraborty, B.~Das, J.~L.~Diaz-Cruz, D.~K.~Ghosh, S.~Moretti and P.~Poulose,
  %``The 125 GeV Higgs signal at the LHC in the CP Violating MSSM,''
  arXiv:1301.2745 [hep-ph].
  %%CITATION = ARXIV:1301.2745;%%
  %2 citations counted in INSPIRE as of 12 Apr 2013   
  
  
  
%\cite{Carena:2001bg}
\bibitem{Carena:2001bg} 
  M.~S.~Carena, H.~E.~Haber, H.~E.~Logan and S.~Mrenna,
  %``Distinguishing a MSSM Higgs boson from the SM Higgs boson at a linear collider,''
  Phys.\ Rev.\ D {\bf 65}, 055005 (2002)
  [Erratum-ibid.\ D {\bf 65}, 099902 (2002)]
  [hep-ph/0106116].
  %%CITATION = HEP-PH/0106116;%%
  %81 citations counted in INSPIRE as of 29 Jul 2013
  

\bibitem{DDGMR} 
  P.~S.~Bhupal Dev, A.~Djouadi, R.~M.~Godbole, M.~M.~Muhlleitner 
and S.~D.~Rindani,
  %``Determining the CP properties of the Higgs Boson,''
  Phys.\ Rev.\ Lett.\  {\bf 100}, 051801 (2008)
  [arXiv:0707.2878 [hep-ph]].
  %%CITATION = ARXIV:0707.2878;%% 
  
  
  

%\cite{Godbole:2011hw}
\bibitem{GHMRS} 
  R.~M.~Godbole, C.~Hangst, M.~Muhlleitner, S.~D.~Rindani and P.~Sharma,
  %``Model-independent analysis of Higgs spin and CP properties 
%in the process $e^+ e^- \to t \bar t \Phi$,''
  Eur.\ Phys.\ J.\ C {\bf 71}, 1681 (2011)
  [arXiv:1103.5404 [hep-ph]].
  %%CITATION = ARXIV:1103.5404;%%

%\cite{Berge:2012wm}
\bibitem{BBS}
  S.~Berge, W.~Bernreuther and H.~Spiesberger,
  %``Determination of the CP parity of Higgs Bosons in their tau decay channels at the ILC,''
  arXiv:1208.1507 [hep-ph].
  %%CITATION = ARXIV:1208.1507;%%

%\cite{Han:2000mi}
\bibitem{HJ} 
  T.~Han and J.~Jiang,
  %``CP violating Z Z H coupling at e+ e- linear colliders,''
  Phys.\ Rev.\ D {\bf 63}, 096007 (2001)
  [hep-ph/0011271].
  %%CITATION = HEP-PH/0011271;%%

%\cite{Rao:2007ce}
\bibitem{RR} 
  K.~Rao and S.~D.~Rindani,
  %``Charged lepton distributions as a probe of contact e+e-HZ interactions at a linear collider with polarized beams,''
  Phys.\ Rev.\ D {\bf 77}, 015009 (2008)
  [Erratum-ibid.\ D {\bf 80}, 019901 (2009)]
  [arXiv:0709.2591 [hep-ph]].
  %%CITATION = ARXIV:0709.2591;%%
  
  
%\cite{AguilarSaavedra:2009mx}
\bibitem{AguilarSaavedra} 
  J.~A.~Aguilar-Saavedra,
  %``A Minimal set of top-Higgs anomalous couplings,''
  Nucl.\ Phys.\ B {\bf 821}, 215 (2009)
  [arXiv:0904.2387 [hep-ph]].
  %%CITATION = ARXIV:0904.2387;%%
  %32 citations counted in INSPIRE as of 08 Mar 2013
  
%\cite{Ananthanarayan:2013cia}
\bibitem{Ananthanarayan:2013cia} 
  B.~Ananthanarayan, S.~K.~Garg, J.~Lahiri and P.~Poulose,
  %``Probing the indefinite CP nature of the Higgs boson through decay distributions in the process $e^+e^- → t\overline{t}Φ$,''
  Phys.\ Rev.\ D {\bf 87}, no. 11, 114002 (2013)
  [arXiv:1304.4414 [hep-ph]].
  %%CITATION = ARXIV:1304.4414;%%
  %1 citations counted in INSPIRE as of 04 Apr 2014  
  
  

%\cite{Kilian:2007gr}
\bibitem{WHIZARD} 
  W.~Kilian, T.~Ohl and J.~Reuter,
  %``WHIZARD: Simulating Multi-Particle Processes at LHC and ILC,''
  Eur.\ Phys.\ J.\ C {\bf 71}, 1742 (2011)
  [arXiv:0708.4233 [hep-ph]].
  %%CITATION = ARXIV:0708.4233;%%
  
 %\cite{Christensen:2010wz}
\bibitem{Interface} 
  N.~D.~Christensen, C.~Duhr, B.~Fuks, J.~Reuter and C.~Speckner,
  %``Introducing an interface between WHIZARD and FeynRules,''
  Eur.\ Phys.\ J.\ C {\bf 72}, 1990 (2012)
  [arXiv:1010.3251 [hep-ph]].
  %%CITATION = ARXIV:1010.3251;%%
  %16 citations counted in INSPIRE as of 24 Feb 2013 
  
 %\cite{Christensen:2008py}
\bibitem{FeynRules} 
  N.~D.~Christensen and C.~Duhr,
  %``FeynRules - Feynman rules made easy,''
  Comput.\ Phys.\ Commun.\  {\bf 180}, 1614 (2009)
  [arXiv:0806.4194 [hep-ph]].
  %%CITATION = ARXIV:0806.4194;%%
  %176 citations counted in INSPIRE as of 24 Feb 2013 

 
  
\bibitem{CPsuperH} 
  J.~S.~Lee, M.~Carena, J.~Ellis, A.~Pilaftsis and C.~E.~M.~Wagner,
  %``CPsuperH2.0: an Improved Computational Tool for Higgs Phenomenology in the MSSM with Explicit CP Violation,''
  Comput.\ Phys.\ Commun.\  {\bf 180}, 312 (2009)
  [arXiv:0712.2360 [hep-ph]].
  %%CITATION = ARXIV:0712.2360;%%
  %77 citations counted in INSPIRE as of 12 Apr 2013
  
 
%\cite{Shu:2013uua}
\bibitem{ZhangCPV} 
  J.~Shu and Y.~Zhang,
  %``Impact of a CP Violating Higgs Sector: From LHC to Baryogenesis,''
  Phys.\ Rev.\ Lett.\  {\bf 111}, no. 9, 091801 (2013)
  [arXiv:1304.0773 [hep-ph]].
  %%CITATION = ARXIV:1304.0773;%%
  %20 citations counted in INSPIRE as of 04 Apr 2014
  
%%%%%%%% CP violating 2HDM references .............



%\cite{Juste:1999af}
\bibitem{Juste} 
  A.~Juste and G.~Merino,
  %``Top Higgs-Yukawa coupling measurement at a linear e+ e- collider,''
  hep-ph/9910301.
  %%CITATION = HEP-PH/9910301;%%
  %72 citations counted in INSPIRE as of 29 Jul 2013
  
%\cite{Kolodziej:2006nr}
\bibitem{Kolodziej:2006nr} 
  K.~Kolodziej and S.~Szczypinski,
  %``Off mass shell effects in associated production of the top quark pair and Higgs boson at a linear collider,''
  Acta Phys.\ Polon.\ B {\bf 38}, 2565 (2007)
  [hep-ph/0612183].
  %%CITATION = HEP-PH/0612183;%%
  %1 citations counted in INSPIRE as of 29 Jul 2013

  %\cite{Brod:2013cka}
\bibitem{Brodetal} 
  J.~Brod, U.~Haisch and J.~Zupan,
  %``Constraints on CP-violating Higgs couplings to the third generation,''
  JHEP {\bf 1311}, 180 (2013)
  [arXiv:1310.1385 [hep-ph]
  %%CITATION = ARXIV:1310.1385;%%
  %6 citations counted in INSPIRE as of 04 Apr 2014oaa
  
%\cite{Cheung:2014oaa}
\bibitem{Higgcision} 
  K.~Cheung, J.~S.~Lee, E.~Senaha and P.~-Y.~Tseng,
  %``Confronting Higgcision with Electric Dipole Moments,''
  arXiv:1403.4775 [hep-ph].
  %%CITATION = ARXIV:1403.4775;%%

  
  
 
  
  


%%%%%%%%%%%%%%%%%%%%YUKAWA COUPLING MEASUREMENT................
 
  
  





\end{thebibliography}
\end{document}